\documentstyle{article}

\def\noheaderplainsetup{%

\topmargin=-10pt \headheight=0pt \headsep=0pt  \oddsidemargin=0pt \evensidemargin=0pt  \textheight=9.23truein \textwidth=6.5truein}

\noheaderplainsetup

\begin{document}


\newcommand{\lmt}{\ell}
\newcommand{\col}[1]{\mbox{$#1$:}}
\newcommand{\chess}{\mbox{\em Chess}}
\newcommand{\checkers}{\mbox{\em Checkers}}
\newcommand{\elzz}[2]{ \langle #1\rangle\hspace{-2pt} \downarrow\hspace{-2pt} #2} 
\newcommand{\elz}[1]{\mbox{$\parallel\hspace{-3pt} #1 \hspace{-3pt}\parallel$}} 
\newcommand{\qelz}[1]{\mbox{$| #1 |$}}
\newcommand{\emptyrun}{\langle\rangle} 
\newcommand{\oo}{\bot}            
\newcommand{\pp}{\top}            
\newcommand{\xx}{\wp}               
\newcommand{\win}[2]{\mbox{\bf Wn}^{#1}_{#2}} 
\newcommand{\seq}[1]{\langle #1 \rangle}           


\newcommand{\plus}{\mbox{\hspace{1pt}\raisebox{0.05cm}{\tiny\boldmath $+$}\hspace{1pt}}}
\newcommand{\mult}{\mbox{\hspace{1pt}\raisebox{0.05cm}{\tiny\boldmath $\times$}\hspace{1pt}}}
\newcommand{\mminus}{\mbox{\hspace{1pt}\raisebox{0.05cm}{\tiny\boldmath $-$}\hspace{1pt}}}
\newcommand{\equals}{\mbox{\hspace{1pt}\raisebox{0.05cm}{\tiny\boldmath $=$}\hspace{1pt}}}
\newcommand{\notequals}{\mbox{\hspace{1pt}\raisebox{0.05cm}{\tiny\boldmath $\not=$}\hspace{1pt}}}

\newcommand{\mless}{\mbox{\hspace{1pt}\raisebox{0.05cm}{\tiny\boldmath $<$}\hspace{1pt}}}
\newcommand{\mgreater}{\mbox{\hspace{1pt}\raisebox{0.05cm}{\tiny\boldmath $>$}\hspace{1pt}}}
\newcommand{\mleq}{\mbox{\hspace{1pt}\raisebox{0.05cm}{\tiny\boldmath $\leq$}\hspace{1pt}}}
\newcommand{\mgeq}{\mbox{\hspace{1pt}\raisebox{0.05cm}{\tiny\boldmath $\geq$}\hspace{1pt}}}

\newcommand{\clt}{\mbox{\bf CL13}}
\newcommand{\clte}{\mbox{\bf CL14}}
\newcommand{\cltee}{\overline{\clte}}
\newcommand{\intimpl}{\mbox{\hspace{2pt}$\circ$\hspace{-0.14cm} \raisebox{-0.043cm}{\Large --}\hspace{2pt}}}
\newcommand{\sintimpl}{\mbox{\hspace{2pt}\raisebox{0.033cm}{\tiny $ | \hspace{-4pt} >$}\hspace{-0.14cm} \raisebox{-0.039cm}{\large --}\hspace{2pt}}}
\newcommand{\ade}{\mbox{\Large $\sqcup$}\hspace{1pt}}      
\newcommand{\ada}{\mbox{\Large $\sqcap$}\hspace{1pt}}      
\newcommand{\sst}{\mbox{\raisebox{-0.07cm}{\scriptsize $-$}\hspace{-0.2cm}$\pst$}}
\newcommand{\scost}{\mbox{\raisebox{0.20cm}{\scriptsize $-$}\hspace{-0.2cm}$\pcost$}}
\newcommand{\sqc}{\mbox{\hspace{2pt}\small \raisebox{0.0cm}{$\bigtriangleup$}\hspace{2pt}}}
\newcommand{\sqci}{\mbox{\scriptsize \raisebox{0.0cm}{$\bigtriangleup$}}}
\newcommand{\sqd}{\mbox{\hspace{2pt}\small \raisebox{0.06cm}{$\bigtriangledown$}\hspace{2pt}}}
\newcommand{\sqdi}{\mbox{\scriptsize \raisebox{0.05cm}{$\bigtriangledown$}}}
\newcommand{\sqe}{\mbox{\large \raisebox{0.07cm}{$\bigtriangledown$}}}
\newcommand{\sqa}{\mbox{\large \raisebox{0.0cm}{$\bigtriangleup$}}}
\newcommand{\mld}{\vee}    
\newcommand{\mlc}{\wedge}  
\newcommand{\tgd}{\mbox{\hspace{2pt}$\vee$\hspace{-1.29mm}\raisebox{0.1mm}{\rule{0.13mm}{2mm}}\hspace{5pt}}}    
\newcommand{\tgc}{\mbox{\hspace{2pt}$\wedge$\hspace{-1.29mm}\raisebox{0.02mm}{\rule{0.13mm}{2mm}}\hspace{5pt}}}    
\newcommand{\tge}{\hspace{1pt}\mbox{\Large $\vee$\hspace{-1.84mm}\raisebox{0.1mm}{\rule{0.13mm}{3.0mm}}\hspace{6pt}}}   
\newcommand{\tga}{\mbox{\hspace{1pt}\Large $\wedge$\hspace{-1.84mm}\raisebox{0.02mm}{\rule{0.13mm}{3.0mm}}\hspace{6pt}}}     
\newcommand{\tgpst}{\mbox{\raisebox{-0.01cm}{\scriptsize $\wedge$}\hspace{-4pt}\raisebox{0.06cm}{\small $\mid$}\hspace{2pt}}}
\newcommand{\tgpcost}{\mbox{\raisebox{0.12cm}{\scriptsize $\vee$}\hspace{-3.8pt}\raisebox{0.04cm}{\small $\mid$}\hspace{2pt}}}
\newcommand{\tgst}{\mbox{\raisebox{-0.05cm}{$\circ$}\hspace{-0.12cm}\raisebox{0.05cm}{\small $\mid$}\hspace{2pt}}} 
\newcommand{\tgcost}{\mbox{\raisebox{0.12cm}{$\circ$}\hspace{-0.12cm}\raisebox{0.04cm}{\small $\mid$}\hspace{2pt}}}
\newcommand{\tgpi}{\mbox{\hspace{2pt}\raisebox{0.033cm}{\tiny $>$}\hspace{-0.28cm} \raisebox{-2.3pt}{\LARGE --}\hspace{2pt}}}
\newcommand{\tgbi}{\mbox{\hspace{2pt}$\circ$\hspace{-0.26cm} \raisebox{-2.3pt}{\LARGE --}\hspace{2pt}}}
\newcommand{\mle}{\mbox{\hspace{1pt}\Large $\vee$}\hspace{1pt}}    
\newcommand{\mla}{\mbox{\hspace{1pt}\Large $\wedge$}\hspace{1pt}}  
\newcommand{\add}{\hspace{0pt}\sqcup}                      
\newcommand{\adc}{\hspace{0pt}\sqcap}                      
\newcommand{\gneg}{\neg}                  
\newcommand{\rneg}{\neg}               
\newcommand{\pneg}{\neg}               
\newcommand{\mli}{\rightarrow}                     
\newcommand{\intf}{\$}               
\newcommand{\tlg}{\bot}               
\newcommand{\twg}{\top}               

\newcommand{\pst}{\mbox{\raisebox{-0.01cm}{\scriptsize $\wedge$}\hspace{-4pt}\raisebox{0.16cm}{\tiny $\mid$}\hspace{2pt}}}
\newcommand{\cla}{\mbox{\large $\forall$}\hspace{1pt}}      
\newcommand{\cle}{\mbox{\large $\exists$}\hspace{1pt}}        
\newcommand{\pintimpl}{\mbox{\hspace{2pt}\raisebox{0.033cm}{\tiny $>$}\hspace{-0.18cm} \raisebox{-0.043cm}{\large --}\hspace{2pt}}}
\newcommand{\pcost}{\mbox{\raisebox{0.12cm}{\scriptsize $\vee$}\hspace{-4pt}\raisebox{0.02cm}{\tiny $\mid$}\hspace{2pt}}}
\newcommand{\st}{\mbox{\raisebox{-0.05cm}{$\circ$}\hspace{-0.13cm}\raisebox{0.16cm}{\tiny $\mid$}\hspace{2pt}}}
\newcommand{\cost}{\mbox{\raisebox{0.12cm}{$\circ$}\hspace{-0.13cm}\raisebox{0.02cm}{\tiny $\mid$}\hspace{2pt}}}


\newtheorem{theoremm}{Theorem}[section]
\newtheorem{conjecturee}[theoremm]{Conjecture}
\newtheorem{exercisee}[theoremm]{Exercise}
\newtheorem{definitionn}[theoremm]{Definition}
\newtheorem{lemmaa}[theoremm]{Lemma}
\newtheorem{propositionn}[theoremm]{Proposition}
\newtheorem{conventionn}[theoremm]{Convention}
\newtheorem{examplee}[theoremm]{Example}
\newtheorem{remarkk}[theoremm]{Remark}
\newtheorem{factt}[theoremm]{Fact}
\newtheorem{claimm}[theoremm]{Claim}

\newenvironment{conjecture}{\begin{conjecturee}}{\end{conjecturee}}
\newenvironment{definition}{\begin{definitionn} \em}{ \end{definitionn}}
\newenvironment{theorem}{\begin{theoremm}}{\end{theoremm}}
\newenvironment{lemma}{\begin{lemmaa}}{\end{lemmaa}}
\newenvironment{proposition}{\begin{propositionn} }{\end{propositionn}}
\newenvironment{convention}{\begin{conventionn} \em}{\end{conventionn}}
\newenvironment{remark}{\begin{remarkk} \em}{\end{remarkk}}
\newenvironment{proof}{ {\bf Proof.} }{\  \rule{2mm}{2mm} \vspace{.15in} }
\newenvironment{example}{\begin{examplee} \em}{\end{examplee}}
\newenvironment{exercise}{\begin{exercisee} \em}{\end{exercisee}}
\newenvironment{fact}{\begin{factt} \em}{\end{factt}}
\newenvironment{claim}{\begin{claimm} \em}{\end{claimm}}

\title{Toggling operators in computability logic}
\author{Giorgi Japaridze\thanks{This material is based upon work supported by the National Science Foundation under Grant No. 0208816}\\ \\ 
{\small School of Computer Science and Technology, Shandong University;}\\
{\small Department of Computing Sciences, Villanova University}}
\date{}
\maketitle
\begin{abstract} 
{\em Computability logic} (CoL) is a recently introduced semantical platform and research program for redeveloping logic as a formal theory of computability, as opposed to the formal theory of truth which it has more traditionally been. Formulas in CoL stand for interactive computational problems, seen as games between a machine and its environment; logical operators represent operations on such entities; and ``truth'' is understood as existence of an effective solution, i.e.,  of an algorithmic  winning strategy. 

The formalism of CoL is open-ended, and may undergo series of extensions as the studies of the subject advance. Propositional connectives and quantifiers in it  come in a variety of indispensable versions. So far three  sorts of conjunction  and disjunction --- {\em parallel}, {\em sequential} and {\em choice} ---  have been studied, with the first and the third sorts being reminiscent of the multiplicative and additive operators of linear logic, respectively. The present paper adds one more natural kind to this collection, termed {\em toggling}. The toggling operations can be characterized as lenient versions of choice operations where  choices are retractable, being allowed to be  reconsidered any finite number of times. This way, they model {\em trial-and-error} style decision steps in interactive computation. The main technical result of this paper is constructing a sound and complete axiomatization for the propositional fragment of computability logic whose vocabulary includes all four  kinds of conjunction and disjunction: parallel, toggling, sequential and choice, together with negation. Along with toggling conjunction and disjunction,  the paper also introduces the toggling versions of quantifiers and recurrence (``exponential'') operations.  

\end{abstract}

\noindent {\em MSC}: primary: 03B47; secondary: 03F50; 03B70; 68Q10; 68T27; 68T30; 91A05 

\

\noindent {\em Keywords}: Computability logic; Game semantics; Interactive computation; Constructive logics; Resource logics

\section{Introduction}\label{sintr}

{\em Computability logic} (CoL), introduced in \cite{Jap03,Japic,Japfin}, is a semantical, mathematical and philosophical platform, and an ambitious program, for redeveloping logic as a formal theory of computability, as opposed to the formal theory of truth which logic has more traditionally been. 

 Under the approach of CoL, formulas represent computational problems, 
and their ``truth'' is seen as algorithmic solvability. In turn, computational problems --- understood in their  most general, {\em interactive} sense --- are defined as games played by a machine against its environment, with ``algorithmic solvability'' meaning existence of a machine that wins the game against any possible behavior of the environment. And a collection of the most basic and natural operations 
on interactive computational problems forms the logical vocabulary of the theory.  With this semantics, CoL provides a systematic answer to the fundamental question ``{\em what can be computed?}\hspace{1pt}'', just as classical logic is a systematic tool for telling what is true. Furthermore, as it turns out, in positive cases ``{\em what} can be computed'' always allows itself to be replaced by ``{\em how} can be computed'', which makes CoL of potential interest in not only theoretical computer science, but many more applied areas as well, including interactive knowledge base systems, resource oriented systems for planning and action, or declarative programming languages. 

On the logical side, CoL promises to be an appealing, constructive and computationally meaningful alternative to classical logic as a basis for applied theories. The first concrete steps towards realizing this potential have been made very recently in \cite{Japtowards,ptarithmetic}, where the CoL-based versions {\bf CLA1} and {\bf CLA4} of Peano arithmetic were elaborated. All theorems of the former express  number-theoretic computational problems with algorithmic solutions, and all theorems of the latter express number-theoretic computational problems with polynomial time solutions. In either case, solutions can be effectively extracted from proofs, which reduces problem-solving to theorem-proving. Furthermore, {\bf CLA4} has also been shown to be complete in the sense that every number-theoretic computational problem with a polynomial time solution is represented by some theorem of the system.   

The formalism of CoL is open-ended, and is expected to undergo series of extensions as the studies of the subject advance. Correspondingly, among the main goals of CoL at the present early stage of development still remains identifying the most natural and potentially interesting operations on computational problems, and finding axiomatizations for the corresponding sets of valid formulas.  Considerable advances have already been made in this direction (\cite{Japtocl1}-\cite{Cirq},\cite{Japjsl}-\cite{Japseq},\cite{Japfour}-\cite{Japtowards},\cite{Ver}), and the present paper tells one more success story. 

The main operations studied so far 
are:
\begin{itemize}
\item {\em Negation}: $\gneg$.
\item {\em Constant elementary games} ($0$-ary operations): 
\begin{itemize}
\item $\twg$ (automatically won game) and $\tlg$ (automatically lost game).
\end{itemize}
\item {\em Parallel operations}: 
\begin{itemize}
\item $\mlc$ (parallel conjunction) and $\mld$ (parallel disjunction);
\item $\mla$ (parallel universal quantifier) and $\mle$ (parallel existential quantifier);
\item $\pst$ (parallel recurrence) and $\pcost$ (parallel corecurrence).
\end{itemize}
\item {\em Choice operations}: 
\begin{itemize}
\item $\adc$ (choice conjunction) and $\add$ (choice disjunction);
\item  $\ada$ (choice universal quantifier) and  $\ade$ (choice existential quantifier).
\end{itemize}
\item {\em Sequential operations:} 
\begin{itemize}
\item $\sqc$ (sequential conjunction) and  $\sqd$ (sequential disjunction);
\item $\sqa$ (sequential universal quantifier) and  $\sqe$ (sequential existential quantifier);
\item $\sst$ (sequential recurrence) and  $\scost$ (sequential corecurrence).
\end{itemize}
\item {\em Blind operations}: 
\begin{itemize}
\item $\cla$ (blind universal quantifier) and  $\cle$ (blind existential quantifier).
\end{itemize}
\item {\em Branching operations}: 
\begin{itemize}
\item $\st$ (branching recurrence) and  $\cost$ (branching corecurrence).
\end{itemize}
The branching operations have a number of natural sharpenings, among which are the {\em finite} and the {\em countable} versions of branching recurrence and  corecurrence. 
\end{itemize} 
There are also various {\em reduction} operations: $\mli$, defined by $A\mli B=\gneg A\mld B$; $\pintimpl$, defined by $A\pintimpl B=\pst A\mli B$; $\intimpl$, defined by $A\intimpl B=\st A\mli B$; $\sintimpl$, defined by $A\sintimpl B=\sst A\mli B$; etc.

The present paper introduces the following new group:
\begin{itemize}
\item {\em Toggling operations}:  
\begin{itemize}
\item $\tgc$ (toggling conjunction) and $\tgd$ (toggling disjunction);
\item  $\tga$ (toggling universal quantifier) and  $\tge$ (toggling existential quantifier);
\item $\tgpst$ (toggling recurrence) and  $\tgpcost$ (toggling corecurrence);
\item $\tgst$ (toggling-branching recurrence) and   $\tgcost$ (toggling-branching corecurrence).
\end{itemize}
\end{itemize}
This group also induces the reduction operations $\tgpi$ and $\tgbi$, defined by $A\tgpi B=\tgpst A\mli B$ and $A\tgbi B=\tgst A\mli B$.

The main technical result of this paper is constructing a sound and complete axiomatization for the propositional fragment of CoL whose logical vocabulary consists of \ $\gneg$, $\twg$, $\tlg$, $\mlc$, $\mld$, \hspace{-1pt}$\tgc$\hspace{-2pt}, $\tgd$\hspace{-2pt},\hspace{-1pt} $\sqc$\hspace{-2pt},\hspace{-2pt} $\sqd$\hspace{-2pt}, $\adc$,\hspace{1pt} $\add$.

\section{A quick tour of the operation zoo of computability logic}\label{s2}

In this section we give a very brief and informal overview of the language of computability logic and the game-semantical meanings of its main operators  for those unfamiliar with the subject. In what follows, $\pp$ and $\oo$ are symbolic names for the players to which we earlier referred as the machine and the environment, respectively. 

First of all, it should be noted that computability logic is a conservative extension of classical logic. Classical propositions --- as well as predicates as generalized propositions --- are viewed as special, {\em elementary} sorts of games that have no moves and are automatically won by the machine if true, and lost if false. The languages of various reasonably expressive fragments of computability logic would typically include two sorts of atoms: {\em elementary} atoms $p$, $q$, $r(x)$, $s(x,y)$, \ldots to represent elementary games, and {\em general atoms} $P$, $Q$, $R(x)$, $S(x,y)$, \ldots to represent any, not-necessarily-elementary, games. The classically-shaped operators $\gneg,\mlc,\mld,\mli,\cla,\cle$ are conservative generalizations of the corresponding classical operations from elementary games to all games. This means that, when applied to elementary games, they again produce elementary games, and their meanings happen to coincide with the classical meanings. 

\subsection{Constant elementary games} These are two $0$-ary ``operations'', for which we use the same symbols $\twg$ and $\tlg$ as for the two players. $\twg$ is an elementary game automatically won by $\pp$, and $\tlg$ is an elementary game won by $\oo$. Just as classical logic, computability logic sees no difference between two true or two false propositions, so that we have ``Snow is white''=``$0\equals 0$''=$\twg$ and ``Snow is black''=``$0\equals 1$''=$\tlg$.    

\subsection{Negation}
Negation $\gneg$ is a role-switch operation: $\gneg A$ is obtained from $A$ by turning $\pp$'s (legal) moves and wins into $\oo$'s (legal) moves and wins, and vice versa. For example, if {\em Chess} means the game of chess from the point of view of the white player, then $\gneg${\em Chess} is the same game from the point of view of the black player.\footnote{Here and later, we consider a version of chess with no draw outcomes --- for instance, the one where draw is declared to be a win for the black player --- so that each play is won by one of the players and lost by the other player.}  And where $0\equals 0$ is an elementary game automatically won by $\pp$, $\gneg 0\equals 0$ is an elementary game automatically won by $\oo$ --- there are no moves to interchange here, so only the winners are interchanged. From this explanation it must be clear that $\gneg$, when applied to elementary games
(propositions or predicates), indeed acts like classical negation, as promised.

\subsection{Choice operations}
The choice operations model decision steps in the course of interaction, with disjunction 
and existential quantifier meaning $\pp$'s choices, and conjunction and universal quantifier 
meaning choices by $\oo$. For instance, where $f(x)$ is a function, $\ada x\ade y \bigl(y\equals f(x)\bigr)$ is a game in which the first move/choice is by the environment, consisting in 
specifying a particular value $m$ for $x$. Such a move, which intuitively can be seen as asking  the question ``{\em what is the value of $f(m)$?}\hspace{1pt}'' brings the game down to the position $\ade y \bigl(y\equals f(m)\bigr)$. 
The next step is by the machine, which should specify a value 
$n$ for $y$, further bringing the game down to the elementary game $n\equals f(m)$,  won by the machine if true and lost if false. $\pp$'s move $n$ can thus be seen as answering/claiming that $n$ is the value of $f(m)$.
From this explanation it must be clear that   
$\ada x\ade y \bigl(y\equals f(x)\bigr)$ represents the problem of computing $f$, with $\pp$ having an algorithmic winning strategy for this game iff $f$ is a computable function. Similarly, where $p(x)$ is a predicate, $\ada x\bigl(\gneg p(x)\add p(x)\bigr)$ represents the problem of deciding $p(x)$: here, again, the first move is by the environment, consisting in choosing a value $m$ for $x$ (asking whether $p(m)$ is true); and the next step is by the machine which, in order to win, should 
choose the true disjunct of  $\gneg p(m)\add p(m)$, i.e. correctly answer the question. Formally, $A\add B$  can be defined as $\gneg(\gneg A\adc \gneg B)$, or 
$A\adc B$ can be defined as $\gneg(\gneg A\add \gneg B)$; furthermore, assuming that the universe of discourse is $\{1,2,3,\ldots\}$,  $\ada xA(x)$ can be defined as 
$A(1)\adc A(2)\adc A(3)\adc\ldots$ and $\ade xA(x)$ as $A(1)\add A(2)\add A(3)\add\ldots$. It should be mentioned that making an initial choice of a component by the corresponding player in a choice combination of games is not only that player's privilege, but also an obligation: the player will be considered to lose the game  if it fails to make a choice.

\subsection{Parallel operations}
The parallel operations combine games in a way that corresponds to the intuition of concurrent computations. Playing $A\mlc B$ or $A\mld B$ means playing, in parallel, the two games $A$ and $B$. In $A\mlc B$, $\pp$ is considered the winner if it wins in both of the components, while in $A\mld B$ it is sufficient to win in one of the components. Then the parallel quantifiers and recurrences are defined by: 
\[\begin{array}{rcl}
\mla xA(x) & = & A(1)\mlc A(2)\mlc A(3)\mlc\ldots\vspace{3pt}\\
\mle xA(x) & = & A(1)\mld A(2)\mld A(3)\mld\ldots\vspace{3pt}\\ 
\pst A & = & A\mlc A\mlc A\mlc\ldots\vspace{3pt}\\
\pcost A & = & A\mld A\mld A\mld\ldots
\end{array}\] 
To appreciate the difference between choice operations and their parallel counterparts, let us compare the games ${\em Chess}\mld\gneg \mbox{\em Chess}$ and $\mbox{\em Chess} \add \gneg \mbox{\em Chess}$. 
The former  is, in fact, a simultaneous play on two boards, where on the left board $\pp$ plays white, and on the right  board $\pp$ plays black. There is a simple strategy for $\pp$ that guarantees success against any adversary. All that $\pp$ needs to do is to mimic, in {\em Chess}, the moves made by $\oo$ in $\gneg \mbox{\em Chess}$, and vice versa. On the other hand, winning the game $\mbox{\em Chess}\hspace{0.02in} \add\gneg \mbox{\em Chess}$ is not easy: here, at the very beginning, $\pp$ has to choose between {\em Chess} and $\gneg \mbox{\em Chess}$, and then win the chosen one-board game. 

While all classical tautologies automatically hold when the classically-shaped operators are applied to elementary games, in the general (nonelementary) case the class of valid principles shrinks. For example,  $\gneg P\mld (P\mlc P)$ is no longer valid. The above ``mimicking strategy" would obviously fail in the three-board game \[\gneg \mbox{\em Chess}\mld (\mbox{\em Chess}\mlc \mbox{\em Chess}),\] for here the best that $\pp$ can do is to pair $\gneg \mbox{\em Chess}$ with one of the two conjuncts of $\mbox{\em Chess}\mlc \mbox{\em Chess}$. It is possible that then $\gneg \mbox{\em Chess}$ and the unmatched {\em Chess} are both lost, in which case the whole game will be lost.  As much as this example may remind us of linear logic, it should be noted that the class of principles with parallel connectives validated by computability logic is not the same as the class of multiplicative formulas provable in linear or affine logic. An example separating CoL from both linear and affine logics is Blass's \cite{Bla92} principle \[\bigl((\gneg P\mld\gneg Q)\mlc(\gneg R\mld\gneg S)\bigr) \mld \bigl((P\mld R)\mlc(Q\mld  S)\bigr),\] not provable in affine logic but valid in CoL.
The same applies to principles containing choice (``additive'') and recurrence (``exponential'') operators.
 
\subsection{Reduction}
The operation $\mli$, defined in the standard way by $A\mli B=\gneg A\mld B$, is perhaps most  interesting from the 
computability-theoretic point of view. Intuitively, $A\mli B$ is the problem of {\em reducing} $B$ to $A$. Putting it in other words, solving $A\mli B$ means solving $B$ while having $A$ as an (external, environment-provided) {\em computational resource}.\label{0cr} ``Computational resource" is symmetric to ``computational problem": what is a problem (task) for the machine, is a resource for the environment, and vice versa.
To get a feel for $\mli$ as a problem reduction operator, let us look at  reducing 
the acceptance problem to the halting problem. The halting problem  can be expressed by 
\[\ada x\ada y \bigl(\mbox{\em Halts}(x,y) \add \gneg \mbox{\em Halts}(x,y)\bigr),\]
 where $\mbox{\em Halts}(x,y)$ is the predicate  
``Turing machine  (encoded by) $x$ halts on input $y$". And the acceptance problem can be expressed by 
\[\ada x\ada y \bigl(\mbox{\em Accepts}(x,y) \add \gneg \mbox{\em Accepts}(x,y)\bigr),\] 
with $\mbox{\em Accepts}(x,y)$ meaning 
``Turing machine  $x$ accepts input $y$''. While the acceptance problem is not decidable, it is algorithmically reducible to the halting problem. In particular, there is a machine that always wins the game
\[
\ada x\ada y \bigl(\mbox{\em Halts}(x,y)\add \gneg \mbox{\em Halts}(x,y)\bigr) \mli\ \ada x\ada y \bigl(\mbox{\em Accepts}(x,y)\add \gneg \mbox{\em Accepts}(x,y)\bigr).
\]
A strategy for solving this problem is to wait till the environment specifies values $m$ and $n$ for $x$ and $y$ in the consequent, thus asking  the question ``does machine $m$ accept input $n$?''. In response, $\pp$ selects the same values $m$ and $n$ for  $x$ and $y$ in the antecedent (where the roles of $\pp$ and $\oo$ are switched), thus asking the counterquestion ``does $m$ halt on $n$?''. The environment will have to correctly answer this counterquestion, or else it loses. If it answers ``No'', then $\pp$ also says ``No'' in the consequent, i.e., selects the right disjunct there, as not halting implies not accepting. Otherwise, if the environment's response in the antecedent is ``Yes'',
$\pp$ simulates machine $m$ on input $n$ until it halts and then selects, in the consequent, the left or the right disjunct depending on whether the simulation accepted or rejected. 

\subsection{Blind operations}
The blind group of operations comprises $\cla$  and its dual $\cle$ ($\cle x =\gneg\cla x\gneg$). The meaning of $\cla xA(x)$ is similar to that of $\ada xA(x)$, with the difference that the particular value of $x$ that the environment ``selects'' is invisible to the machine (more precisely, there is no move signifying such a ``selection''), so that it has to play blindly in a way that guarantees success no matter what that value is. This way, $\cla$ and $\cle$  produce games with {\em imperfect information}. 

Compare the problems
\[\ada x\bigl(\mbox{\em Even$(x)$}\add \mbox{\em Odd$(x)$}\bigr)\] and  \[\cla x\bigl(\mbox{\em Even$(x)$}\add \mbox{\em Odd$(x)$}\bigr).\] 
Both of them are about telling whether an arbitrary given number is even or odd; the difference is only in whether that ``given number'' is communicated to the machine or not. The first problem is an easy-to-win, two-move-deep game of a structure that we have already seen.  The second game, on the other hand, is one-move deep with only  the machine to make a move --- select the ``true''  disjunct, which is hardly possible to do as the value of $x$ remains unspecified. 

As an example of a solvable nonelementary $\cla$-problem, let us look at
\[\cla x\Bigl(\mbox{\em Even$(x)$}\add \mbox{\em Odd$(x)$}\ \mli\ \ada y\bigl(\mbox{\em Even$(x\plus y)$}\add
\mbox{\em Odd$(x\plus y)$}\bigr)\Bigr),\]
solving which means solving what follows ``$\cla x$'' without knowing the value of $x$. Unlike $\cla x\bigl(\mbox{\em Even$(x)$}\add \mbox{\em Odd$(x)$}\bigr)$, this game is certainly winnable: The machine waits till the environment selects  a value $n$ for $y$ in the consequent and also selects one of the $\add$-disjuncts in the antecedent (if either selection is never made, the machine automatically wins). Then: If $n$ is even, in the consequent the machine makes the same selection 
{\em left}  or {\em right} as the environment made in the antecedent, and otherwise, if $n$ is odd, it reverses the environment's selection. 

\subsection{Sequential operations} \label{s2seq} 
One of the ways to characterize the sequential conjunction $A\sqc B$ is to say that this is a game that starts and proceeds as a play of $A$; it will also end as an ordinary play of $A$ unless, at some point, 
$\oo$ decides --- by making a special {\em switch} move --- to abandon $A$ and switch to $B$. In such a case the play restarts, continues and ends as an ordinary play of $B$ without the possibility to go back to $A$. $A\sqd B$ is the same, only here it is $\pp$ who decides whether and when to switch from $A$ to $B$. These generalize to the infinite cases $A_0\sqc A_1\sqc A_2\sqc\ldots$ and $A_0\sqd A_1\sqd A_2\sqd\ldots$: here the corresponding player can make any finite number $n$ of switches, in which case the winner in the play will be the player who wins in $A_n$; and if an infinite number of switches are made, then the player responsible for this is considered the loser.  The sequential quantifiers, as we may guess, are defined by \[\sqa xA(x)=A(1)\sqc A(2)\sqc A(3)\sqc\ldots\] and \[\sqe xA(x)=A(1)\sqd A(2)\sqd A(3)\sqd\ldots,\] and the sequential recurrence and corecurrence are defined by 
\[\sst A=A\sqc A\sqc A\sqc\ldots\] and \[\scost A=A\sqd A\sqd A\sqd\ldots.\] Below are a couple of examples providing insights into the computational intuitions associated with the sequential operations. See \cite{Japseq} for more.

Let $p(x)$ be any predicate. Then the game $\ada x\bigl(\gneg p(x)\sqd p(x)\bigr)$ represents the problem of  {\em semideciding} $p(x)$: it is not hard to see that this game has an effective winning strategy by $\pp$ iff $p(x)$ is semidecidable (recursively enumerable). Indeed, if $p(x)$ is semidecidable, a winning strategy is to wait until $\oo$ selects a particular $m$ for $x$, thus bringing the game down to  $\gneg p(m)\sqd p(m)$. After that, $\pp$ starts looking for a certificate of $p(m)$'s being true. If and when such a certificate is found (meaning that $p(m)$ is indeed true), $\pp$ makes a switch move turning $\gneg p(m)\sqd p(m)$ into the true and hence $\pp$-won $p(m)$; and if no certificate exists (meaning that $p(m)$ is false), then  $\pp$ keeps looking for a non-existent certificate forever and thus never makes any moves, meaning that the game ends as $\gneg p(m)$, which, again, is a true and hence $\pp$-won elementary game. And vice versa: any effective winning strategy for $\ada x\bigl(\gneg p(x)\sqd p(x)\bigr)$ can obviously be seen as a semidecision procedure for $p(x)$, which accepts an input $m$ iff the strategy ever makes a switch move in the scenario where $\oo$'s initial choice of a value for $x$ is $m$.       

Algorithmic solvability (computability) of games has been shown to be closed under modus ponens and a number of other familiar or expected rules, such as ``from $A$ and $B$ conclude $A\mlc B$'', ``from $A$ conclude $\ada x A$'', ``from $A$ conclude $\pst A$'', etc. In view of these closures, the validity (= ``always computability'') of the principles discussed below implies  certain 
known facts from the theory of computation. Needless to say, those examples demonstrate how CoL can be used as a systematic tool for defining new interesting properties and relations between computational problems, and not only reproducing already known theorems but also discovering an infinite variety of new facts.  

The following formula, which can be shown to be valid with respect to our semantics, implies --- in a sense, ``expresses'' --- the well known fact that, if both a predicate $p(x)$ and its negation $\gneg p(x)$ are recursively enumerable (i.e., $p(x)$ is both semidecidable and {\em co-semidecidable}), then $p(x)$ is decidable:
\begin{equation}\label{nov16}
\ada x\bigl(\gneg p(x)\sqd p(x)\bigr)\mlc\ada x\bigl(p(x)\sqd\gneg p(x)\bigr)\mli \ada x\bigl(\gneg p(x)\add p(x)\bigr).
\end{equation}
Actually, the validity of (\ref{nov16}) means something more than just noted: it means that the problem of deciding $p(x)$ is reducible to (the $\mlc$-conjunction of) the problems of semideciding $p(x)$ and $\gneg p(x)$. In fact, a reducibility in an even stronger sense (in a sense that has no name) holds, expressed by the following valid formula:
\begin{equation}\label{nov16a}
\ada x\Bigl(\bigl(\gneg p(x)\sqd p(x)\bigr)\mlc\bigl((p(x)\sqd\gneg p(x)\bigr)\mli \bigl(\gneg p(x)\add p(x)\bigr)\Bigr).
\end{equation}

Computability logic defines computability of a game $A(x)$ as computability of its $\ada$-closure, so the prefix $\ada x$ can be safely removed in the above formula and, after writing simply ``$p$'' instead of ``$p(x)$'', the validity of (\ref{nov16a}) means the same as the validity of the following propositional-level formula, provable in our sound and complete propositional system $\clt$:
 
\begin{equation}\label{nov16b}
(\gneg p\sqd p)\mlc(p\sqd\gneg p)\mli \gneg p\add p.
\end{equation}

Furthermore, the above principle is valid not only for predicates (elementary games), but also for all games that we consider, as evidenced by the provability of the following formula in (the sound) $\clt$:

\begin{equation}\label{nov16c}
(\gneg P\sqd P)\mlc(P\sqd\gneg P)\mli \gneg P\add P.
\end{equation}
Similarly, formula (\ref{nov16}) remains  valid with $P(x)$ instead of $p(x)$:
\begin{equation}\label{nov16d}
\ada x\bigl(\gneg P(x)\sqd P(x)\bigr)\mlc\ada x\bigl(P(x)\sqd\gneg P(x)\bigr)\mli \ada x\bigl(\gneg P(x)\add P(x)\bigr).
\end{equation}

For our next example, remember the relation of {\em mapping reducibility} (more often  called {\em many-one reducibility}) of a predicate $q(x)$ to a predicate $p(x)$, defined as existence of an effective function $f$ such that, for any $n$, $q\bigl(n\bigr)$ is equivalent to $p\bigl(f(n)\bigr)$. It is not hard to see that this relation holds if and only if the game \[\ada x\ade y\Bigl(\bigl(q(x)\mli p(y)\bigr)\mlc \bigl(p(y)\mli q(x)\bigr)\Bigr),\]
which we abbreviate as  $\ada x\ade y\bigl(q(x)\leftrightarrow p(y)\bigr)$, has an algorithmic winning strategy by $\pp$. In this sense, $\ada x\ade y\bigl(q(x)\leftrightarrow p(y)\bigr)$  expresses the problem of mapping reducing $q$ to $p$. Then the  validity  of the following formula implies the known fact that, if $q$ is mapping reducible to $p$ and $p$ is recursively enumerable, then so is $q$:\footnote{By the way, the same principle does not hold with ``Turing reducible'' instead of ``mapping reducible''.} 

\begin{equation}\label{nov16e}
\ada x\ade y\bigl(q(x)\leftrightarrow p(y)\bigr)\mlc \ada x\bigl(\gneg p(x)\sqd p(x)\bigr)\mli \ada x\bigl(\gneg q(x)\sqd q(x)\bigr).\end{equation}
As in the earlier examples, the validity of (\ref{nov16e}), in fact, means something even more: it means that the problem of semideciding $q$ is reducible to the ($\mlc$-conjunction of the) problems of mapping reducing $q$ to $p$ and semideciding $p$.

\subsection{Branching operations}
The branching operations come in the form of branching recurrence $\st$ and its dual branching corecurrence $\cost$, which can be defined by $\cost A=\gneg\st\gneg A$. We have already seen two other --- parallel and sequential --- sorts of recurrences, and it might be a good idea to explain $\st$ by comparing it with them.

What is common to all members of the family of (co)recurrence operations is that, when applied to $A$, they turn it into a game playing which means repeatedly playing $A$. In terms of resources, recurrence operations generate multiple ``copies'' of $A$, thus making $A$ a reusable/recyclable resource. The difference between the various sorts of recurrences is how ``reusage'' is exactly understood.

Imagine a computer that has a program successfully playing $\chess$. The resource that such a computer provides is obviously something stronger than just $\chess$, for it permits to play $\chess$ as many times as the user wishes, while $\chess$, as such, only assumes one play. Even the simplest operating system would allow to start a session of $\chess$, then --- after finishing or abandoning and destroying it --- start a new play again, and so on. The game that such a system plays --- i.e. the resource that it supports/provides --- 
is the already known to us sequential recurrence $\sst \chess$, which assumes an unbounded number of plays of $\chess$ in a sequential fashion.  A more advanced operating system, however, would not require to destroy the old sessions before starting new ones; rather, it would allow to run as many parallel sessions as the user needs. This is what is captured by the parallel recurrence $\pst\chess$. As a resource, $\pst\chess$ is obviously stronger than $\sst\chess$ as it gives the user more flexibility. But $\pst$ is still not the strongest form of reusage. A really good operating system would not only allow the user to start new sessions of $\chess$ without destroying old ones; it would also make it possible to branch/replicate each particular stage of each particular session, i.e. create any number of ``copies" of any already reached position
of the multiple parallel plays of $\chess$, thus giving the user 
the possibility to try different continuations from the same position. What corresponds to this intuition is the branching recurrence $\st\chess$.

So, the user of the resource $\st A$  does not have to restart $A$ from the very beginning every time it wants to reuse it; rather, it is (essentially) allowed to backtrack to any of the previous --- not necessarily starting --- positions and try a new continuation from there, thus depriving the adversary of the possibility to reconsider the moves it has already made in that position. This is in fact the type of reusage every purely software resource allows or would allow in the presence of an advanced operating system and unlimited memory:
one can start running process $A$; then fork it  
at any stage  thus creating two threads  that have a common past but possibly diverging futures  (with the possibility to treat one of the threads as 
a ``backup copy'' and preserve it for backtracking purposes); then further fork any of the branches at any time; and so on. The less flexible type of reusage of $A$ assumed by $\pst A$, on the other hand, is closer to what infinitely many autonomous 
physical resources would naturally offer, such as an unlimited number of independently acting robots each performing task $A$, or an unlimited number of computers with limited memories, each one only capable of and responsible for running a single thread 
of process $A$. Here  the effect of replicating/forking an advanced stage of $A$ cannot be achieved unless, by good luck, 
there are two identical copies of the stage, meaning that the corresponding two robots or computers have so far acted in precisely the same ways. As for $\sst A$, it models the task performed by a single reusable physical resource --- the resource that can perform  task $A$ over and over again any number of times.  

The operation $\st$ also has a series of weaker versions obtained by imposing various restrictions on the quantity and form of reusages. Among the interesting and natural weakenings of $\st$ is the {\em countable branching recurrence} $\st^{\aleph_0}$ in the style of Blass's \cite{Bla72,Bla92} {\em repetition operation} $R$. See \cite{Japfour} for a discussion of such operations. 

Branching recurrence $\st$ stands out as the strongest of all recurrence operations, allowing to reuse $A$ (in $\st A$) in the strongest algorithmic sense possible. 
This makes the associated reduction operation $\intimpl$, defined by $A\intimpl B=\st A\mli B$, the weakest and hence most general form of algorithmic reduction. 
The well known concept of {\em Turing reduction} has the same claims. The latter, however, is only defined for the traditional, non-interactive sorts of computational problems --- two-step, input-output, question-answer sorts of problems that in our terms are written as $\ada x\bigl(p(x)\add\gneg p(x)\bigr)$ (the problem of deciding predicate $p$) or 
 $\ada x\ade y\bigl(y\equals f(x)\bigr)$ (the problem of computing function $f$). And it is no surprise that our $\intimpl$, when restricted to such problems, turns out to be equivalent to Turing reduction. Furthermore, when $A$ and $B$ are traditional sorts of problems, $A\intimpl B$ further turns out to be equivalent to $A\pintimpl B$ (but not to $A\sintimpl B$), as the differences between 
$A\pintimpl B$ and $A \intimpl B$, while substantial in the general (truly interactive) case, turn out to be  too subtle to   be relevant when $A$ is a game that models only a very short and simple potential dialogue between the interacting parties, consisting in just 
asking a question and giving an answer.  The benefits from the greater degree of resource-reusage flexibility offered by $A\intimpl B$ (as opposed to $A\pintimpl B$) are related to the possibility for the machine to try different reactions to the same action(s) by the environment in $A$. But such potential benefits cannot be realized when $A$ is, say, $\ada x\bigl(p(x)\add \gneg p(x)\bigr)$, because here a given individual session of $A$ immediately ends with an environment's move, to which the machine simply has no legal or meaningful responses at all, let alone having multiple possible responses to experiment  with. 

Thus, both $\intimpl$ and $\pintimpl$ are conservative extensions of Turing reduction from traditional sorts of problems to problems of arbitrary degrees and forms of interactivity. Of these two operations, however, only $\intimpl$ has the moral right to be called a legitimate successor of Turing reducibility, in the sense that, just like Turing reducibility (in its limited context), $\intimpl$ rather than $\pintimpl$ is an ultimate formal counterpart of our most general intuition of algorithmic reduction. And perhaps it is no accident that, as shown in \cite{Japjsl,Propint,Ver}, its logical behavior --- along with the choice operations --- is precisely captured by Heyting's intuitionistic calculus. As an aside, this means that CoL offers a good justification  --- in the form of a mathematically strict and intuitively convincing semantics --- of the constructivistic claims of intuitionistic logic, and a materialization of Kolmogorov's \cite{Kol32} well known yet  so far rather abstract thesis, according to which intuitionistic logic is a logic of problems. 

Our recurrence operations, in their logical spirit, are reminiscent of the exponential operators of linear logic. It should be noted that, as shown in \cite{Japfin}, linear --- in fact, affine --- logic is sound but incomplete when its additives are read as our choice operators, multiplicatives as parallel operators, and exponentials as either parallel or branching recurrences. Here the sequential sort of recurrences stands out in that linear logic becomes simply unsound if its exponentials $!,?$ are interpreted as our $\sst,\scost$. The same applies to the toggling sorts $\tgpst,\tgpcost,\tgst,\tgcost$ of recurrences that will be introduced shortly. 

Remember the concept of the Kolmogorov complexity $k(x)$ of a number $x$, which can be defined as the size (logarithm) $|m|$ of the smallest Turing machine (encoded by) $m$ that returns $x$ on input $1$.
Just like the acceptance problem $\ada x\ada y \bigl(\mbox{\em Accepts}(x,y)\add \gneg \mbox{\em Accepts}(x,y)\bigr)$, the Kolmogorov complexity problem 
$\ada x\ade y \bigl(y\equals k(x)\bigr)$ is known to be algorithmically reducible --- specifically, Turing reducible --- to the halting problem. Unlike the former case, however, the reduction in the latter case essentially requires repeated usage of the halting problem as a resource. Namely, the reducibility holds only in the sense of $\intimpl$,  $\pintimpl$ or even $\sintimpl$ but not in the sense of $\mli$. As an exercise, the reader may try to come up with an informal description of an algorithmic winning strategy for any one of the following games:
\[\begin{array}{cc}
\ada x\ada y \bigl(\mbox{\em Halts}(x,y)\add \gneg \mbox{\em Halts}(x,y)\bigr)\intimpl \ada x\ade y \bigl(y\equals k(x)\bigr);\\
\ada x\ada y \bigl(\mbox{\em Halts}(x,y)\add \gneg \mbox{\em Halts}(x,y)\bigr)\pintimpl \ada x\ade y \bigl(y\equals k(x)\bigr);\\
\ada x\ada y \bigl(\mbox{\em Halts}(x,y)\add \gneg \mbox{\em Halts}(x,y)\bigr)\sintimpl \ada x\ade y \bigl(y\equals k(x)\bigr).
\end{array}\]

\subsection{Toggling operations} \label{s2tog}
The new, toggling group of operations forms another natural phylum in this zoo of game operations. 

One of the intuitive ways to characterize the toggling disjunction $A\tgd B$ is the following. This game  starts and proceeds as a play of $A$. It will also end as an ordinary play of $A$ unless, at some point, $\pp$ decides to switch to $B$, after which the game becomes and continues as $B$. It will also end as $B$ unless, at some point, $\pp$ decides to switch back to $A$. In such a case the game again becomes $A$, where $A$ resumes from the position in which it was abandoned (rather than from its start position, as would be the case, say, in $A\sqd B\sqd A$). Later $\pp$ may again switch to (the abandoned position of) $B$, and so on. $\pp$ wins the overall play iff it switches from one component to another (``changes its mind'', or ``corrects its mistake'') at most finitely many times and wins in its final (and hence ``real'') choice, i.e., in the component which was chosen last to switch to. 

An alternative way to characterize $A\tgd B$ is to say that it is played exactly as $A\mld B$, with the only difference that $\pp$ is expected to make a ``choose $A$'' or ``choose $B$'' move some finite number of times. If infinitely many choices are made, $\pp$ loses. Otherwise, the winner in the play will be the player who wins in the component that was chosen last (``the eventual choice''). The case of $\pp$ having made no choices at all is treated as if it had chosen $A$ (thus, as in sequential disjunction, the leftmost component is the ``default'', or ``automatically made'', initial choice). 

It is important to note that the adversary never knows whether a given choice of a component of $A\tgd B$ is the last choice or not (and perhaps $\pp$ itself does not know that either, every time it makes a choice honestly believing  that the present choice is going to be final). Otherwise, if $\pp$ was required to indicate that it has made its final choice, then the resulting operation would simply be the same as --- more precisely, equivalent to --- $A\add B$. Indeed, in the  kind of games that we deal with (called {\em static} games), it never hurts a player to postpone making moves, so the adversary could just inactively wait till the last choice is declared, and start playing the chosen component only after that, as in the case of $A\add B$; under these circumstances, making some temporary choices before making the final choice would not make any sense for $\pp$, either.  

What would happen if we did not require that $\pp$ can change its mind only finitely meany times? There would be no ``final choice'' in this case. So, the only natural winning condition in the case of infinitely many choices would be to say that $\pp$ wins iff it simply wins in one of the components. But then the resulting operation would be the same as --- more precisely, equivalent to --- our kind old friend $\mld$, as a smart $\pp$ would always opt for keeping switching between components forever. That is, allowing infinitely many choices would amount to not requiring any choices at all, as is the case with $A\mld B$. 

One may also ask what would happen if we allowed $\pp$ to make an arbitrary initial choice between $A$ and $B$ and then reconsider its choice only (at most) once? (``$n$ times'' instead of ``once'' for any {\em particular} $n>1$ would not be natural or necessary to consider). Such an operation on games, albeit reasonable,  would not be basic. That is because it can be expressed through our primitives as $(A\sqd B)\add (B\sqd A)$.
  
Thus, we have four basic and natural sorts $A\mld B$, $A\tgd B$, $A\sqd B$, $A\add B$ of disjunctions, and denying any of these full citizenship  would make computability logic unsettlingly incomplete. What is common between these four operations and what warrants the shared qualification ``disjunction'' is that each one is a ``win one out of two'' kind of a combination of games from $\pp$'s perspective. $A \mld B$ is the weakest (easiest for $\pp$ to win) kind of a disjunction, as it does not require any choices at all. Next comes  $A\tgd B$, which {\em does} require a choice but in the weakest sense possible. $A\add B$ is the hardest-to-win 
disjunction, requiring a choice in the strongest sense. $A\sqd B$ is the next-hardest disjunction. It replaces 
the strict choice of $A\add B$ by the next-strictest kind  known in the traditional theory of computation as semidecision.  
 
Does the (very) weak sort of choice captured by $\tgd$ have a meaningful non-mathematical, everyday-life counterpart? Obviously it does. This is the kind of  choice that one would ordinarily call (making a correct) {\em choice after trial and error}. So, an alternative,  sexier  name for our toggling operations could
perhaps be ``trial-and-error operations''.  Indeed, a problem is generally considered to be solved after trial and error (a correct choice/solution/answer found) if, after perhaps coming up with several wrong solutions, a true solution is eventually found. That is, mistakes are tolerated and forgotten as long as they are eventually corrected. It is however necessary that new solutions stop coming at some point, so that there is a last solution whose correctness determines the success of the effort. Otherwise, if answers have kept changing all the time, no answer has really been given after all. Or, imagine Bob has been married and divorced several times. Every time he said ``I do'', he probably honestly believed that this time, at last, his bride was ``the one'', with whom he would live happily ever after.  Bob will be considered to have found his Ms. Right after all if and only if one of his marriages indeed turns out to be happy and final.    

Back from our detour to the layman's world, as we already know, for a predicate $p(x)$, $\ada x\bigl(\gneg p(x)\add p(x)\bigr)$ expresses the problem of deciding $p(x)$, and $\ada x\bigl(\gneg p(x)\sqd p(x)\bigr)$
 expresses the weaker problem of semideciding $p(x)$. What is then expressed by $\ada x\bigl(\gneg p(x)\tgd p(x)\bigr)$? This is also a decision-style problem, but still weaker than the problem of semideciding $p(x)$. This problem has been studied in the literature under several names, most common of which probably is {\bf recursively approximating} $p(x)$ (cf. \cite{Hinman}, Definition 8.3.9). It  means telling whether $p(x)$ is true or not, but doing so in the same style as semideciding does in negative cases: by correctly saying ``Yes'' or ``No'' at some point (after perhaps taking back previous answers several times) and never reconsidering this answer afterwards.  Observe that semideciding $p(x)$ can be seen as always saying ``No'' at the beginning and then, if this answer is incorrect, changing it to ``Yes'' at some later time; so, when the answer is negative, this will be expressed by saying ``No'' and never taking back this answer, yet without ever indicating that the answer is final and will not change.\footnote{Unless, of course, the procedure halts by good luck. Halting without saying ``Yes'' can then be seen as an explicit indication that the original answer ``No'' was final.} Thus, the difference between semideciding and recursively approximating is that, unlike a semidecision procedure, a recursive approximation procedure can reconsider {\em both} negative and positive answers, and  do so several times rather than only once.

According to Sh\"{o}nfield's Limit Lemma,\footnote{Cf. \cite{Hinman}, Lemma 8.3.12.} a predicate $p(x)$ is recursively approximable (i.e., the problem of its recursive approximation has an algorithmic solution)  iff $p(x)$ is of Turing degree $\mleq \emptyset'$, that is, $p(x)$ is Turing reducible to the halting problem. It is known that this, in turn, means nothing but having the arithmetical complexity  $\Delta_2$, i.e., that both $p(x)$ and its negation can be written in the form $\cle z\cla y\hspace{1pt}s(z,y,x)$, where $s(z,y,x)$ is a decidable predicate.\footnote{See  Section 5.1 of \cite{Hinman} for a definition of all classes of the arithmetical hierarchy, including $\Delta_n$ ($n=0,1,2,\ldots$).}   
In the theory of computability-in-principle (as opposed to, say, complexity theory), by importance,  the class of predicates of complexity $\Delta_2$ is only next to the classes of decidable, semidecidable and co-semidecidable predicates. This class also plays a special role in logic: it is known that a formula of classical predicate logic is valid if and only if it is true in every model where  all atoms of the formula are interpreted as predicates of complexity $\Delta_2$.

To see that recursive approximability of a predicate  $p(x)$ is equivalent to this predicate's being of complexity $\Delta_2$, first assume that $p(x)$ is of complexity $\Delta_2$, so that  $p(x)=\cle z\cla y\hspace{1pt}q(z,y,x)$ and $\gneg p(x)=\cle z\cla y\hspace{1pt} r(z,y,x)$ for some decidable predicates $q(z,y,x)$ and $r(z,y,x)$.
Then $\ada x\bigl(\gneg p(x)\tgd p(x)\bigr)$ is solved by the following strategy. Wait till the environment specifies a value $m$ for $x$, thus bringing the game down to $\gneg p(m)\tgd p(m)$. Then initialize both $i$ and $j$ to $1$, choose the $ p(m)$ component, and do the following:
\begin{description}
\item[Step 1:] Check whether $q(i,j,m)$ is true. If yes, increment $j$ to $j\plus 1$ and repeat Step 1. If not, switch to the $\gneg p(m)$ component, reset $j$ to $1$, and go to Step 2.
\item[Step 2:] Check whether $r(i,j,m)$ is true. If yes, increment $j$ to $j\plus 1$ and repeat Step 2. If not, switch to the $p(m)$ component, reset $j$ to $1$, increment $i$ to $i\plus 1$, and go to Step 1.
\end{description}
With a moment's thought, one can see that the above algorithm indeed solves $\ada x\bigl(\gneg p(x)\tgd p(x)\bigr)$. 

For the opposite direction, assume a given algorithm $\cal A$ solves $\ada x\bigl(\gneg p(x)\tgd p(x)\bigr)$. Let $q(z,y,x)$ be the predicate such that $q(i,j,m)$ is true iff, in 
the scenario where the environment specified $x$ as $m$ at the beginning of the play, so that the game was brought down to $\gneg p(m)\tgd p(m)$, we have:
\begin{itemize} 
\item at the $i$th computation step, $\cal A$ chose the $p(m)$ component;
\item at the $j$th computation step, $\cal A$ did not move. 
\end{itemize}
Quite similarly, let $r(z,y,x)$ be the predicate such that $r(i,j,m)$ is true iff, 
in the scenario where the environment specified $x$ as $m$ at the beginning of the play, so that the game was brought down to $\gneg p(m)\tgd p(m)$, we have:
\begin{itemize} 
\item either $i\equals 1$ or, at the $i$th computation step, $\cal A$ chose the $\gneg p(m)$ component;
\item at the $j$th computation step, $\cal A$ did not move. 
\end{itemize}
Of course, both $q(z,y,x)$ and $r(z,y,x)$ are decidable predicates, and hence so are $y\mgreater z\mli q(z,y,x)$ and $y\mgreater z\mli r(z,y,x)$. 
Now, it is not hard to see that $p(x)=\cle z\cla y\bigl(y\mgreater z\mli q(z,y,x)\bigr)$ and $\gneg p(x)=\cle z\cla y \bigl(y\mgreater z\mli r(z,y,x)\bigr)$, so that $p(x)$ is indeed of complexity $\Delta_2$. 

As a real-life example of a predicate which is recursively approximable but neither semidecidable nor co-semidecidable, consider the predicate $k(x)\mless k(y)$, saying that number $x$ is simpler than number $y$ in the sense of Kolmogorov complexity. It is known that $k(z)$ (the Kolmogorov complexity of $z$) is bounded, never exceeding the size (logarithm) $|z|$  of $z$ plus a certain constant $c$. Fix this $c$.   Here is an algorithm which recursively approximates the predicate $k(x)\mless k(y)$, i.e.,  solves the problem \[\ada x\ada y\bigl(\gneg k(x)\mless k(y)\tgd k(x)\mless k(y)\bigr).\] Wait till the environment brings the game down to $\gneg k(m)\mless k(n)\tgd k(m)\mless k(n)$ for some $m$ and $n$. Then start simulating, in parallel, all Turing machines of sizes $\mleq  min(|m|,|n|)\plus c$ on input $1$. Whenever you see that a machine $t$ returns $m$  and the size of $t$ is smaller than that of any other previously found machines that return $m$ or $n$ on input $1$, choose 
$k(m)\mless k(n)$. Quite similarly, whenever you see that a machine $t$ returns $n$  and the size of $t$ is smaller than that of any other previously found machine that returns $n$ on input $1$, as well as smaller or equal to the size of any other previously found machines that return $m$ on input $1$, choose 
$\gneg k(m)\mless k(n)$.
 Obviously, the correct choice between $\gneg k(m)\mless k(n)$  and $k(m)\mless k(n)$ will be made sooner or later and never reconsidered afterwards. This will happen when the procedure hits --- in the role of $t$ --- a smallest machine that returns either $m$ or $n$ on input $1$. 

Once we have toggling disjunction, its dual operation of toggling conjunction $A\tgc B$ can be defined in a fully symmetric way, with the roles of the machine and the environment interchanged. That is, here it is the environment rather than the machine that makes choices. Equivalently, $A\tgc B$ can be defined as $\gneg(\gneg A\tgd\gneg B)$. 
  
The toggling versions of quantifiers and recurrences are defined in the same way as in the case of parallel or sequential operations. Namely:
\[\begin{array}{rcl}
\tga xA(x)& = & A(1)\tgc A(2)\tgc A(3)\tgc\ldots\\
\tge xA(x)& = & A(1)\tgd A(2)\tgd A(3)\tgd\ldots\\
\tgpst A & = & A\tgc A\tgc A\tgc\ldots\\
\tgpcost A & = & A\tgd A\tgd A\tgd\ldots\\
\end{array}\]

There is yet another natural sort $\tgst,\tgcost$ of toggling (co)recurrence operations worth considering.  We call these {\em toggling-branching recurrence} and {\em toggling-branching corecurrence}, respectively. Roughly, $\tgst,\tgcost$ are the same to $\st,\cost$ as $\tgpst,\tgpcost$ are to $\pst,\pcost$. Namely, a play over $\tgcost A$ proceeds as over $\cost A$, with the difference that now $\pp$ is required to make a choice --- which can be reconsidered any finite number of times --- of a particular session/branch of $A$ out of the many sessions that are being played. As with all other toggling operations, if  choices are retracted infinitely many times, $\pp$ loses. Otherwise, the winner in the overall game will be the player which wins in the session of $A$ that was chosen last. $\tgst A$, as expected, is the dual of $\tgcost A$, which can be defined by interchanging the roles of the two players, or by $\tgst A=\gneg \tgcost \gneg A$.

For our last example illustrating CoL operations at work, remember that Kolmogorov complexity $k(x)$ is not a computable function, i.e., the problem $\ada x\ade y\bigl(y\equals k(x)\bigr)$ has no algorithmic solution. However, replacing $\ade y$ with $\tge\hspace{-1pt} y$ in it yields an algorithmically solvable (yet nontrivial) problem. A solution for   $\ada x\tge \hspace{-1pt}y\bigl(y\equals k(x)\bigr)$ goes like this. Wait till the environment chooses a number $m$ for $x$, thus bringing the game down to  $\tge \hspace{-1pt}y\bigl(y\equals k(m)\bigr)$, i.e., to 
\begin{equation}\label{march18}
1\equals k(m)\tgd 2\equals k(m)\tgd 3\equals k(m)\tgd\ldots
\end{equation}
Initialize $i$ to a sufficiently large number, such as $i\equals |m|\plus c$ ($c$ is the constant mentioned earlier), and then do the following routine:

\begin{quote} ROUTINE: Choose the $i$th $\tgd$-disjunct  of (\ref{march18}). Then start simulating on input $1$, in parallel, all Turing machines whose sizes are smaller than $i$. If and when you see that one of such machines returns $m$, update $i$ to the size of that machine, and repeat ROUTINE. 
\end{quote}

A similar argument convinces us that the problems $\ada x\tgpcost \ade y\bigl(y\equals k(x)\bigr)$ and $\ada x\tgcost \ade y\bigl(y\equals k(x)\bigr)$ also have algorithmic solutions. So do the problems $\tga x\tge\hspace{-1pt} y\bigl(y\equals k(x)\bigr)$ and $\mla x\tge\hspace{-1pt} y\bigl(y\equals k(x)\bigr)$, but certainly not the problem $\cla x\tge \hspace{-1pt}y\bigl(y\equals k(x)\bigr)$.

\section{Toggling and sequential operations defined formally}\label{sdef}

In what follows, we rely on the first six sections of \cite{Japfin} as an external source. Although long, \cite{Japfin} is very easy to read and has a convenient glossary\footnote{The glossary for the published version of \cite{Japfin} is given at the end of the {\em book} (rather than {\em article}), on pages 371-376. The reader may instead use the preprint version of \cite{Japfin}, available at http://arxiv.org/abs/cs.LO/0507045 The latter includes both the main text and the glossary.} to look up any unfamiliar terms and symbols. A reader not familiar with \cite{Japfin} or unwilling to do some parallel reading, may want to either stop here or just browse the rest of the paper without attempting to go into the technical details of formal definitions and proofs. Due to the very dynamic recent development, computability logic has already reached a point where it is no longer feasible to reintroduce all relevant concepts all over again in each new paper on the subject --- this would make any significant or fast progress within the project near impossible.  

Here we only provide formal definitions for the  toggling and sequential operations. Definitions of all other relevant operations are found in \cite{Japfin}. Definitions of sequential operations are also given in \cite{Japseq} in a form technically different from our present one, but otherwise yielding equivalent (in every reasonable sense) concepts, and whether to adopt the definitions of \cite{Japseq} or the present definitions of sequential operations is purely a matter of taste or convenience. As for toggling operations, they have never been defined before.  

In the definitions of this section, following  \cite{Japfin}, we use the notation  $\Gamma^\alpha$ --- where $\Gamma$ is a run and $\alpha$ is a string --- for the result of deleting from $\Gamma$ all labmoves (labeled  moves) except those that have the form $\xx\alpha\beta$ for some string $\beta$ and player $\xx$, and then further replacing each such remaining labmove $\xx\alpha\beta$ by $\xx\beta$.

\begin{definition}\label{nov10}
Let $A_1,\ldots,A_n$ ($n\geq 2$) be any constant games. Let us agree that, in the context of a toggling or sequential disjunction or conjunction of these games, a {\bf switch move} --- or just a {\bf switch} ---  means the move/string $i$ for some $i\in \{1,\ldots,n\}$ (here we identify natural numbers with their decimal representations). When there are finitely many switches in a run $\Gamma$, by the {\bf active component} of a toggling or sequential combination of $A_1,\ldots,A_n$ in $\Gamma$ we mean $A_i$  such that $i$ is the last (rightmost) switch move in $\Gamma$; in case there are no switch moves in $\Gamma$ at all,  $A_1$ is considered to be the active component. The components other than the active one are said to be {\bf dormant}.

\begin{enumerate}
\item  The {\bf toggling disjunction} $A_1\tgd\ldots\tgd A_n$ of $A_1,\ldots,A_n$ is defined as follows:
\begin{itemize}
\item A position $\Phi$ is a legal position of $A_1\tgd\ldots\tgd A_n$ iff every move of $\Phi$ is either a switch by $\pp$,  or  the move $i.\alpha$ by either player, where $i\in\{1,\ldots,n\}$ and $\alpha$ is some string, and the following condition is satisfied: for each $i\in\{1,\ldots,n\}$, $\Phi^{i.}$ is a legal position of $A_i$.   
\item Let $\Gamma$ be a legal run of $A_1\tgd\ldots\tgd A_n$. Then  $\Gamma$ is a $\pp$-won run of $A_1\tgd\ldots\tgd A_n$ iff there are finitely many switches (by $\pp$) in  $\Gamma$ and, where $A_i$ is the active component of $A_1\tgd\ldots\tgd A_n$ in $\Gamma$,  $\Gamma^{i.}$ is a $\pp$-won run of $A_i$. 
\end{itemize}
\item  The {\bf toggling conjunction} $A_1\tgc\ldots\tgc A_n$ of $A_1,\ldots,A_n$ is defined as follows:
\begin{itemize}
\item A position $\Phi$ is a legal position of $A_1\tgc\ldots\tgc A_n$ iff every move of $\Phi$ is either a switch by $\oo$,  or is the move $i.\alpha$ by either player, where $i\in\{1,\ldots,n\}$ and $\alpha$ is some string, and the following condition is satisfied: for each $i\in\{1,\ldots,n\}$, $\Phi^{i.}$ is a legal position of $A_i$.   
\item Let $\Gamma$ be a legal run of $A_1\tgc\ldots\tgc A_n$. Then  $\Gamma$ is a $\oo$-won run of $A_1\tgc\ldots\tgc A_n$ iff there are finitely many switches (by $\oo$) in  $\Gamma$ and, where $A_i$ is the active component of $A_1\tgc\ldots\tgc A_n$ in $\Gamma$,  $\Gamma^{i.}$ is a $\oo$-won run of $A_i$.
\end{itemize}
\item  The {\bf sequential disjunction} $A_1\sqd\ldots\sqd A_n$ of $A_1,\ldots,A_n$ is defined exactly as  $A_1\tgd\ldots\tgd A_n$, with the only difference that, in order for a position $\Phi$ to be a legal position of $A_1\sqd\ldots\sqd A_n$, it should satisfy the following {\em additional} condition: Whenever $ i_1,i_2,\ldots,i_j$ is the sequence of the switch moves made (by $\pp$) in $\Phi$, we have $i_1\equals 2,\ i_2\equals 3,\ \ldots,\ i_j\equals j\plus 1$.  
\item  The {\bf sequential conjunction} $A_1\sqc\ldots\sqc A_n$ of $A_1,\ldots,A_n$ is defined exactly as  $A_1\tgc\ldots\tgc A_n$, with the only difference that, in order for a position $\Phi$ to be a legal position of $A_1\sqc\ldots\sqc A_n$, it should satisfy the following {\em additional} condition: Whenever $ i_1, i_2,\ldots,i_j$ is the sequence of the switch moves made (by $\oo$) in $\Phi$, we have $i_1\equals 2,\ i_2\equals 3,\ \ldots,\ i_j\equals j\plus 1$. 
\end{enumerate}
\end{definition}

As we see, a legal run $\Gamma$ of $A_1\tgd\ldots\tgd A_n$ is nothing but a legal run of $A_1\mld\ldots\mld A_n$ with perhaps some switch moves by $\pp$ inserted. As in the case of $\mld$, such a $\Gamma$ is seen as a parallel play in the $n$ components, with the play (run) in each component $A_i$  being $\Gamma^{i.}$. The meaning of a switch move is that of a retractable choice of a disjunct. As we remember from \cite{Japfin}, in order to win a parallel disjunction, it is sufficient for $\pp$ to win in any one of its disjuncts. Winning a toggling disjunction is harder: here it is necessary for $\pp$ to win in the disjunct that was chosen last. The difference between $A_1\tgd\ldots\tgd A_n$ and $A_1\sqd\ldots\sqd A_n$ is that, while in the former switches/choices can go back and forth and the same component can be re-chosen many times, in the latter the chosen component should always be the one next after the previously chosen component. And, as always, either sort of conjunction is a dual of the corresponding sort of disjunction, obtained from it by interchanging the roles of $\pp$ and $\oo$. 

A reasonable behavior in a toggling or sequential combination of static games by either player is to always assume that the latest choice of a component is final, correspondingly play only in the (currently) active component of the combination, and worry about the other components only if and when they become active. This is so because eventually the outcome of the game will be determined by what happened in the (then) active component. If there are strategically useful moves in a dormant component, they can always wait till that component becomes active (if and when this happens), as postponing moves in a static game never hurts a player. Yet, this ``reasonable'' behavior is not enforced by the rules of the game, and ``unreasonable'' yet innocent actions such as making moves in dormant components are legal. This particular design choice in Definition \ref{nov10} has been made purely for the considerations of simplicity. Alternative definitions that yield equivalent operations but are more restrictive/demanding on legal behaviors are not hard to come up with. One example is the definition of sequential operations given in \cite{Japseq}. In any case, however, obtaining such definitions would not be just as straightforward as declaring all moves in dormant components illegal --- doing so could violate the important condition that the operations should preserve the static property of games. As we remember from \cite{Japfin}, computability logic and its static games are asynchronous-communication-friendly. If the communication between the two players is asynchronous, $\oo$ (in the case of $\tgd$ or $\sqd$) or $\pp$ (in the case of $\tgc$ or $\sqc$) cannot be sure that the component it considers ``active'' and in which it wants to make a move is indeed still active; it is possible that the component has already been ``deactivated'' by the adversary, but this information has not arrived yet.  

As an aside, the existence of a variety of alternative definitions for our game operations and their robustness (modulo equivalence) with respect to technical variations is a strong indication of the naturalness of those operations. This is in the same sense as the existence of many equivalent versions of Turing machines and other models of computation, and the robustness of the class of computable functions with respect to those variations, is one of the strongest pieces of empirical evidence in favor of the Church-Turing thesis. Among the interesting alternatives to our present definitions of the four sorts of disjunctions, most clearly showing the similarities and differences between them, is the following. (1) Keep the definition of $\mld$ (from \cite{Japfin}) unchanged. (2) Define $\tgd$ as in Definition \ref{nov10}, with the only difference that not having made any switch/choice moves is now considered a loss for $\pp$ (there is no ``default'' choice, that is). (3) Define  $\sqd$  as the version of (the new) $\tgd$ where the choices are required to be consecutive numbers starting from $1$. (4) Define $\add$ as the version of (the new) $\tgd$ where one single choice is allowed. Thus, the differences between the four disjunctions are in how many (if any) choices, and in what order, are required or allowed.  

For the sake of readability, in the sequel we will often employ relaxed, intuitive and semiformal terminology when referring to moves, typically naming them by their meanings or effects on the game instead of the actual strings that those moves are.  Consider, for example, the game $(A\add B)\mlc \bigl((C\adc D)\tgd (E\adc F)\bigr)$ and the legal run $\seq{\pp 1.1, \oo 2.1.1,\pp 2.2,\oo 2.2.2,\pp 2.1}$ of it. We may say that:
\begin{itemize}
\item The {\bf effect} of the move $1.1$ by $\pp$ is (or such a move {\bf signifies}) choosing $A$ within the $A\add B$ component. Notice that after this move is made in 
$(A\add B)\mlc \bigl((C\adc D)\tgc (E\adc F)\bigr)$, the game is {\bf brought down to} --- in the sense that it continues as --- $A \mlc \bigl((C\adc D)\tgd (E\adc F)\bigr)$. We may also refer to making the move $1.1$ in the overall game as making the move $1$ in its left $\mlc$-conjunct, because this is exactly the meaning of the move $1.1$. 
\item The effect of the next move $2.1.1$ by $\oo$ is choosing $C$ in the $C\adc D$ component. Such a move further brings the game down to $A \mlc \bigl(C \tgd (E\adc F)\bigr)$. 
\item The effect of the next, switch move $2.2$ by $\pp$ is  activating the right component of  $C \tgd (E\adc F)$, or {\bf switching} from $C$ to $E\adc F$ in it. Remember that initially the active component of a toggling combination is always the leftmost component. Such a component in our case was $C\adc D$, which later became ({\bf evolved to}) $C$. 
\item The effect of the next move $2.2.2$ by $\oo$ is choosing $F$ in the $E\adc F$ component. This brings the overall game down to  $A \mlc (C \tgd F)$, where the active (sub)component  within the $C\tgd F$ component is $F$.
\item The effect of the last move $2.1$ by $\pp$ is switching from $F$ back to $C$ in the $C \tgd F$ component, i.e. making the right component dormant and the left component active, as was the case when the game started. 
\end{itemize}

Definition \ref{nov10} straightforwardly extends from the $n$-ary cases to the infinite cases \ $A_1\tgd A_2\tgd\ldots$, \ \ $A_1\tgc$ $ A_2\tgc\ldots$, \ \ $A_1\sqd A_2\sqd\ldots$ \ and \ $A_1\sqc A_2\sqc\ldots$ \ by just changing ``$i\in\{1,\ldots,n\}$'' to ``$i\in\{1,2,\ldots\}$''. 
  
Even though we have officially defined  $\tgd\hspace{-2pt},\tgc\hspace{-2pt},\sqd\hspace{-2pt},\sqc$  only for constant games, our definitions extend to all games in the standard way, as explained in the second paragraph of Section 4 of \cite{Japfin}. Namely, for any not-necessarily-constant  games $A_1,\ldots,A_n$, $A_1\tgd\ldots\tgd A_n$ is the unique game such that, for any valuation (assignment of constants to variables) $e$, we have  $e[A_1\tgd\ldots\tgd A_n]=e[A_1]\tgd\ldots\tgd e[A_n]$. Similarly for $\tgc\hspace{-2pt},\sqd\hspace{-2pt},\sqc$ and the infinite versions of $\tgd\hspace{-2pt},\tgc\hspace{-2pt},\sqd\hspace{-2pt},\sqc\hspace{-2pt}$. (The meaning of the notation $e[\ldots]$, just like the meanings of any other unfamiliar terms or notations, as already noted, can  and should be looked up in \cite{Japfin}.) 

Whenever new game operations are introduced, one needs to make sure that they preserve the static property of games, for otherwise many things can go wrong: 
\begin{theorem}\label{static}
The class of static games is closed under (both the $n$-ary and the infinite versions of) $\tgd,\tgc,\sqd,\sqc$. 
\end{theorem}
\begin{proof} Given in Appendix A. \end{proof}

The remaining definitions of this section are not relevant to the subsequent sections of the paper, and we include them here just for the purposes of officially registering all toggling operations (and re-registering all sequential operations) as full-fledged inhabitants of the operation zoo of computability logic.

The {\bf toggling} and {\bf sequential} sorts of {\bf quantifiers} and {\bf (co)recurrences},  as we already know from Section \ref{s2}, are defined as follows:
\begin{definition}\label{nov102}
For any games $A$ or $A(x)$:
\begin{enumerate}
\item $\tgpst A\ = A\tgc A\tgc A\tgc\ldots$ \ and \ $\tgpcost A\ = A\tgd A\tgd A\tgd\ldots$;
\item $\sst A\ =\ A\sqc A\sqc A\sqc\ldots$ \ and  \ $\scost A\ =\ A\sqd A\sqd A\sqd\ldots$;
\item $\tga x A(x)\ =\ A(1)\tgc A(2)\tgc A(3)\tgc\ldots$ \ and \ $\tge\hspace{-1pt} x A(x)\ =\ A(1)\tgd A(2)\tgd A(3)\tgd\ldots$;
\item $\sqa xA(x)\ =\ A(1)\sqc A(2)\sqc A(3)\sqc\ldots$ \ and \ $\sqe xA(x)\ =\ A(1)\sqd A(2)\sqd A(3)\sqd\ldots$.
\end{enumerate} 
\end{definition}

It is not hard to see that the DeMorgan dualities hold for the toggling and sequential operations, just as they do for all other groups of operations (parallel, choice, branching, blind). Namely, we have:
\[\begin{array}{l}
\gneg(A_1\tgc\ldots\tgc A_n)\ =\ \gneg A_1\tgd\ldots\tgd \gneg A_n,\\
\gneg(A_1\tgd\ldots\tgd A_n)\ =\ \gneg A_1\tgc\ldots\tgc \gneg A_n,
\end{array}\]
and similarly for sequential conjunction and disjunction, as well as infinite versions of toggling and sequential operations, including $\tgst A$, $\tgcost A$, $\tga xA(x)$, $\tge\hspace{-1pt} xA(x)$, $\sst A$, $\scost A$, $\sqa xA(x)$, $\sqe xA(x)$.

Defining the {\bf toggling-branching} group $\tgst,\tgcost$ of {\bf (co)recurrences} is not as easy as defining the (just) toggling group $\tgpst,\tgpcost$. Let us first reproduce --- or, perhaps rather rephrase --- from \cite{Japfin}, in a compact form, the definition of branching recurrence $\st$ and the associated intuitions.  

  In semiformal terms, a play of $\st A$  starts as an ordinary play of game $A$. At any time, however, player $\oo$ is allowed to make a ``replicative move'', which creates two copies of the current position $\Phi$ of $A$. From that point on, the game turns into two games played in parallel, each continuing 
from position $\Phi$. We use the bits $0$ and $1$ to denote those two threads --- threads that have a common past (position $\Phi$) but possibly diverging futures. Again, at any time, $\oo$ can further branch either thread, creating two copies of the current position in that thread. If thread $0$ was branched, the resulting two threads will be denoted by $00$ and $01$; and if the branched thread was $1$, then the resulting threads will be denoted by $10$ and $11$. And so on: at any time, $\oo$ may split any of the existing threads $w$ into two threads $w0$ and $w1$. Each thread in the eventual run of the game will be thus denoted by a (possibly infinite) bit string. The game is considered won by $\pp$ if it wins $A$ in each of the threads; otherwise the winner is $\oo$.

In fully formal terms,  there are two types of legal moves in (legal) positions  of $\st A$: (1) replicative and (2) non-replicative. To define these, let us agree that by an {\em actual node}\footnote{Intuitively, an actual node is (the name of) an already existing thread of a play over $A$.} of a position $\Phi$ we mean a bit string $w$ such that $w$ either is empty,\footnote{Intuitively, the empty string is the name/address of the initial thread; all other threads will be descendants of that thread.} or else is $u0$ or $u1$ for some bit string $u$ such that $\Phi$ contains the move $\col{u}$. A replicative move can only be made by (is only legal for) $\oo$, and such a move in a given position $\Phi$ should be $\col{w}$, where $w$ is an actual node of $\Phi$ and $\Phi$ does not already contain the same move $\col{w}$.\footnote{The intuitive meaning of move $\col{w}$ is splitting thread $w$ into $w0$ and $w1$, thus ``actualizing'' these two new nodes/threads.} As for non-replicative moves, they can be made by either player. Such a move by a player $\xx$ in a given position $\Phi$ should be $w.\alpha$, where   $w$ is an actual node of $\Phi$ and $\alpha$ is a move such that, for any infinite bit string $v$, $\alpha$ is a legal move by $\xx$ in position $\Phi^{\preceq wv}$ of $A$.\footnote{The intuitive meaning of such a move $w.\alpha$ is making move $\alpha$ in thread $w$ and all of its (current or future) descendants.} Here  and later, for a run $\Theta$ and bit string $x$, $\Theta^{\preceq x}$ means the result of deleting from $\Theta$ all moves except those that look like $u.\beta$ for some initial segment $u$ of $x$, and then further deleting the prefix ``$u.$'' from such moves.\footnote{Intuitively, $\Theta^{\preceq x}$ is the run of $A$ that has been played in thread $x$, if such a thread exists (has been  generated); otherwise, $\Theta^{\preceq x}$ is the run of $A$ that has been played in (the unique) existing thread which (whose name, that is) is
some initial segment of $x$.} A legal run $\Gamma$ of $\st A$ is considered won by $\oo$ iff, for some infinite bit string $v$, $\Gamma^{\preceq v}$ is a $\oo$-won run of $A$.   This completes our definition of $\st$.

Now we are ready to define $\tgst A$. The legal runs of this game are defined as those of $\st A$, with the only difference that the former can now include any number of ``switch'' moves made by $\oo$ (and only $\oo$). Each switch move made in a position $\Phi$ is nothing but one of the actual nodes of $\Phi$. Next, a legal run $\Gamma$ is considered won by $\oo$ iff there are only finitely many switch moves made in it and, where $w$ is the last one of such moves and $v$ is $w$ with infinitely many $0$s appended to it, $\Gamma^{\preceq v}$ is a $\oo$-won run of $A$. If no switch moves were made at all, the above $w$ is assumed to be the empty bit string, so that $v$ is the infinite string of $0$s. This completes our definition of $\tgst$. The operation $\tgcost$ is defined by interchanging $\oo$ with $\pp$ in this definition. Equivalently, it can be defined by $\tgcost A=\gneg \tgst \gneg A$.   

We claim that, just like all other operations defined in this section, the operations $\tgst$ and $\tgcost$ preserve the static property of games. This claim can be proven by adapting to $\tgst$ the proof of the similar theorem about $\st$ given in \cite{Jap03}. The latter is fairly technical, for which reason we let the present claim remain without a  proof --- no results of this paper depend on it, anyway.

\section{Logic $\clt$}\label{intr}
In this section we introduce the new propositional system $\clt$.\footnote{In this name, ``CL'' stands for ``computability logic'', and ``$13$'' indicates that this  is the 13th formal system for CoL introduced so far.}   The building blocks of its language are:
\begin{itemize}
\item Infinitely many nonlogical {\bf elementary atoms}, for which we use the metavariables $p,q,r,s$.
\item Infinitely many nonlogical {\bf general atoms}, for which we use the metavariables $P,Q,R,S$.
\item The 0-ary operators $\twg$ and $\tlg$. They can as well be called {\bf logical atoms}.
\item The unary operator $\gneg$.
\item The operators $\mlc,\hspace{1pt}\mld,\tgc\hspace{-2pt},\tgd\hspace{-2pt},\sqc\hspace{-2pt},\sqd\hspace{-2pt},\hspace{2pt}\adc,\hspace{2pt}\add$. Their arities are not fixed and can be any $n\geq 2$.\footnote{The reason why we do not restrict these operators to their strictly binary versions is that doing so could lengthen rather than shorten many definitions and proofs. This is related to the fact that, while the games  $(A\mlc B)\mlc C$  and $A\mlc B\mlc C$ are certainly {\em equivalent}, they are still   not the {\em same} unless $A,B,C$ are elementary. Similarly for $\mld,\adc,\add$.}
\end{itemize}

{\bf Formulas}, to which we refer as {\bf $\clt$-formulas}, are built from atoms and operators in the standard way, with the requirement (yielding no loss of expressiveness) that $\gneg$ can only be applied to nonlogical atoms. A {\bf literal} means $L$ or $\gneg L$, where $L$ is an atom. Such a literal is said to be elementary, general, nonlogical or logical if $L$ is so. When $F$ is not a nonlogical atom, $\gneg F$ is understood as an abbreviation defined by:
\begin{itemize}
\item $\gneg\gneg E  =  E$;
\item $\gneg \twg  =  \tlg$ and $\gneg \tlg =  \twg$;
\item $\gneg(E_1\mlc\ldots\mlc E_n)  =  \gneg E_1\mld\ldots\mld \gneg E_n$ and $\gneg(E_1\mld\ldots\mld E_n)  =  \gneg E_1\mlc\ldots\mlc \gneg E_n$;
\item $\gneg(E_1\tgc\ldots\tgc E_n)  =  \gneg E_1\tgd\ldots\tgd \gneg E_n$ and $\gneg(E_1\tgd\ldots\tgd E_n)  =  \gneg E_1\tgc\ldots\tgc \gneg E_n$;
\item $\gneg(E_1\sqc\ldots\sqc E_n)  =  \gneg E_1\sqd\ldots\sqd \gneg E_n$ and $\gneg(E_1\sqd\ldots\sqd E_n)  =  \gneg E_1\sqc\ldots\sqc \gneg E_n$;
\item $\gneg(E_1\adc\ldots\adc E_n)  =  \gneg E_1\add\ldots\add \gneg E_n$ and $\gneg(E_1\add\ldots\add E_n)  =  \gneg E_1\adc\ldots\adc \gneg E_n$.
\end{itemize}
Also, if we write $E\mli F$, it is to be understood as an abbreviation of $\gneg E\mld F$. Simplicity is the only reason for our choice to treat $\mli$ and $\gneg$ as abbreviations: doing otherwise would lengthen the definition of system $\clt$, as well as the related proofs, due to the necessity to additionally/separately consider the cases of negative occurrences of operators in formulas. Note that our design choice does not yield any loss of expressive power: the games $A\mli B$ and $\gneg A\mld B$ are not only equivalent, but simply {\em equal}; similarly for $\gneg\gneg A$ vs.\hspace{-3pt} $A$, $\gneg (A\mlc B)$ vs.\hspace{-3pt} $\gneg A\mld \gneg B$, etc. 

Parentheses will often be omitted for the sake of readability. When doing so, we agree that $\gneg$ has the highest precedence, and $\mli$ has the lowest precedence. So, for instance, $\gneg P\add Q\mli R\tgc S$ means $((\gneg P)\add Q)\mli (R\tgc S)$. 

Formulas that do not contain general atoms 
we call {\bf elementary-base}. This terminology also extends to the corresponding fragment of $\clt$. In particular,  the 
{\em elementary-base fragment} of $\clt$ is the set of all elementary-base theorems of $\clt$. 

An {\bf interpretation}\label{z6} for the language of $\clt$ is a function $^*$ that sends each nonlogical elementary atom $p$ to an elementary game $p^*$, and sends each general atom $P$ to any, not-necessarily-elementary, static game $P^*$. This mapping extends to all formulas by letting it respect all logical operators as the corresponding game operations. That is, $\twg^*=\twg$, $(E\tgc F)^*=E^*\tgc F^*$, etc. When $F^*=A$, we say that {\bf $^*$ interprets $F$ as $A$}.

Recall, from \cite{Japfin}, the models HPM (``hard-play machine'') and EPM (``easy-play machine'') of interactive computation. Very briefly, an HPM is Turing machine with the capability of making moves and with full reading access to all relevant information on the game played, such as its ``current'' position at any given time, as well as the ``actual'' values of the parameters (variables) on which the game may depend; the environment can make any number of moves at any computation step of an HPM, with such moves being the only nondeterministic events from the HPM's perspective. An EPM means the same, with the difference that the environment can move only when the machine explicitly ``grants permission'' to do so, with the machine required to grant permission infinitely many times during the play. CoL defines {\em computability} of a game as existence of an HPM that wins the game against every possible behavior of the environment. It has been shown that, when the  games under question are static, replacing ``HPM'' 
with ``EPM'' in this definition yields (extensionally) the same concept of computability. Since our interpretations are required to interpret atoms as static games and all of our game operations preserve the static property of games, the two --- HPM and EPM --- models of computation are equivalent in every relevant respect.   

A formula $F$ is said to be {\bf valid} iff, for every interpretation $^*$, the game $F^*$ is computable. And $F$ is {\bf uniformly valid} iff there is an EPM (or, equivalently, HPM) $\cal S$, called a {\bf uniform solution} of $F$, such that $\cal S$ wins (computes) $F^*$ for every interpretation $^*$. See Sections 7 and 8 of \cite{Japfin} for a discussion of the intuitions associated with these concepts.

Our treatment further relies of the following terminology, notation and conventions:

\begin{itemize}
\item We often need to differentiate between {\em subformulas} as such, and particular {\em occurrences} of subformulas. We will be using the term {\bf osubformula} (``o'' for ``occurrence'') to mean a subformula together with a particular occurrence.  The prefix ``o'' will be used with a similar meaning in terms such as {\bf oatom}, {\bf oliteral}, etc. So, for instance, the formula $P\mlc Q\mli P$ has two {\em atoms} but three {\em oatoms}. Yet,  we may still say ``the oatom $P$'', assuming that it is clear from the context which of the possibly many occurrences of $P$ is meant. Similarly for osubformulas and oliterals. 
\item An occurrence of an osubformula is {\bf positive} iff it is not in the scope of $\gneg$. Otherwise it is {\bf negative}. According to our conventions regarding the usage of $\gneg$, only oatoms may be negative.  
\item An occurrence of an osubformula is {\bf positive} iff it is not in the scope of $\gneg$. Otherwise it is {\bf negative}. According to our conventions regarding the usage of $\gneg$, only oatoms may be negative.  
\item A {\bf politeral} is a positive oliteral. 
\item A {\bf $\tgc$-(sub)formula} is a (sub)formula of the form $F_1\tgc\ldots\tgc F_n$. Similarly for $\tgd\hspace{-2pt},\hspace{1pt}\mlc,\mld,\hspace{-1pt}\sqc\hspace{-2pt},\hspace{-2pt}\sqd\hspace{-2pt},\hspace{1pt}\adc,\hspace{1pt}\add$.
\item A {\bf sequential (sub)formula} is one of the form $F_1\sqc\ldots\sqc F_n$ or $F_1\sqd\ldots\sqd F_n$. We say that $F_1$ is the {\bf head} of such a (sub)formula, 
and $F_2,\ldots, F_n$ form its {\bf tail}. 
\item Similarly, a {\bf parallel (sub)formula} is one of the form $F_1\mlc\ldots\mlc F_n$  or $F_1\mld\ldots\mld F_n$, a {\bf toggling (sub)formula} is one of the form $F_1\tgc\ldots\tgc F_n$  or $F_1\tgd\ldots\tgd F_n$, and a {\bf choice (sub)formula} is one of the form $F_1\adc\ldots\adc F_n$  or $F_1\add\ldots\add F_n$.

\item A formula is said to be  {\bf quasielementary} iff it contains no general atoms and no operators other than $\gneg,\twg,\tlg,\mlc,\mld,\tgc\hspace{-2pt},\tgd\hspace{-2pt}$.  
 
\item A formula is said to be  {\bf elementary} iff it is a formula of classical propositional logic, i.e., contains no general atoms and no operators other than $\gneg,\twg,\tlg,\mlc,\mld$.

\item A {\bf semisurface osubformula} (or {\bf occurrence}) is an osubformula (or occurrence) that is not in the scope of any choice connectives (i.e., $\adc$ and $\add$). 

\item A {\bf surface osubformula} (or {\bf occurrence}) is an osubformula (or occurrence) that is not in the scope of any connectives other than $\gneg,\mlc,\mld$.

\item The {\bf quasielementarization} of a formula $F$, denoted by $\qelz{F}$, is the result of replacing in $F$ every sequential osubformula by its head, every $\adc$-osubformula   by $\twg$, every  $\add$-osubformula   by $\tlg$, and every general politeral by $\tlg$ (the order of these replacements does not matter). For instance, \[\qelz{((P\tgd q)\mld ((p\mlc\gneg P)\sqc (Q\mlc R)))\tgc (q\adc (r\add s))}= ((\tlg\tgd q) \mld (p\mlc \tlg))\tgc \twg.\]

\item The {\bf elementarization} of a quasielementary formula $F$,  denoted by $\elz{F}$, is the result of replacing in  $F$ every $\tgc$-osubformula  by $\twg$ and every  $\tgd$-osubformula by $\tlg$ (again, the order of these replacements does not matter). For instance, 
\[\elz{(s\mlc(p\tgc(q\tgd r)))\mld (\gneg s\mld (p\tgd r))}= (s\mlc\twg)\mld (\gneg s\mld \tlg).\]
 
\item A  quasielementary formula $F$ is said to be {\bf stable} iff its elementarization $\elz{F}$ is a tautology of classical logic. Otherwise $F$ is {\bf instable}.   
\end{itemize}

\begin{definition}\label{defcl9}
We define our deductive system $\clt$ by the following six {\bf rules of inference}, where ${\cal P}\mapsto F$ means ``from premise(s) $\cal P$ conclude $F$'':

\begin{description}
\item[$(\hspace{-2pt}\tgc\hspace{-2pt})$:] $\vec{H}\mapsto F$, where $F$ is a stable quasielementary formula, and  $\vec{H}$ the smallest set of formulas satisfying the following condition: 
\begin{itemize}
\item Whenever $F$ has a surface osubformula  $E_1\tgc\ldots\tgc E_n$,  for each $i\in\{1,\ldots,n\}$, $\vec{H}$ contains the result of replacing  in $F$ that osubformula by $E_i$. 
\end{itemize}
\item[$(\hspace{-2pt}\tgd\hspace{-2pt})$:] $H\mapsto F$, where $F$ is a quasielementary formula, and $H$ is the result of replacing in $F$ a surface osubformula $E_1\tgd\ldots\tgd E_n$  by $E_i$ for some $i\in\{1,\ldots, n\}$.
\item[$(\hspace{-2pt}\sqc\hspace{-2pt}\adc)$:]  $\qelz{F},\vec{H}\mapsto F$, where 
$F$ is a non-quasielementary formula (note that otherwise $F=\qelz{F}$), and  
$\vec{H}$ is the smallest set of formulas satisfying the following two conditions: 
\begin{itemize} 
\item Whenever $F$ has a semisurface osubformula $E_1\adc\ldots\adc E_n$, for each 
$i\in\{1,\ldots,n\}$, $\vec{H}$ contains the result of replacing in $F$ that osubformula  by $E_i$.
\item Whenever $F$ has a semisurface osubformula $E_1\sqc E_2\sqc\ldots\sqc E_n$, $\vec{H}$ contains the result of replacing in $F$ that osubformula  by $E_2\sqc\ldots\sqc E_n$.\footnote{In this definition and later, if $n\equals 2$, $E_2\sqci\ldots\sqci E_n$ or $E_2\sqdi\ldots\sqdi E_n$ is simply understood as $E_2$.} 
\end{itemize}
We will be referring to the above formula $\qelz{F}$ as the {\bf senior premise} of the rule (or its conclusion), and referring to the formulas of $\vec{H}$ as {\bf junior premises}.
\item[$(\add)$:]  $H\mapsto F$, where $H$ is the result of replacing in $F$ a semisurface osubformula $E_1\add\ldots\add E_n$  by $E_i$ for some $i\in\{1,\ldots, n\}$.
\item[$(\hspace{-2pt}\sqd\hspace{-2pt})$:]  $H\mapsto F$, where $H$ is the result of replacing in $F$ a semisurface osubformula $E_1\sqd E_2\sqd\ldots\sqd E_n$  by 
$E_2\sqd\ldots\sqd E_n$.$^{14}$
\item[$(\mbox{M})$:] $H\mapsto F$, where $H$ is the result of replacing in $F$ two --- one positive and one negative ---
semisurface occurrences of some general atom $P$ by a nonlogical elementary atom $p$ that does not occur in $F$.
\end{description}
\end{definition}

Axioms are not explicitly stated, but the set of premises of the $(\hspace{-2pt}\tgc\hspace{-2pt})$ rule can be empty, in which case (the conclusion of) this rule acts like an axiom. 

A {\bf $\clt$-proof} of a formula $F$ can be defined as a sequence of formulas ending with $F$ such that every formula follows from some (possibly empty) set of earlier formulas by one of the rules of the system.

\begin{example}\label{oct17aa} 
In view of the discussions of Section \ref{s2tog}, the formula $(\gneg p\tgd p)\mlc (\gneg q\tgd q)\mli \gneg (p\mlc q)\tgd (p\mlc q)$ expresses --- or rather implies --- the known fact that, if two predicates $p$ and $q$ are recursively approximable, then so is their intersection $p\mlc q$. The following sequence is a $\clt$-proof of this formula, re-written into its official form $\bigl((p\tgc \gneg p)\mld (q\tgc \gneg q)\bigr)\mld \bigl((\gneg p\mld \gneg q)\tgd (p\mlc q)\bigr)$:\vspace{7pt}

\noindent 1.  $(p\mld q)\mld (\gneg p\mld \gneg q)$  \ \ \ \ \ $(\hspace{-2pt}\tgc\hspace{-2pt})$: \{\}

\noindent 2.  $(p\mld q)\mld \bigl((\gneg p\mld \gneg q)\tgd (p\mlc q)\bigr)$ \ \ \ \  \ $(\hspace{-2pt}\tgd\hspace{-2pt})$: 1

\noindent 3.  $(p\mld \gneg q)\mld (\gneg p\mld \gneg q)$ \ \ \ \ \ $(\hspace{-2pt}\tgc\hspace{-2pt})$: \{\}

\noindent 4.  $(p\mld \gneg q)\mld \bigl((\gneg p\mld \gneg q)\tgd (p\mlc q)\bigr)$ \ \ \ \ \ $(\hspace{-2pt}\tgd\hspace{-2pt})$: 3

\noindent 5.  $\bigl(p\mld (q\tgc \gneg q)\bigr)\mld \bigl((\gneg p\mld \gneg q)\tgd (p\mlc q)\bigr)$ \ \ \ \ \ $(\hspace{-2pt}\tgc\hspace{-2pt})$: \{2,4\}

\noindent 6.  $(\gneg p\mld q)\mld (\gneg p\mld \gneg q)$ \ \ \ \ \ $(\hspace{-2pt}\tgc\hspace{-2pt})$: \{\}

\noindent 7.  $(\gneg p\mld q)\mld \bigl((\gneg p\mld \gneg q)\tgd (p\mlc q)\bigr)$ \ \ \ \ \ $(\hspace{-2pt}\tgd\hspace{-2pt})$: 6

\noindent 8.  $(\gneg p\mld  \gneg q)\mld  (p\mlc q)$ \ \ \ \ \ $(\hspace{-2pt}\tgc\hspace{-2pt})$: \{\}

\noindent 9.  $(\gneg p\mld  \gneg q)\mld \bigl((\gneg p\mld \gneg q)\tgd (p\mlc q)\bigr)$ \ \ \ \ \ $(\hspace{-2pt}\tgd\hspace{-2pt})$: 8

\noindent 10.  $\bigl(\gneg p\mld (q\tgc \gneg q)\bigr)\mld \bigl((\gneg p\mld \gneg q)\tgd (p\mlc q)\bigr)$ \ \ \ \ \ $(\hspace{-2pt}\tgc\hspace{-2pt})$: \{7,9\}

\noindent 11.  $\bigl((p\tgc \gneg p)\mld q\bigr)\mld \bigl((\gneg p\mld \gneg q)\tgd (p\mlc q)\bigr)$ \ \ \ \ \ $(\hspace{-2pt}\tgc\hspace{-2pt})$: \{2,7\}

\noindent 12.  $\bigl((p\tgc \gneg p)\mld \gneg q\bigr)\mld \bigl((\gneg p\mld \gneg q)\tgd (p\mlc q)\bigr)$ \ \ \ \ \ $(\hspace{-2pt}\tgc\hspace{-2pt})$: \{4,9\}

\noindent 13.  $\bigl((p\tgc \gneg p)\mld (q\tgc \gneg q)\bigr)\mld \bigl((\gneg p\mld \gneg q)\tgd (p\mlc q)\bigr)$ \ \ \ \ \ $(\hspace{-2pt}\tgc\hspace{-2pt})$: \{5,10,11,12\}\vspace{7pt}

\end{example}

\begin{example}\label{oct17}
 Pick any two distinct connectives $\mbox{\scriptsize {\bf AND}}_1$ and $\mbox{\scriptsize {\bf AND}}_2$ from the list  $\mlc,\tgc\hspace{-2pt},\sqc\hspace{-2pt},\hspace{2pt}\adc$. Then $\clt$ proves the formula $P\mbox{\scriptsize {\bf AND}}_1Q\mli P\mbox{\scriptsize {\bf AND}}_2 Q$ if and only if $\mbox{\scriptsize {\bf AND}}_1$ is to the left of $\mbox{\scriptsize {\bf AND}}_2$ in the list. Similarly for the list $\add,\sqd\hspace{-2pt},\tgd\hspace{-2pt},\hspace{2pt}\mld$. Below we verify this fact only for the case $\{\mbox{\scriptsize {\bf AND}}_1,\mbox{\scriptsize {\bf AND}}_2\}=\{\hspace{-2pt}\tgc\hspace{-2pt},\hspace{-2pt}\sqc\hspace{-2pt}\}$. The reader may want to try some other combinations as exercises.

Here is a $\clt$-proof of $P\tgc Q\mli P\sqc Q$:\vspace{7pt}

\noindent 1.  $\gneg p\mld p$ \ \ \ \ \ $(\hspace{-2pt}\tgc\hspace{-2pt})$: \{\} 

\noindent 2.  $(\gneg p\tgd \tlg)\mld p$ \ \ \ \ \ $(\hspace{-2pt}\tgd\hspace{-2pt})$: 1

\noindent 3.  $\gneg q\mld  q$ \ \ \ \ \ $(\hspace{-2pt}\tgc\hspace{-2pt})$: \{\} 

\noindent 4.  $(\gneg p\tgd \gneg q)\mld  q$ \ \ \ \  \ $(\hspace{-2pt}\tgd\hspace{-2pt})$: 3 

\noindent 5.  $(\gneg p\tgd \gneg Q)\mld  Q$ \ \ \ \ \ (M): 4 

\noindent 6.  $(\gneg p\tgd \gneg Q)\mld (p\sqc Q)$ \ \ \ \ \ $(\hspace{-2pt}\sqc\hspace{-2pt}\adc)$: \{2,5\}

\noindent 7.  $(\gneg P\tgd \gneg Q)\mld (P\sqc Q)$ \ \ \ \ \ (M): 6\vspace{7pt}

On the other hand, the formula $P\sqc Q\mli P\tgc Q$, i.e. $(\gneg P\sqd \gneg Q)\mld (P\tgc Q)$, has no proof. This can be shown through attempting and failing to construct, bottom-up, a purported $\clt$-proof of the formula. Here we explore one of the branches of a proof-search tree.  $(\gneg P\sqd \gneg Q)\mld (P\tgc Q)$ is not quasielementary, so it could not be derived by (be the conclusion of) the $(\hspace{-2pt}\tgd\hspace{-2pt})$ or $(\hspace{-2pt}\tgc\hspace{-2pt})$ rule.  The $(\add)$ rule does not apply either, as there is no $\add$ in the formula. This leaves us with one of the rules $(\hspace{-2pt}\sqd\hspace{-2pt})$, $(\hspace{-2pt}\sqc\hspace{-2pt}\adc)$ and (M). Let us  see what happens if our target formula is derived by $(\hspace{-2pt}\sqd\hspace{-2pt})$. In this case the premise should be  $\gneg Q\mld (P\tgc Q)$. The latter can be derived only by  $(\hspace{-2pt}\sqc\hspace{-2pt}\adc)$  or (M). Again, let us  try (M). The premise in this subcase should be $\gneg q\mld (P\tgc q)$ for some elementary atom $q$. But the only way $\gneg q\mld (P\tgc q)$  can be derived is by $(\hspace{-2pt}\sqc\hspace{-2pt}\adc)$ from the premise $\gneg q\mld (\tlg\tgc q)$. This formula, in turn, could only be derived by  $(\hspace{-2pt}\tgc\hspace{-2pt})$, in which case $\gneg q\mld \tlg$ is one of the premises. Now we are obviously stuck, as $\gneg q\mld \tlg$ is not the conclusion of any of the rules of the system. We thus hit a dead end.  All remaining possibilities can be checked in a similar routine/analytic way, and the outcome in each case will be a dead end. 
\end{example}

\begin{exercise}\label{jan327}
Construct $\clt$-proofs of formulas (\ref{nov16b}) and (\ref{nov16c}) from Section \ref{s2seq}.
\end{exercise}

In  similar analytic exercises, one can verify that the provabilities ($\vdash$) and unprovabilities ($\not\vdash$) shown in the following table hold. From the table we can see that each sort of disjunction exhibits a unique logical behavior even when taken in isolation from the other sorts. The same, of course, is the case for conjunctions.

\[\begin{array}{ccccc}
 & \mbox{\scriptsize {\bf OR}}=\mld & \mbox{\scriptsize {\bf OR}}=\tgd & \mbox{\scriptsize {\bf OR}}=\sqd & \mbox{\scriptsize {\bf OR}}=\add \\
\gneg P \hspace{2pt}\mbox{\scriptsize {\bf OR}}\hspace{2pt} P & \vdash & \not\vdash & \not\vdash & \not\vdash \\
P\hspace{2pt}\mbox{\scriptsize {\bf OR}}\hspace{2pt}Q\mli Q\hspace{2pt}\mbox{\scriptsize {\bf OR}}\hspace{2pt} P & \vdash & \vdash & \not\vdash & \vdash \\
P\hspace{2pt}\mbox{\scriptsize {\bf OR}}\hspace{2pt} P\mli P & \not\vdash & \not\vdash & \not\vdash & \vdash \\
p\hspace{2pt}\mbox{\scriptsize {\bf OR}}\hspace{2pt} p\mli p & \vdash & \vdash & \vdash & \vdash 

\end{array} \]

The present version of $\clt$ is not syntactically optimal in the sense that certain alternative design choices could offer considerably shorter proofs. Namely, the $(\hspace{-2pt}\tgc\hspace{-2pt})$ rule and, especially, the $(\hspace{-2pt}\sqc\hspace{-2pt}\adc)$ rule are generally quite expensive in that they may sometimes require too many premises. Optimizing (at the expense of simplicity) the system or finding various shortcuts (admissible rules) to reduce proof sizes would be a purely syntactic job which is beyond the scope of the present paper. Below we only point out --- without a proof --- one of the ways to improve the efficiency of proofs and analytic proof-search procedures: 

\begin{claim}\label{clai}
The set of $\clt$-provable formulas would remain the same if we redefined the concept of a semisurface occurrence  by imposing the additional condition that 
such an  occurrence should not be in the tail of any sequential osubformula. 
\end{claim}

Below comes our main theorem. Its soundness part will be proven in Section \ref{ssound}, and its completeness part will be proven in Section \ref{s9}.  
 
\begin{theorem}\label{thcl2}
$\clt\vdash F$ iff $F$ is valid (for any $\clt$-formula $F$).
Furthermore:

a) There is an effective procedure that takes a $\clt$-proof of an arbitrary formula $F$ and 
constructs an EPM $\cal S$ such that,  for every interpretation $^*$, \ $\cal S$ computes $F^*$.

b) If $\clt\not\vdash F$, then $F^*$ is not computable  for some  interpretation 
$^*$ that interprets all elementary atoms of $F$ as finitary predicates of arithmetical complexity $\Delta_3$, and interprets all general atoms of $F$ as problems of the form
\((A^{1}_{1}\add\ldots\add A_{m}^{1})\adc\ldots\adc (A_{1}^{m}\add\ldots\add A_{m}^{m}),\)
where each $A_{i}^{j}$ is a finitary predicate of arithmetical complexity $\Delta_3$.
\end{theorem}

The following facts are immediate corollaries of Theorem \ref{thcl2}, so we state them without proofs:

\begin{fact}\label{fact1}
A $\clt$-formula is valid iff it is uniformly valid.
\end{fact}

\begin{fact}\label{fact2}
$\clt$ is a conservative extension of classical logic. That is, an elementary formula is provable in $\clt$ iff it is a classical tautology. 
\end{fact}

In addition, one can easily see that $\clt$ is closed under the standard rules of logic, such as Modus Ponens and Substitution, provided that, when applying Substitution, elementary atoms are only replaced by elementary formulas. 
  
It is also worth noting that $\clt$ is decidable, with a brute force decision algorithm 
obviously running in at most polynomial space. Whether there are more efficient algorithms is unknown. 

The remaining  sections of this paper are devoted to a proof of Theorem \ref{thcl2}. Two immediate predecessors of that theorem are similar results proven in \cite{Japtocl2} for the $\sqc\hspace{-2pt},\sqd\hspace{-2pt},\tgc\hspace{-2pt},\tgd$-free fragment {\bf CL2} of $\clt$, and in \cite{Japseq} for the $\tgc\hspace{-2pt},\tgd$-free fragment {\bf CL9} of $\clt$. The proofs of both soundness and completeness given in \cite{Japtocl2} were very detailed, to the extent of being pedantically so. As a result, they were fairly long, even though concerning a relatively modest and simple fragment of the logic. The soundness and completeness proofs given in \cite{Japseq} were less detailed, yet still pretty formal. The proofs given in the present paper are far less detailed and quite informal. Continuing proving things in full technical detail in the earlier style would make both writing and reading papers in CoL increasingly difficult, as the studies of the subject advance and the depth and breadth of the results keep increasing. The proofs given in \cite{Japtocl2,Japseq} can serve as  illustrations  of the fact that relaxed and semiformal arguments like the ones relied upon in the present paper can always be turned into strict formal proofs if necessary. The proofs given in the following sections are meant for an advanced reader who has well internalized the concepts and techniques of computability logic.  

\section{The soundness of $\clt$}\label{ssound}

Very briefly, the idea behind our soundness proof for $\clt$ can be outlined as follows. Every $\clt$-proof of a formula $F$, in fact, encodes an input- and interpretation-independent winning strategy (EPM) $\cal S$ for $F^*$. This is a recursive strategy, which acts depending on by which of the rules of $\clt$ the formula was obtained from its 
 its premise(s). 

If $F$ is derived by $(\hspace{-2pt}\tgd\hspace{-2pt})$, $\cal S$ makes a move signifying a switch to the corresponding $\tgd$-disjunct in the corresponding subgame of $F$, this way essentially bringing the game $F$ down to its premise. For instance, if  $F$ is $E\mlc (G_1\tgd G_2)$ and it is derived from $E\mlc G_2$,  then $2.2$ is such a move. After this, $\cal S$ calls itself on the premise, which, by the induction hypothesis, it knows how to win. Similarly, if $F$ is derived by $(\add)$, $\cal S$ makes a move signifying a choice of the corresponding $\add$-disjunct in the corresponding subgame of $F$; and if $F$ is derived by $(\hspace{-2pt}\sqd\hspace{-2pt})$, $\cal S$ makes a move signifying a switch to the next (second) $\sqd$-disjunct in the corresponding subgame of $F$.   

Next, suppose   $F$ is derived by the (M) rule  from the premise $H$, namely, $H$ is the result of replacing, in $F$, two general politerals $P,\gneg P$ by some elementary politerals $p,\gneg p$.  This is a signal for $\cal S$ that, from now on, through applying copycat, its should ``match'', i.e. ``synchronize'', the corresponding two subgames (the ones now represented by $p$ and $\gneg p$) to ensure that they evolve in essentially the same ways, so that, eventually, one of them will be won.  This allows us to safely pretend (at least in our present, very relaxed description/analysis) that the games represented by $p$ and $\gneg p$ are elementary (with one being the negation of the other) and thus the game played by $\cal S$ from now on is, in fact, $H$. The latter, by the induction hypothesis, will be won.  

Next, suppose $F$ is derived by the $(\hspace{-2pt}\sqc\hspace{-2pt}\adc)$ rule.  Then  $\cal S$ simply plays $|F|$ (the senior premise of $F$) until the environment makes a move signifying a choice of a $\adc$- or $\sqc$-conjunct in some $\adc$- or $\sqc$-subgame of $F$, in which case $\cal S$ quits its play of $|F|$ and calls itself on the corresponding junior premise of $F$, which it wins according to the induction hypothesis. If, however, the above event never occurs, the overall play will finish (or continue forever) as a play of $|F|$ which, by the induction hypothesis, will be won by $\cal S$. This, in turn, can be seen to imply that $F$ will also be won, because playing $F$ in the scenario where no moves associated with its choice and sequential components are made, essentially means nothing but simply playing the quasielementarization $|F|$ of $F$. 

Finally, suppose     $F$ is derived by $(\hspace{-2pt}\tgc\hspace{-2pt})$. This is the hardest case to explain in non-technical terms and, for simplicity, let us   assume that $F$ has a single surface occurrence of a $\tgc$-subformula, namely of $E_1\tgc\ldots\tgc E_n$. As we remember, the default active (``chosen'') $\tgc$-conjunct of such a component is $E_1$. Let us write $E_i$ instead, because the original value $1$ of $i$ may change later, anyway. Let $H_i$ be the premise of $F$ resulting from the latter by replacing $E_1\tgc\ldots\tgc E_n$ with its active component $E_i$. What $\cal S$ does in this case is that it simply plays $H_i$, which it knows how to win (by the induction hypothesis), and winning which means winning $F$ as long as $E_i$ remains active. 
Such a play may, however, be interrupted by one of the following events. One possible event is that, while $\cal S$ is playing $H_i$, the environment ``activates'' some new $\tgc$-conjunct $E_j$   of the $E_1\tgc\ldots\tgc E_n$ component of $F$. Then $\cal S$ abandons its play of $H_i$ and switches to a play of $H_j$ instead, where $H_j$ is the corresponding premise of $F$. Another possibility is that the current play of $F$, itself,  is taking place within a   procedure call by a previous similar procedure associated with a $(\hspace{-2pt}\tgc\hspace{-2pt})$-derived formula $F'$ (which, for simplicity, we also assume to have a single surface occurrence of a $\tgc$-subformula), and the environment suddenly ``changes its mind'' and activates a new $\tgc$-conjunct of the $\tgc$-subcomponent of that $F'$. Then, again, $\cal S$ abandons its current play of $H_i$ and switches to a play of $H'_j$, where $H'_j$ is the corresponding premise of $F'$. Yet another possibility is that the current play of $F$  is taking place within a   procedure call by a previous  procedure associated with a $(\hspace{-2pt}\sqc\hspace{-2pt}\adc)$-derived formula $F'$, and the environment makes a move that signifies a  choice of a $\adc$-conjunct or a (new) $\sqc$-conjunct in some $\adc$- or $\sqc$-subcomponent of that $F'$. 
In this case, $\cal S$ abandons $F$ and calls itself on the corresponding premise of $F'$. After properly taking care of  certain details that we have suppressed  
and (over)simplified, this strategy can be shown to guarantee a win for $\cal S$.

But, again, the above was just a rough preliminary description of the proof idea. What follows is a relatively detailed materialization of that outline. We start by stating the following general fact, which is a reproduction of Lemma 4.7 of \cite{Jap03}:

\begin{lemma}\label{may19}
Assume $A$ is a constant static game, $\xx$ is either player, $\gneg \xx$ is $\xx$'s adversary (the other player), and $\Upsilon,\Omega$ are runs such that $\Omega$ is a $\xx$-delay of $\Upsilon$. Then:
\begin{enumerate}
\item If $\Omega$ is a $\xx$-illegal run of $A$, then so is $\Upsilon$.
\item If $\Upsilon$ is a $\pneg\xx$-illegal run of $A$, then so is $\Omega$.
\end{enumerate}
\end{lemma}

Throughout  this section, we fix $F$ as an arbitrary $\clt$-provable formula. Our goal is to (show how to) construct a uniform solution for $F$.

We fix a $\clt$-proof $T$ of $F$. In the preceding section, such a proof would be presented in the form of a sequence of formulas. But now we prefer to see  $T$  as a {\em tree}. Namely, $T$ is a tree where with each node $b$ is associated --- let us say {\bf $\cal F$-associated} --- a formula, which we denote  by ${\cal F}(b)$, such that the formula associated with the root is $F$ and, for each node $b$ of $T$, whenever $c_1,\ldots,c_n$ are the children of $b$, ${\cal F}(b)$ follows from ${\cal F}(c_1),\ldots,{\cal F}(c_n)$ by one of the rules of inference. We will refer to such a (fixed for each $b$) rule as the {\bf justification} of $b$. Note that the justification of a leaf (childless node) can only be $(\hspace{-2pt}\tgc\hspace{-2pt})$, as any other rule requires at least one premise. 

When the justification of a given node is the $(\hspace{-2pt}\sqc\hspace{-2pt}\adc)$ rule, the child with which the senior premise is $\cal F$-associated is said to be the {\bf senior child} of the node, and the children with which junior premises are associated are said to be {\bf junior children}.  

We may safely assume that, in the $\clt$-proof that we consider, the (M) rule never introduces into the premise an elementary atom that is already introduced this way elsewhere in the proof, or that had occurrences in the original formula $F$. Under this assumption,  we  agree to call each elementary atom $p$ that was introduced in 
$T$ by an application of (M) (in the bottom-up view of this rule), as well as the corresponding literals $p$ and $\gneg p$, {\bf pseudoelementary}.   The general atom $P$ that was replaced by $p$ in this process will be said to be the {\bf origin} of $p$. We also say that the above $\gneg p$ (resp. $p$) is the {\bf matching} literal of $p$ (resp. $\gneg p$).

As will be  seen, intuitively, the formulas $\cal F$-associated  with the nodes of $T$  are partial, compressed representations of the games (``positions'') to which the original game $F$ evolves at different stages of a play according to the strategy that we are going to construct.  To make both a description and an analysis of that strategy easier, we need fuller representations of those games. For this purpose we use additional expressions  called {\bf hyperformulas}. These are nothing but formulas with perhaps some osubformulas {\bf underlined} (such underlines can be iterated), and containing\footnote{Here ``containing'' refers to the whole formula rather than the underlined parts.} no nonlogical (general, elementary or pseudoelementary) atoms other than those occurring in  formulas ${\cal F}$-associated with the nodes of $T$. In addition,  each hyperformula satisfies the following conditions: 
\begin{itemize}  
\item In every parallel osubformula $E_1\mld\ldots\mld E_n$ or $E_1\mlc\ldots\mlc E_n$, {\em no} component $E_i$ is underlined. 
\item In every  toggling osubformula $E_1\tgd\ldots\tgd E_n$ or $E_1\tgc\ldots\tgc E_n$,  {\em at most one} component $E_i$ is underlined. If no component is underlined, such an osubformula is said to be  {\bf virgin}. 
\item In every choice  osubformula $E_1\add\ldots\add E_n$ or $E_1\adc\ldots\adc E_n$,  {\em at most one} component $E_i$ is underlined. If no component is underlined, such an osubformula is said to be {\bf virgin}. The non-underlined components of a non-virgin choice osubformula, as well as all osubformulas of such components, are said to be {\bf abandoned}. 
\item In every sequential osubformula $E_1\sqd\ldots\sqd E_n$ or $E_1\sqc\ldots\sqc E_n$, {\em exactly one} component $E_i$ is underlined. The occurrence of a component $E_j$ which is to the left of such $E_i$ (i.e., $j\mless i$), as well as all  osubformulas of $E_j$, are said to be {\bf abandoned}.  
\end{itemize}

As will be seen later, underlines in a hyperformula $H$ carry certain information about the ``current position'' of the game. Namely, an underlined component of a choice osubformula is one that has already been selected by the corresponding player in a choice move associated with that osubformula, and an underlined component of a non-abandoned sequential or toggling osubformula is its ``currently active'' component.    

With each node $b$ of $T$, in addition to the formula ${\cal F}(b)$, we also associate --- let us say {\bf $\cal H$-associate} --- a hyperformula, which will be denoted by ${\cal H}(b)$. Such a hyperformula is always the original formula $F$ with some osubformulas underlined and some (pairs of  positive and  negative) occurrences of general atoms replaced by pseudoelementary atoms. The association will be such that ${\cal F}(b)$ is always essentially a ``fragment'' of ${\cal H}(b)$, obtained from the latter by deleting certain parts, and abbreviating certain other parts through $\twg$ and $\tlg$. 
Hence, to every occurrence of a subformula  in ${\cal F}(b)$ {\bf corresponds} --- in an obvious sense almost requiring no explanation but which will still be explained shortly for safety --- an occurrence of a subformula in ${\cal H}(b)$.  As we are going to see, this correspondence will be such that the following conditions are satisfied for each node $b$ of the tree:
\begin{description}
\item[Condition (i):] To an occurrence of a subformula $E_1\add\ldots\add E_n$ in ${\cal F}(b)$ always corresponds an occurrence of the form $G_1\add\ldots\add G_n$ (no $G_i$ is underlined) in ${\cal H}(b)$. Similarly for $\adc$.
\item[Condition (ii):] To an occurrence of a subformula $E_1\sqd E_2\sqd \ldots\sqd E_n$ in ${\cal F}(b)$ always corresponds an occurrence of the form $H_1\sqd\ldots\sqd H_m\sqd \underline{G_1}\sqd G_2\sqd \ldots\sqd G_n$ in ${\cal H}(b)$. Similarly for $\sqc$.
\item[Condition (iii):] To an occurrence of a subformula $E_1\tgd  \ldots\tgd E_n$ in ${\cal F}(b)$ always corresponds an occurrence of the form $G_1\tgd \ldots\tgd G_n$  (no $G_i$ is underlined) in ${\cal H}(b)$. Similarly for $\tgc$.
\item[Condition (iv):] To an occurrence of a nonlogical (general, elementary or pseudoelementary) literal in ${\cal F}(b)$ always corresponds an occurrence of the same literal in  ${\cal H}(b)$. We extend the concept of pseudoelementary atoms or literals of formulas, and the ``matching'' relation between them, to the corresponding atoms or literals of hyperformulas.  
\end{description}

Here is a description of the association $\cal H$, together with the meaning of the above-mentioned relation ``corresponds''. In each case, it is immediately obvious (by induction) that the above conditions (i)-(iv) are satisfied.
\begin{enumerate}
\item With the root node   of $T$ we $\cal H$-associate the hyperformula 
$\tilde{F}$ 
obtained from $F$ by underlining the leftmost component of each sequential osubformula (and nothing else). To each occurrence of each subformula of $F$ corresponds the same occurrence of the same subformula of $\tilde{F}$ (the latter may only differ from the former in that it is in the scope of some underlines). 

\item Suppose ${\cal F}(b)$ is obtained from ${\cal F}(c)$ by the (M) rule, namely, ${\cal F}(c)$ is the result of replacing in ${\cal F}(b)$ two semisurface general politerals $P$ and $\gneg P$ by some (pseudo)elementary politerals $p$ and $\gneg p$. Then ${\cal H}(c)$ is the result of replacing in ${\cal H}(b)$ the corresponding two occurrences of $P$ and $\gneg P$ (remember Condition (iv)) by $p$ and $\gneg p$. To the occurrences $p$ and $\gneg p$ in ${\cal F}(c)$ correspond the occurrences of the same $p$ and $\gneg p$ in ${\cal H}(c)$; all other correspondences (also) remain the same as before --- as in ${\cal F}(b)$ versus ${\cal H}(b)$, that is. 

Thus, the effect of the (M) rule  on ${\cal H}(b)$ is the same as its effect on ${\cal F}(b)$.

\item Suppose ${\cal F}(b)$ is obtained from ${\cal F}(c)$ by the $(\add)$ rule, namely, ${\cal F}(c)$ is the result of replacing in ${\cal F}(b)$ a semisurface osubformula $E_1\add\ldots\add E_n$ by $E_i$. Then ${\cal H}(c)$ is the result of underlining in ${\cal H}(b)$ the $i$th component $G_i$ of  the corresponding  osubformula $G_1\add \ldots\add G_n$ (remember Condition (i)). To the occurrence of $E_i$ in   ${\cal F}(c)$ corresponds the occurrence of $G_i$ in 
 ${\cal H}(c)$; all other correspondences  (also) remain the same as before. 

Thus, while the effect of the $(\add)$ rule (seen bottom-up) on ${\cal F}(b)$ is replacing a $\add$-osubformula by one of its components, the effect of the same rule on  ${\cal H}(b)$ is simply underlining that (the corresponding) component, without  deleting anything.

\item Suppose ${\cal F}(b)$ is obtained from ${\cal F}(c)$ by the $(\hspace{-2pt}\sqd\hspace{-2pt})$ rule, namely, ${\cal F}(c)$ is the result of replacing in ${\cal F}(b)$ a semisurface osubformula $E_1\sqd E_2\sqd\ldots\sqd E_n$ by $E_2\sqd\ldots\sqd E_n$. Then ${\cal H}(c)$ is the result of  replacing in ${\cal H}(b)$ the corresponding 
 osubformula $H_1\sqd\ldots\sqd H_m\sqd \underline{G_1}\sqd G_2\sqd\ldots\sqd G_n$ (remember Condition (ii))  by $H_1\sqd\ldots\sqd H_m\sqd G_1\sqd \underline{G_2}\sqd\ldots\sqd G_n$.    To the occurrence of $E_2\sqd\ldots\sqd E_n$ in   ${\cal F}(c)$ corresponds the occurrence of $H_1\sqd\ldots\sqd H_m\sqd G_1\sqd \underline{G_2}\sqd\ldots\sqd G_n$ in 
 ${\cal H}(c)$ unless $n\equals 2$, in which case to the occurrence of $E_2$ in   ${\cal F}(c)$ corresponds simply the occurrence of $G_2$ in  ${\cal H}(c)$; all other correspondences remain the same as before. 

Thus, while the effect of the $(\hspace{-2pt}\sqd\hspace{-2pt})$ rule (seen bottom-up) on ${\cal F}(b)$ is deleting the head of a $\sqd$-osubformula, the effect of the same rule on ${\cal H}(b)$ is simply moving the underline one position to the right, otherwise preserving all earlier ---  abandoned --- components of the  $\sqd$-osubformula.

\item Suppose ${\cal F}(b)$ is obtained from ${\cal F}(c_0)$,  ${\cal F}(c_1),\ldots,{\cal F}(c_k)$ by the $(\hspace{-2pt}\sqc\hspace{-2pt}\adc)$ rule, where $c_0$ is the senior child of $b$ and $c_1,\ldots,c_k$ ($k\geq 0$) are the junior children. Then:
\begin{itemize} 
\item ${\cal H}(c_0)={\cal H}(b)$. If $G'$ is an osubformula of ${\cal F}(c_0)$ that replaced an old osubformula $G$ of ${\cal F}(b)$ when transferring from ${\cal F}(b)$ to ${\cal F}(c_0)=|{\cal F}(b)|$ , then to $G'$ corresponds the same osubformula of ${\cal H}(c_0)$ as the osubformula of ${\cal H}(b)$ that corresponded to $G$; all other correspondences remain the same as before.  

Remember that, on the other hand, ${\cal F}(c_0)=\qelz{{\cal F}(b)}$. Thus, while ${\cal F}(c_0)$ is obtained from ${\cal F}(b)$ through ``abbreviating'' each general politeral and each $\add$-osubformula as $\tlg$, ``abbreviating'' each $\adc$-osubformula as $\twg$ and ``abbreviating'' each sequential osubformula as its head, ${\cal H}(c_0)$ is just an exact copy of  ${\cal H}(b)$. 

\item Suppose ${\cal F}(c_j)$ ($1\mleq  j\mleq  k$)  is the result of replacing in ${\cal F}(b)$ a semisurface osubformula $E_1\adc\ldots\adc E_n$ by $E_i$. Then ${\cal H}(c_j)$ is the result of underlining  in ${\cal H}(b)$ the $i$th component $G_i$ of the corresponding osubformula  $G_1\adc \ldots\adc G_n$ (remember Condition (i)). To the occurrence of $E_i$ in   ${\cal F}(c_j)$ corresponds the occurrence of $G_i$ in 
 ${\cal H}(c_j)$; all other correspondences  (also) remain the same as before. 

Thus,  while ${\cal F}(c_j)$ is the result of replacing   in ${\cal F}(b)$ a $\adc$-osubformula by one of its components, the transition from  ${\cal H}(b)$ to ${\cal H}(c_j)$ simply underlines that component.  

\item Suppose ${\cal F}(c_j)$ ($1\mleq  j\mleq  k$)  is the result of replacing in ${\cal F}(b)$ a semisurface osubformula $E_1\sqc E_2\sqc$ $\ldots\sqc E_n$ by $E_2\sqc\ldots\sqc E_n$. Then ${\cal H}(c_j)$ is the result of  replacing in ${\cal H}(b)$ the corresponding  osubformula  $H_1\sqc\ldots \sqc H_m\sqc
\underline{G_1}\sqc G_2\sqc\ldots\sqc G_n$ by $H_1\sqc\ldots \sqc H_m\sqc G_1\sqc \underline{G_2}\sqc \ldots\sqc G_n$ (remember Condition (ii)). To the occurrence of $E_2\sqc\ldots\sqc E_n$ in   ${\cal F}(c_j)$ corresponds the occurrence of $H_1\sqc\ldots\sqc H_m\sqc G_1\sqc \underline{G_2}\sqc\ldots\sqc G_n$ in 
 ${\cal H}(c_j)$ unless $n\equals 2$, in which case to the occurrence of $E_2$ in   ${\cal F}(c_j)$ corresponds the occurrence of $G_2$ in  ${\cal H}(c_j)$; all other correspondences remain the same as before. 

Thus, while ${\cal F}(c_j)$ is the result of deleting  in ${\cal F}(b)$ the head of a $\sqc$-osubformula, a transition from 
${\cal H}(b)$ to ${\cal H}(c_j)$ simply moves the underline in the corresponding $\sqc$-osubformula one position to the right, without otherwise deleting the  abandoned component(s).  
\end{itemize}

\item Suppose ${\cal F}(b)$ is obtained from ${\cal F}(c)$ by the $(\hspace{-2pt}\tgd\hspace{-2pt})$ rule, namely, ${\cal F}(c)$ is the result of replacing in ${\cal F}(b)$ a surface osubformula $E_1\tgd\ldots\tgd  E_n$ by $E_i$. Then ${\cal H}(c)$ is the result of underlining  in  ${\cal H}(b)$ the $i$th component $G_i$  of the 
corresponding  osubformula  $G_1\tgd\ldots\tgd G_n$ (remember Condition (iii)). To the occurrence of $E_i$ in   ${\cal F}(c)$ corresponds the occurrence of $G_i$ in 
 ${\cal H}(c)$; all other correspondences  (also) remain the same as before. 

Thus, while the transition  from ${\cal F}(b)$ to ${\cal F}(c)$ in the present case picks $E_i$ and deletes all other components of $E_1\tgd\ldots\tgd  E_n$, the transition from ${\cal H}(b)$ to ${\cal H}(c)$ simply underlines the ``picked'' component without  deleting anything.

\item Suppose ${\cal F}(b)$ is obtained from   ${\cal F}(c_1),\ldots,{\cal F}(c_k)$ by the $(\hspace{-2pt}\tgc\hspace{-2pt})$ rule. Then:
\begin{itemize}  
\item Suppose ${\cal F}(c_j)$ ($1\mleq  j\mleq  k$)  is the result of replacing in ${\cal F}(b)$ a surface osubformula $E_1\tgc\ldots\tgc E_n$ by $E_i$. Then ${\cal H}(c_j)$ is the result of underlining   in ${\cal H}(b)$ the $i$th component $G_i$ of the corresponding osubformula $G_1\tgc\ldots\tgc G_n$ (remember Condition (iii)). To the occurrence of $E_i$ in   ${\cal F}(c_j)$ corresponds the occurrence of $G_i$ in 
 ${\cal H}(c)$; all other correspondences  (also) remain the same as before.

Thus,  while ${\cal F}(c_j)$ is the result of picking $E_i$ and deleting all other components of $E_1\tgc\ldots\tgc  E_n$ in ${\cal F}(b)$, the transition from ${\cal H}(b)$ to ${\cal H}(c_j)$ simply underlines the ``picked'' component without  deleting anything.
\end{itemize}
\end{enumerate}

The following lemma can be verified through a straightforward analysis of the above construction by induction on the distance of a given node from the root:

\begin{lemma}\label{march31b}
Let $b$ be any node of $T$. 
\begin{enumerate}
\item Suppose the justification of $b$ is one of the rules {\em (M)}, $(\add)$, $(\hspace{-2pt}\sqd\hspace{-2pt})$ or $(\hspace{-2pt}\sqc\hspace{-2pt}\adc)$. Then ${\cal F}(b)$ is the result of replacing in ${\cal H}(b)$ every non-virgin choice osubformula by its underlined component,  removing from every sequential osubformula the abandoned components (the order of these changes does not matter), and deleting all underlines.
\item Suppose the justification of $b$ is one of the rules $(\hspace{-2pt}\tgd\hspace{-2pt})$ or  $(\hspace{-2pt}\tgc\hspace{-2pt})$. Then ${\cal F}(b)$ is the result of replacing in ${\cal H}(b)$ every general politeral by $\tlg$, every virgin $\add$-osubformula by $\tlg$, every virgin $\adc$-osubformula by $\twg$, every non-virgin choice osubformula, every sequential osubformula and every non-virgin toggling osubformula by its underlined component, and deleting all underlines (again, the order of replacements does not matter). 
\end{enumerate}
\end{lemma}

As we remember, our goal is to construct a machine's strategy that wins $e[F^*]$ for any interpretation $^*$ and valuation $e$. Let us fix some arbitrary $^*$ and $e$ for the rest of this section. 
We prefer to define and analyze our strategy  in terms of hyperformulas rather than formulas. When $H$ is a hyperformula, $H^*$ will be simply understood as $G^*$, where $G$ is the formula resulting from $H$ by removing all underlines and replacing each pseudoelementary atom by its origin.\footnote{We can see that, this way,  for any node $b$, the game $({\cal H}(b))^*$ is identical to $F^*$.}   Hence, rephrasing our goal, it is constructing a machine's interpretation-independent strategy that wins $\tilde{F}^*$, where  $\tilde{F}$ is the hyperformula $\cal H$-associated with the root of $T$.  Such a strategy will be presented in the form of an EPM, which we call $\cal S$. 

The work of our machine $\cal S$ will rely on neither $^*$ nor $e$, so we shall usually omit these  parameters and write a {\em hyperformula} $E$ where, strictly speaking, the {\em game} $E^*$ or the constant game $e[E^*]$ is meant. That is, we identify hyperformulas or their subformulas with the corresponding games. Accordingly, any terminology that we could use for games and their subgames we can as well use for hyperformulas and their subformulas, and vice versa. This includes the usage of our ``o'' terminology: we may, for instance, say ``the osubgame $G$ of $G\mlc H$'' to mean the subgame of the game/formula $G\mlc H$ represented by the osubformula $G$. 

For simplicity and convenience, in our construction of $\cal S$ and our analysis of its work, we implicitly rely on what in \cite{Japfin} is called the {\bf clean environment assumption} --- the assumption according to which the adversary never makes illegal moves. Indeed, if the adversary makes an illegal move, $\cal S$ automatically wins no matter what happens later, so such cases are not worth considering.

\begin{definition}\label{march31}
Let $E$ be a hyperformula. We say that a run $\Gamma$ is {\bf $E$-adequate} iff the following conditions are satisfied:
\begin{enumerate}
\item $\Gamma$ is a legal run of $E$.
\item Whenever a given non-abandoned choice osubformula of $E$ is virgin, no moves have been made in $\Gamma$ signifying a choice of a component in that osubformula. 
\item Whenever a component $G_i$ of a given choice osubformula of $E$ is underlined, $\Gamma$ contains a move signifying a choice of $G_i$ in  that osubformula. 
\item Whenever a given component of a non-abandoned sequential or toggling osubformula of $E$ is underlined, that component is the active component\footnote{Remember from Definition \ref{nov10} that the active component is the last-chosen component, or the default first component if there were no choices at all; if infinitely many choices occured, then no component is considered active.} of the corresponding subgame of $E$ in the run $\Gamma$.\footnote{More precisely, in the run $\Gamma'$ which is the subsequence of those labmoves of $\Gamma$ --- with certain prefixes deleted --- that signify labmoves within the sequential or toggling osubformula of $E$. For instance, if $E$ is $G\mld(H\sqc K)$ and the osubformula we are talking about is $H\sqc K$, then such a $\Gamma'$ is $\Gamma^{2.}$.} 
\item $\Gamma$ does not contain any $\pp$-labeled moves signifying moves within general oliterals of $E$.
\item Assume $p$ and $\gneg p$ are  non-abandoned  pseudoelementary politerals of $E$, $\Theta^+$ is the sequence of labmoves made (during playing $\Gamma$) within the $p$ component of $E$, and  $\Theta^-$ is the sequence of labmoves made within the $\gneg p$ component. Then $\Theta^+$ is a $\pp$-delay of $\gneg \Theta^-$ (remember from \cite{Japfin} that, for a run $\Omega$, $\gneg \Omega$ means the result of changing all labels to their opposites in $\Omega$). 
\end{enumerate}
\end{definition}

The above technical concept will play a central role in our soundness proof. The main intuition here is that, when $\Gamma$ is $E$-adequate, the hyperformula 
$E$ accurately represents all relevant information about the position resulting from playing $\Gamma$, so that $E$ can be seen as an ``adequate'' description of that position; also, $\pp$ has played ``wisely'' so far in a way that allows it to further maintain this ``adequacy'' and eventually win.    Namely, ignoring the abandoned components that are no longer relevant, we have the following. The absence of an underline in a choice osubformula of $E$ indicates that no choice of a component has been made in the corresponding subgame (condition 2). On the other hand, if an underline is present in such an osubformula, it shows which component has been chosen (condition 3). The underline in a sequential or toggling osubformula indicates the active component of the corresponding subgame (condition 4).  Also, the subgames in the ``matched'' occurrences of 
atoms have evolved to --- in a sense --- the same games  (condition 6), and $\pp$ has not made any hasty moves in unmatched atoms (condition 5), so that, if and when at some later point such an atom finds a match, $\pp$ will still have a chance to ``even out'' the corresponding two subgames.  

During its work, our machine $\cal S$ maintains a record $b$ which, intuitively, represents the node of $T$ which is currently being ``visited'' by $\cal S$.  Initially, $b$ is the root of $T$. From any given current(ly visited) non-leaf node $b$, $\cal S$ always moves (``{\bf jumps}'') to one of its children $c$. This way, it always follows one of the branches of the tree until it hits a leaf. Once it reaches a leaf, it will either stay there forever, or will backtrack by jumping to one of the siblings of one of the predecessors of that leaf. Then it starts another upward journey from that node towards a leaf, and so on. 

Our description of the work of $\cal S$, i.e. of the procedure WORK that it follows, as well as our further analysis of it, will rely --- more often implicitly than explicitly --- on the earlier-listed Conditions (i)-(iv), the description of the association $\cal H$, Lemma \ref{march31b} and the following lemma which, with Lemma \ref{may19} in mind, can be verified by a straightforward analysis of the steps of WORK (left to the reader):

\begin{lemma}\label{march31a}
At any time $\cal S$ jumps to a given node $b$ of $T$ (here the time of initializing $b$ counts as the time of ``jumping'' to the root), the then-current position $\Phi$  of the play is ${\cal H}(b)$-adequate. 
\end{lemma} 

How $\cal S$  jumps from one node to another, and what moves (if any) it makes on each such transition, depends on the justification of the node, and also --- when the justification is $(\hspace{-2pt}\tgc\hspace{-2pt})$ --- on whether the node is a leaf or not.  So, below we describe each possible case separately. In each case, $b$ stands for the ``current node'', and $\Phi$  stands for the position reached ``by now'' in the overall play of the original game. This is exactly how the procedure WORK followed by $\cal S$ goes:\vspace{5pt}

{\bf Procedure} WORK: $\cal S$ jumps to the root of $T$ (i.e., initializes the record $b$ to the root). Then it acts depending on which of the conditions of the following seven cases  is satisfied by $b$:

\begin{description}
\item[Case 0:]  {\em The justification of $b$ is the {\em (M)} rule.}  Let $c$  be the child of $b$. So, ${\cal H}(c)$ is the result of replacing in ${\cal H}(b)$ two  general politerals $P$ and $\gneg P$ by some pseudoelementary politerals $p$ and $\gneg p$. Then $\cal S$ looks up in $\Phi$ all moves made by the adversary within the politeral $P$, and makes/copies those moves --- in the same order as they are made by the adversary --- within the politeral $\gneg P$. Similarly, it mimics, in $\gneg P$, all moves made by the adversary within $P$. After that, $\cal S$ jumps to $c$  (that is, $\cal S$ updates the content of the record $b$ to $c$).  
\item[Case 1:]  {\em The justification of $b$ is the $(\add)$ rule.} Let $c$ be the child of $b$. So, ${\cal H}(b)$ has a virgin osubformula $G_1\add\ldots\add G_n$, and ${\cal H}(c)$ results from ${\cal H}(b)$ by underlining $G_i$ for some $1\mleq  i\mleq  n$.  Then $\cal S$ makes the move that signifies a choice of the $i$th component in the osubgame $G_1\add\ldots\add G_n$,  and jumps to $c$.
\item[Case 2:]  {\em The justification of $b$ is the $(\hspace{-2pt}\sqd\hspace{-2pt})$ rule.} Let $c$ be the  child of $b$. So, ${\cal H}(b)$ has an  osubformula $G_1\sqd\ldots\sqd G_{i-1}\sqd\underline{G_i}\sqd G_{i+1}\sqd\ldots\sqd G_n$, and ${\cal H}(c)$ results from ${\cal H}(b)$ by moving the underline from $G_i$ to $G_{i+1}$.  Then $\cal S$ makes the move that signifies switching (from $G_i$) to $G_{i+1}$ in the corresponding osubgame, and jumps to $c$. 
\item[Case 3:]  {\em The justification of $b$ is the $(\hspace{-2pt}\sqc\hspace{-2pt}\adc)$ rule.} 
Then $\cal S$ simply jumps to the senior child of $b$. 
\item[Case 4:]  {\em The justification of $b$ is the $(\hspace{-2pt}\tgd\hspace{-2pt})$ rule.} Let $c$ be the child of $b$. So, ${\cal H}(b)$ has a virgin osubformula $G_1\tgd\ldots\tgd G_n$, and ${\cal H}(c)$ results from ${\cal H}(b)$ by underlining one of the $n$ components $G_i$.  Then $\cal S$ makes the move that signifies switching  to $G_{i}$ in $G_1\tgd\ldots\tgd G_n$, and jumps to $c$. 
\item[Case 5:]  {\em The justification of $b$ is the $(\hspace{-2pt}\tgc\hspace{-2pt})$ rule, and $b$ is not a leaf.} Let $E_1\tgc\ldots\tgc E_n$ be the leftmost surface  $\tgc$-osubformula of ${\cal F}(b)$, and let $G_1\tgc\ldots\tgc G_n$ be the corresponding osubformula of ${\cal H}(b)$.  Using $\Phi$, $\cal S$ looks up the currently (in the position $\Phi$, that is) active component $G_i$ of the  osubgame $G_1\tgc\ldots\tgc G_n$. Note that among the children of $b$ is a node $c$ such that ${\cal H}(c)$ is the result of underlining in ${\cal H}(b)$ the $i$th component $G_i$ of the osubformula $G_1\tgc\ldots\tgc G_n$. Then  $\cal S$ jumps to that $c$. 
\item[Case 6:]  {\em The justification of $b$ is the $(\hspace{-2pt}\tgc\hspace{-2pt})$ rule, and $b$ is a leaf.} Let us say that a pseudoelementary politeral of ${\cal H}(b)$ is {\bf widowed}  iff its matching politeral is abandoned. 
This is how $\cal S$ acts. It keeps granting permission until the adversary makes a move. If such a move is a move within an abandoned component of ${\cal H}(b)$, within a general oliteral or within a widowed pseudoelementary oliteral, $\cal S$ does not react and continues granting permission. If the adversary makes a move within a non-abandoned and non-widowed pseudoelementary politeral $p$ or $\gneg p$, then $\cal S$ makes the same move in the matching  politeral $\gneg p$ or $p$, and continues granting permission. If no events other than the above kinds of events ever happen, $\cal S$ will remain at $b$ forever. However, a typical development during a visit of a leaf will be that, sooner or later, one of the following three events  takes place:
\begin{description}
\item[Event 1:] The environment makes a move that signifies a choice of $G_i$ in some non-abandoned osubformula $G_1\adc\ldots\adc G_n$ of ${\cal H}(b)$. Let $d$ be the nearest predecessor of $b$ whose justification is the $(\hspace{-2pt}\sqc\hspace{-2pt}\adc)$ rule.\footnote{A little thought convinces us that such a $d$ indeed exists.} It is not hard to see that, ignoring underlines in toggling osubformulas, ${\cal H}(b)$ and ${\cal H}(d)$ are the same. Also, among the  junior children   of $d$ is a node $c$ such that ${\cal H}(c)$ results from ${\cal H}(d)$ by underlining $G_i$ in the osubformula 
$G_1\adc\ldots \adc G_n$. In this case, $\cal S$ jumps to that $c$.
\item[Event 2:] The environment makes a move that signifies a switch from a component $G_i$ to the component $G_{i+1}$ in some non-abandoned osubformula $G_1\sqc\ldots \sqc G_{i-1}\sqc\underline{G_i}\sqc G_{i+1}\sqc \ldots \sqc G_n$ of ${\cal H}(b)$. Again, let $d$ be the nearest predecessor of $b$ whose justification is the $(\hspace{-2pt}\sqc\hspace{-2pt}\adc)$ rule.$^{19}$ Ignoring underlines in toggling osubformulas, ${\cal H}(b)$ and ${\cal H}(d)$ are the same. Also, among the junior children of $d$ is a node $c$ such that ${\cal H}(c)$  results from ${\cal H}(d)$ by moving the underline from $G_i$ to $G_{i+1}$ in the osubformula $G_1\sqc\ldots \sqc G_{i-1}\sqc\underline{G_i}\sqc G_{i+1}\sqc \ldots \sqc G_n$.  In this case, $\cal S$ jumps to that $c$.
\item[Event 3:] The environment makes a move that signifies a switch to $H_i$ in some non-abandoned, non-virgin osubformula $H_1\tgc\ldots \tgc H_n$ (one of the components is underlined) of ${\cal H}(b)$.  Let $d$ be the nearest predecessor of $b$ whose justification is the $(\hspace{-2pt}\tgc\hspace{-2pt})$ rule and where the osubformula $H_1\tgc\ldots \tgc H_n$ is (still) virgin.$^{19}$  Let $E_{1}^{1}\tgc\ldots\tgc E_{m_1}^{1}$, \ldots, $E^{k}_{1}\tgc\ldots\tgc E_{m_k}^{k}$ be all of the surface $\tgc$-osubformulas of ${\cal F}(d)$ in the left to right order, and let $G_{1}^{1}\tgc\ldots\tgc G_{m_1}^{1}$, \ldots, $G^{k}_{1}\tgc\ldots\tgc G_{m_k}^{k}$ be the corresponding osubformulas of  ${\cal H}(d)$. Then, obviously, $H_1\tgc\ldots \tgc H_n$ is $G_{1}^{i}\tgc\ldots\tgc G_{m_i}^{i}$ for one of $1\mleq  i\mleq  k$. Let $j\equals i\plus 1$ if $i\mless k$, and $j\equals 1$ if $i\equals k$. Let $G_{h}^{j}$ be the currently active component of $G_{1}^{j}\tgc\ldots\tgc G_{m_j}^{j}$ ($\cal S$ finds this component from $\Phi$). And let $c$ be the child of $d$ such that ${\cal H}(c)$ is the result of underlining $G_{h}^{j}$ in ${\cal H}(d)$. Then $\cal S$ jumps to $c$.
\end{description}   

\end{description}

Our remaining task now is to show that $\cal S$ wins $F$. Let us fix any run $\Gamma$ that could have been generated   in the process of $\cal S$ playing the game
against a clean environment. 
From now on, $\Gamma$ and the play/scenario  in which it was generated will be the context of our discourse, so, for instance, when we say ``$\cal S$ has made move $\alpha$'', this is to be understood as that $\cal S$ has made move $\alpha$ in the branch of computation that spells   $\Gamma$. 

We say that a given node $b$ is {\bf established} iff, in the context of our scenario, it satisfies the following conditions:
\begin{itemize}
\item $b$ has been visited a finite, nonzero number of times;
\item any node visited after the last visit of $b$ is a descendant of $b$ in the tree. 
\end{itemize}
Note that the set of established nodes is nonempty as (at least) the root is established. Indeed, all other nodes are descendants of the root. And the root is visited exactly once: it cannot be re-visited after the initial visit because backtracking  always happens to a node that is a sibling of another, already visited, node, but the root is not a sibling of anything. Note also that established nodes form a descendancy chain that starts at the root and goes upward following one of the branches. That is because of any two established nodes, one --- the one that was visited last --- is a descendant of the other. 

Let then, throughout the rest of this section, $\lmt$ be the established node that was visited last among all established nodes. In other words, $\lmt$ is the uppermost established node --- the node at which the upward chain of established nodes ends, so that $\lmt$ is the unique established node that {\em has no established descendants}.  Call such a node the {\bf limit node} of the play (of $\Gamma$, that is). 

\begin{lemma}\label{apr17}
The justification of the limit node $\lmt$ is the $(\hspace{-2pt}\tgc\hspace{-2pt})$ rule. 
\end{lemma}

\begin{proof} Suppose, for a contradiction, that the justification of $\lmt$  is (M), $(\add)$, $(\hspace{-2pt}\sqd\hspace{-2pt})$ or $(\hspace{-2pt}\tgd\hspace{-2pt})$. Then $\lmt$ has a single child $c$, which is always visited immediately after a visit of $\lmt$, and which can be visited only through ascending from $\lmt$. Therefore, $c$ is also visited a finite, nonzero  number of times. Also, all non-$c$ descendants of $\lmt$ are among the descendants of $c$, so any node visited after the last visit of $c$ is a descendant of $c$. To summarize, $c$ is an established child of $\lmt$.  But this is impossible because $\lmt$ is the limit node of the play.

Now suppose the justification of $\lmt$ is $(\hspace{-2pt}\sqc\hspace{-2pt}\adc)$.  Let $c$ be the senior child of $\lmt$. Consider the last visit of $\lmt$ (as an aside, such a visit would also be the first visit). It will be immediately followed by a visit of $c$. And $c$ cannot be re-visited after that, because the only way to get to $c$ is to ascend to it from $\lmt$. So, $c$ is visited a finite, nonzero number of times. If all nodes visited after the visit of $c$ are  descendants of $c$, then $c$ is an established child of $\lmt$, which is impossible because $\lmt$ is the limit node. So, at some time after the last visit of ($\lmt$ and then) $c$, a backtracking jump occurs from a descendant leaf of $c$ to a node $d$ which is not a descendant of $c$. With a moment's thought, such a $d$ can be seen to be a sibling of $c$, i.e., a junior child of $\lmt$. With some more thought, one can see that, after that, the siblings of $d$ will be no longer available for backtracking jumps\footnote{Hint: We may assume that the justification of $d$ is not (M), $(\add)$ or $(\sqdi)$ (otherwise, instead of $d$, consider its nearest descendant satisfying this condition). If ${\cal F}(d)$ is quasielementary,  Events 1 and 2 of Case 6 of WORK will never occur after the visit of $d$, so $\cal S$ will not be able to jump to a sibling of $d$. And if  ${\cal F}(d)$ is not quasielementary, then $d$ is $(\sqci\adc)$-justified. Hence $\lmt$ is not the nearest $(\sqci\adc)$-justified predecessor of any descendant leaf of $d$, and thus no backtracking jump is possible from such a leaf  to a child of $\lmt$, i.e. to a sibling of $d$.}  and, for this reason, any node ever visited after the jump to $d$ will be a descendant of $d$. This makes $d$ an established child of $\lmt$ which, again, is a contradiction.
 
From the above contradictions, we conclude that the justification of  $\lmt$ can only be $(\hspace{-2pt}\tgc\hspace{-2pt})$.    
\end{proof}

For the rest of this section, let $E_{1}^{1}\tgc\ldots\tgc E_{m_1}^{1}$, \ldots, $E^{k}_{1}\tgc\ldots\tgc E_{m_k}^{k}$ be all of the surface $\tgc$-osubformulas of ${\cal F}(\lmt)$ in the left to right order, and let \[G_{1}^{1}\tgc\ldots\tgc G_{m_1}^{1}, \ \ldots, \ G^{k}_{1}\tgc\ldots\tgc G_{m_k}^{k}\] be the corresponding osubformulas of  ${\cal H}(\lmt)$. Thus, 
$\lmt$  has $m_1\plus \ldots \plus m_k$ children (if $\lmt$ is a leaf, then $k\equals 0$), which we correspondingly denote by 
\[c_{1}^{1},\ldots , c_{m_1}^{1}, \ \ \ldots, \ \ c^{k}_{1},\ldots , c_{m_k}^{k}.\]
 We divide these children into $k$ {\bf groups}, according to their superscripts.

\begin{lemma}\label{apr12}
The environment has lost in each osubformula $G_{1}^{i}\tgc\ldots\tgc G_{m_i}^{i}$ ($1\mleq  i\mleq  k$) of  ${\cal H}(\lmt)$.
\end{lemma}

\begin{proof} We assume that $k\not=0$ ($\lmt$ is not a leaf, that is), for otherwise there is nothing to prove. 
Let us try to understand what will be happening after the last visit of $\lmt$. This event will be immediately followed by a visit of one of the children $c^{1}_{i}$ ($1\mleq  i\mleq  m_1$) of group $\#1$. After that, for a while, all new (if any) nodes visited by  $\cal S$ will be among the descendants of $c^{1}_{i}$.
This ``while'', however, cannot last forever. Indeed, if no other nodes are visited, then $c^{1}_{i}$ would be 
an established child of $\lmt$, which is impossible due to $\lmt$'s being the limit node. Thus, sooner or later, $\cal S$ will jump 
from a descendant leaf of $c^{1}_{i}$ to a node $d$ which is not a descendant of $c^{1}_{i}$. With a little thought, we can see that such a $d$ should be a node of the next ($\#2$) group of children of $\lmt$, that is, $d\equals c^{2}_{j}$ for some $1\mleq  j\mleq  m_2$; also, the reason of this jump was that {\em the environment made a 
switch move within the  $G_{1}^{1}\tgc\ldots\tgc G_{m_1}^{1}$ osubformula of ${\cal H}(\lmt)$}. Now we repeat the same analysis for $c^{2}_{j}$ and find that, again, sooner or later, the environment makes a switch move within the  $G_{2}^{1}\tgc\ldots\tgc G_{m_2}^{2}$ osubformula of ${\cal H}(\lmt)$, which causes a jump to one of the nodes of group $\#3$ of children of $\lmt$. And so on. Once a node of the last group $\#k$ is reached, the next group will again be group $\#1$, 
which starts a new, similar to what we have already seen, cycle. This process will go on forever, so that, for each $1\mleq  h\mleq  k$, a switch move in the osubformula  
$G_{1}^{h}\tgc\ldots\tgc G_{m_h}^{h}$  of ${\cal H}(\lmt)$ 
will be made by the environment infinitely many times. But we know that making infinitely many switches in a toggling (sub)game causes the corresponding player --- the environment in the present case --- to lose in that (sub)game. 
\end{proof}

\begin{lemma}\label{apr19}
The run $\Gamma$ is ${\cal H}(\lmt)$-adequate.
\end{lemma}

\begin{proof} According to Lemma \ref{march31a}, at the time of the last jump to $\lmt$, the corresponding initial segment  of $\Gamma$ (the then-current position) was ${\cal H}(\lmt)$-adequate. We need to verify that the ${\cal H}(\lmt)$-adequacy of the run is not lost afterwards. This can be done by examining each of the six conditions of Definition \ref{march31}.

Condition 6 of Definition \ref{march31} obviously remains satisfied after the last visit of $\lmt$. This can be seen from an analysis of the work of $\cal S$ during its visits of leaves.  

For condition 1 of Definition \ref{march31}, note that, if $\cal S$ makes infinitely many jumps, then the legality of $\Gamma$ is guaranteed by Lemma \ref{march31a}, because we have legal positions reached at arbitrarily late times.  Otherwise, it is not hard to see that $\lmt$ is a leaf, which $\cal S$ never leaves after the last visit.  During its visit of a leaf, $\cal S$ only mimics the environment's moves made in pseudoelementary literals. In view of the already verified condition 6 of Definition \ref{march31} and Lemma \ref{may19}, together with some thought, one can see that such moves have to be legal, for otherwise it is the environment who first made an illegal move, which contradicts our clean environment assumption. 
 
For condition 2 of Definition \ref{march31}, note that, while visiting $(\hspace{-2pt}\tgd\hspace{-2pt})$- or $(\hspace{-2pt}\tgc\hspace{-2pt})$-justified nodes (obviously only such nodes are visited after the last visit of the $(\hspace{-2pt}\tgc\hspace{-2pt})$-justified $\lmt$), $\cal S$ does not make any moves signifying a choice of a component of a (virgin and non-abandoned) choice subformula.  And, if such a move was made by the environment (after the last visit of $\lmt$), it would cause $\cal S$ to make a backtracking jump to a node which is not a descendant of $\lmt$, which is impossible because $\lmt$ is established.

Condition 3 of Definition \ref{march31}, by Lemma \ref{march31a}, was satisfied at the time of the last jump to $\lmt$. It trivially remains so afterwards.

Condition 4 of Definition \ref{march31}, again by Lemma \ref{march31a}, was satisfied at the time of the last jump to $\lmt$. It will remain so if no moves are made afterwards by either player signifying switches in non-abandoned sequential and non-virgin toggling osubformulas of ${\cal H}(\lmt)$. But indeed, such moves are not made.  While visiting $(\hspace{-2pt}\tgd\hspace{-2pt})$- or $(\hspace{-2pt}\tgc\hspace{-2pt})$-justified nodes, $\cal S$ does not make any moves signifying a switch in a (non-abandoned) sequential osubformula.   And, if such a move was made by the environment, it would cause $\cal S$ to make a backtracking jump to a node which is not a descendant of $\lmt$, which is impossible because $\lmt$ is established. Similarly, while visiting $(\hspace{-2pt}\tgd\hspace{-2pt})$- or $(\hspace{-2pt}\tgc\hspace{-2pt})$-justified nodes, $\cal S$ does not make any moves signifying a switch in a non-virgin toggling osubformula. And, if such a move was made by the environment, it would cause $\cal S$ to make a backtracking jump to a node which is not a descendant of $\lmt$ which, again, is impossible. 

Finally, that condition 5 of Definition \ref{march31} also remains satisfied after the last visit of $\lmt$ is evident.
\end{proof}

Let $H_1$ be the result of replacing in ${\cal H}(\lmt)$ each non-abandoned virgin $\add$-osubformula by $\tlg$.  In view of the ${\cal H}(\lmt)$-adequacy of $\Gamma$ (Lemma \ref{apr19}), virginity of such an osubformula means that no choices have been made within the corresponding osubgame by the machine, so the machine has lost it. It is then obvious that $\Gamma$ remains  $H_1$-adequate, and $\Gamma$ is a $\pp$-won run of ${\cal H}(\lmt)$ iff it is a $\pp$-won run of $H_1$. 

Let $H_2$ be the result of further replacing in $H_1$ each  non-abandoned virgin $\adc$-osubformula by $\twg$.   Then, for reasons similar (or symmetric) to the ones in the previous case,  $\Gamma$ remains $H_2$-adequate, and $\Gamma$ is a $\pp$-won run of $H_1$ iff it is a $\pp$-won run of $H_2$. 

Let $H_3$ be the result of further replacing in $H_2$ each non-virgin choice osubformula by its underlined component (without including the underline --- it should be removed from under the component). And let $\Gamma_3$ be the result of deleting from $\Gamma$ all moves signifying choices in such choice osubformulas. It is not hard to see that $\Gamma_3$ remains $H_3$-adequate, and $\Gamma_2$ is a $\pp$-won run of $H_2$ iff $\Gamma_3$ is a $\pp$-won run of $H_3$.

Let $H_4$ be the result of further replacing in $H_3$ each sequential osubformula by its underlined component (without including the underline). And let $\Gamma_4$ be the result of removing from $\Gamma_3$ all moves that signified a switch in a sequential osubformula of $H_3$, further removing all moves that signified moves within the non-underlined components of such osubformulas, and then further removing from the remaining moves each substring ``$i.$'' indicating that the move was made within (the underlined) component $\#i$ of a sequential osubformula. For instance, if $H_3=P\mld(Q\sqc \underline{R})$ and $\Gamma_3=\seq{\pp 1.\alpha,\oo 2.1.\beta,\pp 2.2,\pp 2.2.\gamma}$, then $H_4=P\mld R$ and $\Gamma_4=\seq{\pp 1.\alpha,\pp 2.\gamma}$. It is not hard to see that $\Gamma_4$ remains $H_4$-adequate, and $\Gamma_3$ is a $\pp$-won run of $H_3$ iff $\Gamma_4$ is a $\pp$-won run of $H_4$.

Let $H_5$ be the result of further replacing in $H_4$ each non-virgin toggling osubformula by its underlined component (without including the underline). And let $\Gamma_5$ be the result of removing from $\Gamma_4$ all moves that signified a switch in a non-virgin toggling osubformula of $H_4$, further removing all  moves that signified moves within the non-underlined components of such osubformulas, and then further removing from the remaining moves  each substring ``$i.$'' indicating that the move was made within (the underlined) component $\#i$ of a toggling osubformula. For instance, if $H_4=P\mld\bigl(\underline{(Q\tgd \underline{R})}\tgc S\bigr)$ and $\Gamma_4=\seq{\pp 1.\alpha,\oo 2.1.1.\beta,\pp 2.1.2,\pp 2.1.2.\gamma}$, then $H_5=P\mld R$ and $\Gamma_5=\seq{\pp 1.\alpha,\pp 2.\gamma}$.  It is not hard to see that $\Gamma_5$ remains $H_5$-adequate, and $\Gamma_4$ is a $\pp$-won run of $H_4$ iff $\Gamma_5$ is a $\pp$-won run of $H_5$.

Let $H_6$ be the result of replacing in $H_5$ all general politerals  by $\tlg$. And let $\Gamma_6$ be the result of removing from $\Gamma_5$ all moves signifying moves within those politerals.  $\Gamma_6$ clearly remains  $H_6$-adequate. It is also obvious that $\Gamma_5$ is a $\pp$-won run of $H_5$ if (but not necessarily {\em only if}) $\Gamma_6$ is a $\pp$-won run of $H_6$. This is so because we may assume that $\pp$ has lost in all general politerals, so that replacing them with $\oo$ was legitimate; and if $\pp$ did not really lose in some of those politerals, that is ``even better'' due to the monotonicity of the winning conditions associated with all our game operations other than $\gneg$. 

Now, notice that, in view of clause 2 of Lemma \ref{march31b}, $H_6$ coincides with ${\cal F}(\lmt)$. Also, as noted earlier, the games $F$ and ${\cal H}(\lmt)$ are the same (that is, $F^*=\bigl({\cal H}(\lmt)\bigr)^*$). Thus, summarizing our observations of the preceding six paragraphs, we have:

\begin{lemma}\label{apr14a}
$\Gamma_6$ is ${\cal F}(\lmt)$-adequate (with ${\cal F}(\lmt)$ here seen as a hyperformula rather than formula), and it is a $\pp$-won run of  ${\cal F}(\lmt)$ only if $\Gamma$ is a $\pp$-won run of $F$.
\end{lemma}

  So, from now on, we can focus on  ${\cal F}(\lmt)$ and its run $\Gamma_6$. We continue seeing ${\cal F}(\lmt)$ --- as well as $H_7$ and $H_8$ defined shortly --- as hyperformulas rather than formulas, so that, when identifying them with the corresponding games, their pseudoelementary atoms are interpreted as their origins are. We want to show that $\Gamma_6$ is a $\pp$-won run of ${\cal F}(\lmt)$ --- of the game $({\cal F}(\lmt))^*$, to be more precise.   

Let $H_7$ be the result of replacing in ${\cal F}(\lmt)$ (in $H_6$, that is) all surface $\tgc$-osubformulas by $\twg$. And let $\Gamma_7$ be the result of removing from $\Gamma_6$ all moves that signified moves within those $\tgc$-osubformulas. Obviously $\Gamma_7$ remains $H_7$-adequate. Also, in view of 
Lemma \ref{apr12},  it is clear that  $\Gamma_6$ is a $\pp$-won run of ${\cal F}(\lmt)$ iff $\Gamma_7$ is a $\pp$-won run of $H_7$. 

Let $H_8$ be the result of replacing in $H_7$ all surface $\tgd$-osubformulas by $\tlg$. And let $\Gamma_8$ be the result of removing from $\Gamma_7$ all moves that signified moves within those $\tgd$-osubformulas. $\Gamma_8$ clearly remains $H_8$-adequate. It is also obvious that $\Gamma_7$ is a $\pp$-won run of $H_7$ if (but not necessarily {\em only if}) $\Gamma_8$ is a $\pp$-won run of $H_8$. This is so because we may assume that $\pp$ has lost in all 
$\tgd$-osubformulas, so that replacing them with $\oo$ was legitimate; and if $\pp$ did not really lose in some of those osubformulas, that is ``even better'' due to the earlier pointed out monotonicity of the winning conditions associated with our game operations. 

Now observe that $H_8$ coincides with $\elz{{\cal F}(\lmt)}$. Summarizing the observations of the previous two paragraphs and Lemma \ref{apr14a}, we thus have:

\begin{lemma}\label{apr14b}
$\Gamma_8$ is  $\elz{{\cal F}(\lmt)}$-adequate (with $\elz{{\cal F}(\lmt)}$ here seen as a hyperformula rather than formula),  and  it is a $\pp$-won run of  $\elz{{\cal F}(\lmt)}$ only if $\Gamma$ is a $\pp$-won run of $F$.
\end{lemma}

In view of the above lemma, all that now remains to show is that $\Gamma_8$ is a $\pp$-won run of  $\elz{{\cal F}(\lmt)}$. For this purpose, we may safely assume that, 
whenever $p$ and $\gneg p$ are (matching) pseudoelementary politerals  of  ${\cal F}(\lmt)$,  the machine has won in exactly one of these two politerals. To see the legitimacy of this assumption, remember condition 6 of Definition \ref{march31}. The latter, together with the fact of the $\elz{{\cal F}(\lmt)}$-adequacy of $\Gamma_8$, implies that  the run $\Theta^+$ played in $p$ is a $\pp$-delay of $\gneg \Theta^-$, where $\Theta^-$ is the run played in  $\gneg p$. If we now assume that the machine has lost in $\gneg p$, meaning that  $\Theta^-$ is a $\oo$-won run of $\gneg p$, then, by the definition of game negation,  $\gneg \Theta^-$ is a $\pp$-won run of $p$. But then, since $\Theta^+$ is a $\pp$-delay of $\gneg \Theta^-$ and $p$ is a static game, the property of static games implies that $\Theta^+$ is also a $\pp$-won run
 of $p$. To summarize, if the machine has lost in $\gneg p$, then it has  won in $p$. In other words, the machine has won in at least one of the osubgames $p,\gneg p$.  If our assumption about having won in exactly one of the two osubgames is false and both of these osubgames are won, that is only better from the machine's perspective, again due to the earlier mentioned monotonicity of winning conditions.  

 Let us now consider the {\em true/false} assignment $\tau$ for the politerals of $\elz{{\cal F}(\lmt)}$ that sends an (elementary or pseudoelementary) politeral to {\em true} iff  $\pp$ has won the play within that politeral. In view of the assumption of the preceding paragraph, obviously $\tau$ can also be seen as a truth assignment in the classical sense, i.e. as a {\em true/false}  assignment for the atoms (rather than politerals) of $\elz{{\cal F}(\lmt)}$ which extends to all subformulas of $\elz{{\cal F}(\lmt)}$ in the standard classical way.  Since $\elz{{\cal F}(\lmt)}$ is a tautology, it is true under $\tau$. This, in turn, obviously implies that $\Gamma_8$ is a $\pp$-won run of $\elz{{\cal F}(\lmt)}$. Hence, by Lemma \ref{apr14b}, $\Gamma$ is a $\pp$-won run of the original $F$. 
 
\section{Preliminaries for the completeness proof}

\subsection{Machines against machines}

This subsection borrows a discussion from \cite{Japtocl1}, providing certain background information necessary for our completeness proof but missing in \cite{Japfin}, the only external source on computability logic on which the present paper was promised to rely.

 For a run $\Gamma$ and a computation branch $B$ of an HPM or EPM, we say that $B$ {\bf cospells} $\Gamma$ iff
$B$ spells $\rneg\Gamma$ ($\Gamma$ with all labels reversed) in the sense of Section 6 of \cite{Japfin}. 
Intuitively, when a machine $\cal C$ plays as  $\oo$ (rather than $\pp$), then the run that is generated by  a given computation branch $B$ of $\cal C$ is the run cospelled (rather than spelled) by $B$, for the moves that $\cal C$ makes 
get the label $\oo$, and the moves that its adversary makes get the label $\pp$.

We say that an EPM $\cal C$ is {\bf fair} iff, for every valuation $e$, every $e$-computation branch 
of $\cal C$ is fair in the sense of Section 6 of \cite{Japfin}. 
 
\begin{lemma}\label{lem}
Assume $\cal E$ is a fair EPM, $\cal H$ is any HPM, and $e$ is any valuation. There are a uniquely defined 
 $e$-computation branch $B_{\cal E}$ of $\cal E$ and a uniquely defined $e$-computation branch $B_{\cal H}$ of $\cal H$
--- which we respectively call {\bf the $({\cal E},e,{\cal H})$-branch} and {\bf the $({\cal H},e,{\cal E})$-branch}
 --- such that the run spelled by $B_{\cal H}$, called  {\bf the $\cal H$ vs. $\cal E$ run on $e$}, 
 is the run cospelled by $B_{\cal E}$.\end{lemma}
 
When ${\cal H},{\cal E},e$ are as above, $\Gamma$ is the $\cal H$ vs.\hspace{-4pt} $\cal E$ run on $e$ and $A$ is a game such that $\Gamma$ is a $\pp$-won (resp. $\oo$-won) run of $e[A]$,  we say that $\cal H$ {\bf wins}
(resp. {\bf loses}) $A$ {\bf against $\cal E$ on $e$}. 

A strict proof of the above lemma can be found in \cite{Jap03} (Lemma 20.4), and we will not reproduce  
the formal proof here.  Instead, the following intuitive explanation should suffice:\vspace{7pt}

{\bf Proof idea.} Assume $\cal H$, $\cal E$, $e$ are as in Lemma \ref{lem}. The play that we are going to describe is the unique 
play generated when the two machines play against each other, with $\cal H$ in the role of $\pp,$  $\cal E$ in the role of $\oo$, and $e$ spelled on the valuation tapes of both machines. 
We can visualize this play as follows.
Most of the time during the process $\cal H$ remains inactive (sleeping); it is woken up only when $\cal E$ enters a permission state, on which event $\cal H$ makes a (one single) transition to its next computation step --- that may or may not result in making a move --- and goes back into a sleep that will continue until $\cal E$ enters  a permission state again, and so on. From ${\cal E}$'s perspective, $\cal H$ acts as a patient adversary who makes one or zero move only when granted permission, just as the EPM-model assumes.  And from $\cal H$'s perspective, which, like a person in a coma,  has no sense of time during its sleep and hence can think that the wake-up events that it calls the beginning of a clock cycle happen at a constant rate, $\cal E$ acts as an adversary who can make any finite number of moves during a clock cycle (i.e. while $\cal H$ was sleeping), just as the HPM-model assumes. This scenario uniquely determines an $e$-computation branch $B_{\cal E}$ of $\cal E$ that we call
the $({\cal E},e,{\cal H})$-branch, and an $e$-computation branch $B_{\cal H}$ of $\cal H$ that we call
the $({\cal H},e,{\cal E})$-branch. What we call the $\cal H$ vs. $\cal E$ run on $e$ is the run generated 
in this play. In particular --- since we let $\cal H$ play in the role of $\pp$ --- this is the run spelled by $B_{\cal H}$. $\cal E$, who plays in the role of $\oo$, sees the same run, only it sees the labels of the moves of that run in negative colors. 
That is, $B_{\cal E}$ cospells rather than spells that run. This is exactly what Lemma \ref{lem} asserts.

\subsection{Logic $\clte$   and its dual}

Our proof of the completeness part of Theorem \ref{thcl2} employs the conservative, elementary-base fragment $\clte$ of $\clt$, obtained by restricting the language of the latter to elementary-base formulas --- we refer to formulas of this restricted language as {\bf $\clte$-formulas} --- and (correspondingly) deleting the (M) rule.  Logic {\bf CL1}, historically the first system for computability logic proven (in \cite{Japtocl1}) to be sound and complete, is a $\sqc\hspace{-2pt},\sqd\hspace{-2pt},\tgc\hspace{-2pt},\tgd$-free counterpart of $\clte$. 

Of course, $\clte$ inherits soundness from $\clt$. In this section we are going to prove the completeness of $\clte$. 

Our completeness proof for $\clte$, in turn, employs the other logic $\cltee$ 
 which is a ``dual'' of $\clte$:

\begin{definition}\label{nov27}
The language of $\cltee$ is the same as that of $\clte$, and the rules of inference are:

\begin{description}
\item[$(\overline{\hspace{-2pt}\tgd\hspace{-2pt}})$:] $\vec{H}\mapsto F$, where $F$ is an instable quasielementary formula, and  $\vec{H}$ the smallest set of formulas satisfying the following condition: 
\begin{itemize}
\item Whenever $F$ has a surface osubformula  $E_1\tgd\ldots\tgd E_n$,  for each $i\in\{1,\ldots,n\}$, $\vec{H}$ contains the result of replacing  in $F$ that osubformula by $E_i$. 
\end{itemize}
\item[$(\overline{\hspace{-2pt}\tgc\hspace{-2pt}})$:] $H\mapsto F$, where $F$ is a quasielementary formula, and $H$ is the result of replacing in $F$ a surface osubformula $E_1\tgc\ldots\tgc E_n$  by $E_i$ for some $i\in\{1,\ldots, n\}$.
\item[$(\overline{\hspace{-2pt}\sqd\hspace{-2pt}\add})$:]  $\qelz{F},\vec{H}\mapsto F$, where $F$ is a non-quasielementary formula, and   
$\vec{H}$ is the smallest set of formulas satisfying the following two conditions: 
\begin{itemize} 
\item Whenever $F$ has a semisurface osubformula $E_1\add\ldots\add E_n$, for each 
$i\in\{1,\ldots,n\}$, $\vec{H}$ contains the result of replacing in $F$ that osubformula  by $E_i$.
\item Whenever $F$ has a semisurface osubformula $E_1\sqd E_2\sqd\ldots\sqd E_n$, $\vec{H}$ contains the result of replacing in $F$ that osubformula  by $E_2\sqd\ldots\sqd E_n$.
\end{itemize}
The above formula $\qelz{F}$ is said to be the {\bf senior premise} of the rule or its conclusion, and  the formulas of $\vec{H}$ are the {\bf junior premises}.
\item[$(\overline{\adc})$:]  $H\mapsto F$, where $H$ is the result of replacing in $F$ a semisurface osubformula $E_1\adc\ldots\adc E_n$  by $E_i$ for some $i\in\{1,\ldots, n\}$.
\item[$(\overline{\hspace{-2pt}\sqc\hspace{-2pt}})$:]  $H\mapsto F$, where $H$ is the result of replacing in $F$ a semisurface osubformula $E_1\sqc E_2\sqc\ldots\sqc E_n$  by 
$E_2\sqc\ldots\sqc E_n$.
\end{description}
\end{definition}

\begin{lemma}\label{nov4}
$\clte\not\vdash F$ iff $\cltee\vdash F$ (for any $\clte$-formula $F$).
\end{lemma}

\begin{proof} We prove this lemma by induction on the complexity of $F$. It is sufficient to verify the `only if' part, as the `if' part (which we do not need anyway) can be handled in a fully symmetric way. So, assume $\clte\not\vdash F$ and let us see that then $\cltee\vdash F$. There are the following cases to consider:

{\em Case 1}: $F$ is quasielementary.

{\em Subcase 1.1}: $F$ is stable. Then there must be a $\clte$-unprovable formula $H$ satisfying the following condition, for otherwise $F$ would be $\clte$-derivable by the $(\hspace{-2pt}\tgc\hspace{-2pt})$ rule: 
\begin{itemize}
\item $H$ is the result of replacing in $F$ a surface  osubformula  $E_1\tgc\ldots\tgc E_n$  by $E_i$ for some $i\in\{1,\ldots, n\}$.
\end{itemize}
But, if so,  by the induction hypothesis, $\cltee\vdash H$, whence, by $(\overline{\hspace{-2pt}\tgc\hspace{-2pt}})$,  $\cltee\vdash F$.

{\em Subcase 1.2}: $F$ is instable. Let $\vec{H}$ be the smallest set of formulas such that the following  condition is 
satisfied: 
\begin{itemize}
\item Whenever $F$ has a surface osubformula $E_1\tgd\ldots\tgd E_n$, for each 
$i\in\{1,\ldots,n\}$, $\vec{H}$ contains the result of replacing in $F$ that osubformula  by $E_i$.
\end{itemize}
None of the elements of $\vec{H}$ is provable in $\clte$, for otherwise $F$ would also be derivable in $\clte$ by $(\hspace{-2pt}\tgd\hspace{-2pt})$.  Therefore, by the induction hypothesis, each element of $\vec{H}$ is $\cltee$-provable.  Hence,  by  $(\overline{\hspace{-2pt}\tgd\hspace{-2pt}})$, we have  $\cltee\vdash F$. 
 
{\em Case 2}: $F$ is not quasielementary. 

{\em Subcase 2.1}: $\clte\vdash\qelz{F}$. Then there must be a $\clte$-unprovable formula $H$ satisfying one of the following two conditions, for otherwise $F$ would be $\clte$-derivable by the $(\hspace{-2pt}\sqc\hspace{-2pt}\adc)$ rule: 
\begin{itemize}
\item $H$ is the result of replacing in $F$ a semisurface  osubformula  $E_1\adc\ldots\adc E_n$  by $E_i$ for some $i\in\{1,\ldots, n\}$.
\item $H$ is the result of replacing in $F$ a semisurface osubformula \(E_1\sqc E_2\sqc\ldots\sqc E_{n}\)   by \(E_2\sqc\ldots\sqc {E_n}.\)
\end{itemize}
In either case, by the induction hypothesis, $\cltee\vdash H$, whence, by $(\overline{\adc})$ (if the first condition is satisfied) or $(\overline{\hspace{-2pt}\sqc\hspace{-2pt}})$ (if the second condition is satisfied), $\cltee\vdash F$.

{\em Subcase 2.2}: $\clte\not\vdash\qelz{F}$.  Let $\vec{H}$ be the smallest set of formulas such that the following two conditions are 
satisfied: 
\begin{itemize}
\item Whenever $F$ has a semisurface osubformula $E_1\add\ldots\add E_n$, for each 
$i\in\{1,\ldots,n\}$, $\vec{H}$ contains the result of replacing in $F$ that osubformula  by $E_i$.
\item Whenever $F$ has a semisurface osubformula $E_1\sqd E_2\sqd \ldots\sqd E_n$, $\vec{H}$ contains the result of replacing in $F$ that osubformula  by $E_2\sqd \ldots\sqd E_n$.
\end{itemize}
None of the elements of $\vec{H}$ is provable in $\clte$, for otherwise $F$ would also be derivable in $\clte$ by $(\add)$ or $(\hspace{-2pt}\sqd\hspace{-2pt})$.  Therefore, by the induction hypothesis, each element of $\vec{H}$ is $\cltee$-provable. Since $\clte\not\vdash\qelz{F}$, by the induction hypothesis, we also have 
$\cltee\vdash\qelz{F}$. Hence,  by  $(\overline{\hspace{-2pt}\sqd\hspace{-2pt}\add})$,  $\cltee\vdash F$. 
\end{proof}

\subsection{The completeness of $\clte$}\label{scompleteness}

\begin{lemma}\label{oct5}
If $\clte\not\vdash F$, then $F$ is not valid (for any $\clte$-formula $F$).

In fact, if $\clte\not\vdash F$, then $F^*$ is not computable for some interpretation $^*$ that interprets all atoms as finitary predicates of complexity $\Delta_3$.
\end{lemma}

{\bf Idea}. For the purposes of the present explanation, let us fix a variable $x$, agree to identify any valuation $e$ with the value assigned by it to $x$ (as no other variables are going to be relevant), and call that value ``input''. Our proof of the above lemma  rests on a technique which, 
for a $\clte$-unprovable and hence $\cltee$-provable formula $F$, constructs an EPM $\cal C$ and an interpretation $^*$ such that every HPM $\cal H$ loses the game $F^*$ against
$\cal C$ when $\cal H$ receives (on the valuation tape)  itself --- its own code, to be more precise --- as an input.  

Revisiting our soundness proof for $\clt$, the central idea in it was to design  $\pp$'s strategy --- the 
EPM ${\cal S}$ --- 
following which guaranteed that the game would be eventually ``brought down to'' and ``stabilize at''  $L^*$ (in the sense that the rest of the play would essentially be just a play of $L^*$) for some $(\hspace{-2pt}\tgc\hspace{-2pt})$-justified and hence stable quasielementary formula $L={\cal F}(\lmt)$. 
Such a strategy was directly extracted from a $\clt$-proof of $F$. And it was shown that, as long as (or rather because) the elementarization $\elz{L}^*$ was true, $\cal S$ would be the winner in the game. In a  symmetric (yet simpler due to the absence of general atoms) way, a $\cltee$-proof 
of $F$ allows us to extract $\oo$'s strategy --- the EPM $\cal C$ --- that makes sure that, in every legal scenario,
$\pp$'s play against $\cal C$ over  $F^*$ will be brought down to and stabilize   
at game $L^*$ for some $(\overline{\hspace{-2pt}\tgd\hspace{-2pt}})$-justified and hence instable formula $L$ --- let us call such a formula $L$ the {\em limit formula} of the play --- such that $\cal C$ wins as long as $\elz{L}^*$ is false.  The instability of $L$ means that, whatever  the valuation/input $e$ is,  ${\elz{L}}^*$ is indeed false at $e$ for some $^*$, so that $\cal C$ wins $F^*$ on $e$ for such $^*$. However, the trouble is that $\pp$'s different strategies --- as well as different inputs $e$ of course --- may yield different limit formulas $L$ and hence require different,  perhaps conflicting, interpretations $^*$ to falsify ${\elz{L}}^*$ at $e$. Since we are trying to show the non-validity rather than non-uniform-validity of $F$, the whole trick now is to find a {\em one} common interpretation $^*$ that, for $\pp$'s 
arbitrary strategy (HPM) $\cal H$, would falsify ${\elz{L}}^*$ at some input $e$, where $L$ is the limit formula that $\cal H$ and that very input $e$ yield.  This is where a diagonalization-style idea comes in. We manage to define an interpretation $^*$  that makes 
the elementarization of every instable quasielementary formula $G$ imply the following: ``When HPM $\cal H$ plays against $\cal C$ on input $\cal H$, the limit formula is not $G$''. Now, if we let any given HPM $\cal H$ play on input $\cal H$ against $\cal C$, the elementarization of the limit formula
$L$ of the play --- which, under interpretation $^*$, claims that $L$ is not really the limit formula --- is guaranteed to be false at $\cal H$. This eventually  translates 
into $\cal H$'s having lost the game $F^*$ on input ${\cal H}$.\vspace{7pt} 

\begin{proof} Assume $\clte\not\vdash F$, which, by Lemma \ref{nov4}, means that $\cltee\vdash F$. Fix a $\cltee$-proof $T$ of $F$. From such a proof, we can extract an environment's interpretation- and valuation-independent EPM-counterstrategy $\cal C$ for $F$ in a way  symmetric to the way we extracted the machine's EPM-strategy $\cal S$ from a  $\clt$-proof in Section \ref{ssound}. $\cal C$ is a {\em counterstrategy} in the sense that $\cal C$ plays in the role of $\oo$ rather than $\pp$.\footnote{If we want to see $\cal C$ as a strategy in the ordinary sense, then it is a strategy for $\gneg F$.}   In fact, $\cal C$ is much simpler than $\cal S$, because in the present case we only deal with elementary-base formulas. Below we continue using the terminology and notation adopted in Section \ref{ssound}. 

As in Section \ref{ssound}, we see the $\cltee$-proof $T$ as a tree, with each node $b$ of which is associated a formula ${\cal F}(b)$, such that $F$ is associated with the root and, whenever $c_1,\ldots,c_n$ are the children of a node $b$, ${\cal F}(b)$ follows from 
${\cal F}(c_1),\ldots,{\cal F}(c_n)$ by one of the rules of $\cltee$, said to be the {\bf justification} of $b$. Note that the justification of a leaf can only be $(\overline{\hspace{-2pt}\tgd\hspace{-2pt}})$, as any other rule of $\cltee$ requires at least one premise. 

As in Section \ref{ssound}, with each node $b$ of $T$ we additionally associate a hyperformula ${\cal H}(b)$.  This association goes like this:
 
\begin{enumerate}
\item With the root node of $T$ we $\cal H$-associate the hyperformula 
$\tilde{F}$ 
obtained from $F$ by underlining the leftmost component of each sequential osubformula.

\item Suppose ${\cal F}(b)$ is obtained from ${\cal F}(c)$ by the $(\overline{\adc})$ rule, namely, ${\cal F}(c)$ is the result of replacing in ${\cal F}(b)$ a semisurface osubformula $E_1\adc\ldots\adc E_n$ by $E_i$. Then ${\cal H}(c)$ is the result of underlining in ${\cal H}(b)$ the $i$th component $G_i$ of  the corresponding  osubformula $G_1\adc \ldots\adc G_n$.

\item Suppose ${\cal F}(b)$ is obtained from ${\cal F}(c)$ by the $(\overline{\hspace{-2pt}\sqc\hspace{-2pt}})$ rule, namely, ${\cal F}(c)$ is the result of replacing in ${\cal F}(b)$ a semisurface osubformula $E_1\sqc E_2\sqc \ldots\sqc E_n$ by $E_2\sqc\ldots\sqc E_n$. Then ${\cal H}(c)$ is the result of  replacing in ${\cal H}(b)$ the corresponding 
 osubformula $H_1\sqc\ldots\sqc H_m\sqc \underline{G_1}\sqc G_2\sqc\ldots\sqc G_n$   by $H_1\sqc\ldots\sqc H_m\sqc G_1$ $\sqc \underline{G_2}\sqc\ldots\sqc G_n$.

\item Suppose ${\cal F}(b)$ is obtained from  senior premise ${\cal F}(c_0)$ and junior premises ${\cal F}(c_1),\ldots,{\cal F}(c_k)$ by the $(\overline{\hspace{-2pt}\sqd\hspace{-2pt}\add})$ rule. Then:
\begin{itemize} 
\item ${\cal H}(c_0)={\cal H}(b)$. 
\item Suppose ${\cal F}(c_j)$ ($1\mleq  j\mleq  k$)  is the result of replacing in ${\cal F}(b)$ a semisurface osubformula $E_1\add\ldots\add E_n$ by $E_i$. Then ${\cal H}(c_j)$ is the result of underlining  in ${\cal H}(b)$ the $i$th  component $G_i$ of the corresponding osubformula  $G_1\add \ldots\add G_n$.

\item Suppose ${\cal F}(c_j)$ ($1\mleq  j\mleq  k$)  is the result of replacing in ${\cal F}(b)$ a semisurface osubformula $E_1\sqd E_2\sqd$ $\ldots\sqd E_n$ by $E_2\sqd\ldots\sqd E_n$. Then ${\cal H}(c_j)$ is the result of  replacing in ${\cal H}(b)$ the corresponding  osubformula  $H_1\sqd\ldots \sqd H_m\sqd
\underline{G_1}\sqd G_2\sqd \ldots\sqd G_n$ by $H_1\sqd\ldots \sqd H_m\sqd G_1\sqd \underline{G_2}\sqd \ldots\sqd G_n$.

\end{itemize}

\item Suppose ${\cal F}(b)$ is obtained from ${\cal F}(c)$ by the $(\overline{\hspace{-2pt}\tgc\hspace{-2pt}})$ rule, namely, ${\cal F}(c)$ is the result of replacing in ${\cal F}(b)$ a surface osubformula $E_1\tgc\ldots\tgc  E_n$ by $E_i$. Then ${\cal H}(c)$ is the result of underlining  in  ${\cal H}(b)$ the $i$th component $G_i$ of the 
corresponding  osubformula  $G_1\tgc\ldots\tgc G_n$.

\item Suppose ${\cal F}(b)$ is obtained from   ${\cal F}(c_1),\ldots,{\cal F}(c_k)$ by the $(\overline{\hspace{-2pt}\tgd\hspace{-2pt}})$ rule. Then:
\begin{itemize}  
\item Suppose ${\cal F}(c_j)$ ($1\mleq  j\mleq  k$)  is the result of replacing in ${\cal F}(b)$ a surface osubformula $E_1\tgd\ldots\tgd E_n$ by $E_i$. Then ${\cal H}(c_j)$ is the result of underlining   in ${\cal H}(b)$ the $i$th component $G_i$ of the corresponding osubformula $G_1\tgd\ldots\tgd G_n$. 

\end{itemize}
\end{enumerate}

Our goal is to construct an environment's (counter)strategy $\cal C$ and an interpretation $^*$ such that any HPM in the role of  $\pp$ loses $e[F^*]$ against $\cal C$ for an appropriately selected valuation $e$. As in Section \ref{ssound}, 
we define and analyze $\cal C$ in terms of hyperformulas rather than formulas. 
The work of $\cal C$ will rely on neither (the yet to be defined) $^*$ nor $e$, so, as before, we shall usually omit these  parameters and identify hyperformulas with the corresponding games.  
We shall also implicitly rely on the dual of the clean environment assumption, according to which $\cal C$'s adversary never makes illegal moves. And we also rely on certain observations --- perhaps in a symmetric, dual form --- made in Section \ref{ssound}, such as Conditions (i)-(iv), Lemma \ref{march31b} or Lemma \ref{march31a}.

During its work, just like $\cal S$, our present EPM $\cal C$ maintains a record $b$, which represents the ``currently visited node'' of the tree $T$. This is exactly how the procedure COUNTERWORK followed by $\cal C$ goes:\vspace{5pt}

{\bf Procedure} COUNTERWORK: $\cal C$ jumps to the root of $T$. Then it acts depending on which of the conditions of the following six cases is satisfied by $b$:

\begin{description}
\item[Case 1:]  {\em The justification of $b$ is the $(\overline{\adc})$ rule.} Let $c$ be the child of $b$. So, ${\cal H}(b)$ has a virgin osubformula $G_1\adc\ldots\adc G_n$, and ${\cal H}(c)$ results from ${\cal H}(b)$ by underlining $G_i$ for some $1\mleq  i\mleq  n$.  Then $\cal C$ makes the move that signifies a choice of the $i$th component $G_i$ in the osubgame $G_1\adc\ldots\adc G_n$,  and jumps to $c$.
\item[Case 2:]  {\em The justification of $b$ is the $(\overline{\hspace{-2pt}\sqc\hspace{-2pt}})$ rule.} Let $c$ be the  child of $b$. So, ${\cal H}(b)$ has an  osubformula $G_1\sqc\ldots\sqc G_{i-1}\sqc\underline{G_i}\sqc G_{i+1}\sqc\ldots\sqc G_n$, and ${\cal H}(c)$ results from ${\cal H}(b)$ by moving the underline from $G_i$ to $G_{i+1}$.  Then $\cal C$ makes the move that signifies switching (from $G_i$) to $G_{i+1}$ in the corresponding osubgame, and jumps to $c$.   
\item[Case 3:]  {\em The justification of $b$ is the $(\overline{\hspace{-2pt}\sqd\hspace{-2pt}\add})$ rule.} 
Then $\cal C$ simply jumps to the senior child of $b$ --- i.e., the child with which the senior premise of ${\cal F}(b)$ is associated. 
\item[Case 4:]  {\em The justification of $b$ is the $(\overline{\hspace{-2pt}\tgc\hspace{-2pt}})$ rule.} Let $c$ be the child of $b$. So, ${\cal H}(b)$ has a virgin osubformula $G_1\tgc\ldots\tgc G_n$, and ${\cal H}(c)$ results from ${\cal H}(b)$ by underlining one of the $n$ components $G_i$.  Then $\cal C$ makes the move that signifies switching  to $G_{i}$ in $G_1\tgc\ldots\tgc G_n$, and jumps to $c$. 
\item[Case 5:]  {\em The justification of $b$ is the $(\overline{\hspace{-2pt}\tgd\hspace{-2pt}})$ rule, and $b$ is not a leaf.} Let $E_1\tgd\ldots\tgd E_n$ be the leftmost surface  $\tgd$-osubformula of ${\cal F}(b)$, and let $G_1\tgd\ldots\tgd G_n$ be the corresponding osubformula of ${\cal H}(b)$.   Let  $G_i$ be the currently  active component of the  osubgame $G_1\tgd\ldots\tgd G_n$. Note that among the children of $b$ is a node $c$ such that ${\cal H}(c)$ is the result of underlining in ${\cal H}(b)$ the $i$th component $G_i$ of the osubformula $G_1\tgd\ldots\tgd G_n$. Then  $\cal C$ jumps to that $c$. 
\item[Case 6:]  {\em The justification of $b$ is the $(\overline{\hspace{-2pt}\tgd\hspace{-2pt}})$ rule, and $b$ is a leaf.} $\cal C$  keeps granting permission until one of the following three events takes place, and then acts accordingly: 
\begin{description}
\item[Event 1:] The adversary makes a move that signifies a choice of $G_i$ in some non-abandoned osubformula $G_1\add\ldots\add G_n$ of ${\cal H}(b)$. Let $d$ be the nearest predecessor of $b$ whose justification is the $(\overline{\hspace{-2pt}\sqd\hspace{-2pt}\add})$ rule. It is not hard to see that, ignoring underlines in toggling osubformulas, ${\cal H}(b)$ and ${\cal H}(d)$ are the same. Also, among the junior children  --- i.e., children $\cal F$-associated with the junior premises --- of $d$ is a node $c$ such that ${\cal H}(c)$ results from ${\cal H}(d)$ by underlining $G_i$ in the osubformula 
$G_1\add\ldots \add G_n$. In this case, $\cal C$ jumps to that $c$.
\item[Event 2:] The adversary makes a move that signifies a switch from a component $G_i$ to the component $G_{i+1}$ in some non-abandoned osubformula $G_1\sqd\ldots\sqd G_{i-1}\sqd\underline{G_i}\sqd G_{i+1}\sqd\ldots\sqd  G_n$ of ${\cal H}(b)$. Again, let $d$ be the nearest predecessor of $b$ whose justification is the $(\overline{\hspace{-2pt}\sqd\hspace{-2pt}\add})$ rule. Ignoring underlines in toggling osubformulas, ${\cal H}(b)$ and ${\cal H}(d)$ are the same. Also, among the junior children of $d$ is a node $c$ such that ${\cal H}(c)$ results from ${\cal H}(d)$ by moving the underline from $G_i$ to $G_{i+1}$ in the osubformula $G_1\sqd\ldots\sqd G_{i-1}\sqd\underline{G_i}\sqd G_{i+1}\sqd\ldots\sqd  G_n$.  In this case, $\cal C$ jumps to that $c$.
\item[Event 3:] The adversary makes a move that signifies a switch to $H_i$ in some non-abandoned, non-virgin osubformula $H_1\tgd\ldots \tgd H_n$ (one of the components is underlined) of ${\cal H}(b)$.  Let $d$ be the nearest predecessor of $b$ whose justification is the $(\overline{\hspace{-2pt}\tgd\hspace{-2pt}})$ rule and where the osubformula $H_1\tgd\ldots \tgd H_n$ is (still) virgin.  Let $E_{1}^{1}\tgd\ldots\tgd E_{m_1}^{1}$, \ldots, $E^{k}_{1}\tgd\ldots\tgd E_{m_k}^{k}$ be all of the surface $\tgd$-osubformulas of ${\cal F}(d)$ in the left to right order, and let $G_{1}^{1}\tgd\ldots\tgd G_{m_1}^{1}$, \ldots, $G^{k}_{1}\tgd\ldots\tgd G_{m_k}^{k}$ be the corresponding osubformulas of  ${\cal H}(d)$. Then, obviously, $H_1\tgd\ldots \tgd H_n$ is $G_{1}^{i}\tgd\ldots\tgd G_{m_i}^{i}$ for one of $1\mleq  i\mleq  k$. Let $j\equals i\plus 1$ if $i\mless k$, and $j\equals 1$ if $i\equals k$. Let $G_{h}^{j}$ be the currently active component of $G_{1}^{j}\tgd\ldots\tgd G_{m_j}^{j}$. And let $c$ be the child of $d$ such that ${\cal H}(c)$ is the result of underlining $G_{h}^{j}$ in ${\cal H}(d)$. Then $\cal C$ jumps to $c$.
\end{description}   

\end{description}

It is obvious that $\cal C$ grants permission infinitely many times (even if the adversary makes illegal moves in violation of our assumption), so that it is a  fair EPM.  

Note the symmetry between our present $\cal C$ and the machine $\cal S$ from Section \ref{ssound}, as well as between the systems $\cltee$ and $\clt$, and the corresponding proof trees. The only difference beyond this symmetry is that, in the present case, we do not have to deal with general or pseudoelementary atoms, and the associated (M) rule. This only makes things simpler. So, further we can reason in a way 
 symmetric to (but somewhat simpler than) the way we reasoned in Section \ref{ssound}, and establish that, in any play with environment's counterstrategy $\cal C$, a node $\lmt$ --- the limit node of the play (defined exactly as before) --- will be reached  such that:
\begin{itemize}
\item The justification of $\lmt$ is the $(\overline{\hspace{-2pt}\tgd\hspace{-2pt}})$ rule, and hence ${\cal F}(\lmt)$ is an instable quasielementary formula, so that $\elz{{\cal F}(\lmt)}$ is a non-tautological elementary formula.
\item The overall run generated in the play is ${\cal H}(\lmt)$-adequate (``adequacy'', again, defined as before, albeit Conditions 5 and 6 of Definition \ref{march31} are now vacuous and should be ignored). 
\end{itemize}

Next, remember from Section \ref{ssound}  how we constructed the (hyper)formula $H_8=\elz{{\cal F}(\lmt)}$ and the run $\Gamma_8$ from ${\cal H}(\lmt)$ and an arbitrary run $\Gamma$ --- let us fix it throughout this paragraph  --- of the original game generated in the scenario where $\cal S$ played in the role of $\pp$.  We can apply a similar (but simpler because some old steps will be redundant due to the absence of general and pseudoelementary atoms) here and, after reasoning in a way symmetric to the one that took us to Lemma \ref{apr14b}, find that Lemma \ref{apr14b} holds in the present case in the following, symmetric form: 
\begin{equation}\label{apr14bb}
\mbox{\em $\Gamma_8$ is $\elz{{\cal F}(\lmt)}$-adequate,   and it is a  $\oo$-won run of $\elz{{\cal F}(\lmt)}$  only if $\Gamma$ is a $\oo$-won run of $F$.}
\end{equation}
But the $\elz{{\cal F}(\lmt)}$-adequacy of $\Gamma_8$   in our present case means that $\Gamma_8$ is simply the empty run $\emptyrun$ 
--- unlike the situation in Section \ref{ssound}, where it could contain moves in (the now absent) pseudoelementary oliterals. Thus, rephrasing (\ref{apr14bb}), we have:
\begin{equation}\label{apr15a}
\mbox{\em If $\emptyrun$ is a $\oo$-won run of  $\elz{{\cal F}(\lmt)}$, then $\Gamma$ is a $\oo$-won run of $F$.}
\end{equation}
But $\elz{{\cal F}(\lmt)}$ is an elementary formula with all of its atoms interpreted as elementary games. Hence it is obvious that the empty run $\emptyrun$ is a $\oo$-won run of it if and only if $\elz{{\cal F}(\lmt)}$ is simply false. Thus, (\ref{apr15a}) can be further rephrased as follows:
\begin{equation}\label{apr15b}
\mbox{\em If $\elz{{\cal F}(\lmt)}$ is false, then $\Gamma$ is a $\oo$-won run of $F$.}
\end{equation}

What we have observed so far and what is relevant to our further argument, after restoring the so far usually suppressed parameters $^*$ and $e$, can be summarized as the following lemma:

\begin{lemma}\label{apr15c}
Let $\Gamma$ be an arbitrary run that could have been generated when $\cal C$ plays in the role of $\oo$. Then there is a node $\lmt$ of $T$, called the {\bf limit node} of $\Gamma$,  such that the following two conditions are satisfied:

1. The justification of $\lmt$ is $(\overline{\hspace{-2pt}\tgd\hspace{-2pt}})$.

2. For any interpretation $^*$ and valuation $e$, if $e[\elz{{\cal F}(\lmt)}^*]$ is false, then $\Gamma$ is a $\oo$-won run of $e[F^*]$.
\end{lemma}

Of course, different $\pp$'s strategies (HPMs) $\cal N$ and different valuations $e$ may yield different runs  and hence induce different limit nodes. So, our goal now is    
 to select an interpretation $^*$ such that, for any HPM $\cal N$, there is a valuation $e$ at which $\elz{{\cal F}(\lmt)}^*$ is false, where 
$\lmt$ is the limit node of the play of $\cal N$ against $\cal C$ on valuation $e$. In view of Lemma \ref{apr15c}, this would mean that no HPM can win $F^*$ against $\cal C$. 

Let us fix some standard way of describing HPMs, and let 
\[{\cal N}_1,\ {\cal N}_2,\ {\cal N}_3,\ \ldots\]
be the list of all HPMs arranged according to the lexicographic order of their descriptions, so that each constant $c$ can be considered the code of ${\cal N}_c$. Next, we fix a variable $x$ and agree that, for each constant $c$, \[e_c\] is the valuation with $e_c(x)\equals c$ (and, say,  $e_c(y)\equals 1$ for any other variable $y\not=x$). 
Further, \[\lmt_c\] will denote the limit node of the game over $F$ between ${\cal N}_c$, in the role of $\pp$, and our $\cal C$, in the role of $\oo$, on valuation $e_c$. In more precise terms, $\lmt_c$ is the limit formula of (induced by) the ${\cal N}_c$ vs. $\cal C$ run on $e_c$ (remember Lemma \ref{lem}). 

Finally, let \[d_1,\ldots,d_k\] be all nodes of $T$ whose justification is the $(\overline{\hspace{-2pt}\tgd\hspace{-2pt}})$ rule. For each such $d_i$, we fix a classical model (true/false assignment for atoms) $M_i$ such that 
\[\mbox{\em $M_i$ makes the elementarization of ${\cal F}(d_i)$ false.}\]
 And, for each $i\in\{1,\ldots,k\}$, we define the predicate $K_i$ by 
\[\mbox{\em $K_i$ is true at a valuation $e$ iff $\lmt_{e(x)}=d_i$.}\]  
Now we define the interpretation $^*$ by stipulating that, for each atom $p$, 
\[p^*=\mld\{K_i\ |\ 1\mleq  i\mleq  k, \ \mbox{\em $p$ is true in $M_i$}\}.\]
($\mld {\cal L}$ means the $\mld$-disjunction of all elements of ${\cal L}$, understood as $\tlg$ when the set $\cal L$ is empty.)

Consider an arbitrary $c\in\{1,2,\ldots\}$. In view of clause 1 of Lemma \ref{apr15c}, we must have $\lmt_c=d_j$ for one of the $j\in\{1,\ldots,k\}$. Fix this $j$. Observe that, at valuation $e_c$, $K_j$ is true and all other $K_i$ ($i\not=j$, $1\mleq  i\mleq  k$) are false. With this fact in mind, it is easy to see that, for every atom $p$, 
$p$ is true in $M_j$ iff the predicate $p^*$ is true at $e_c$. This obviously extends from atoms to their $\gneg,\mlc,\mld$-combinations, 
so that $\elz{{\cal F}(d_j)}$ is true in $M_j$ iff the predicate ${\elz{{\cal F}(d_j)}}^*$ is true at $e_c$. And, by our choice 
of the models $M_i$, the formula $\elz{{\cal F}(d_j)}$, i.e. $\elz{{\cal F}(\lmt_c)}$, is false in $M_j$. Consequently, ${\elz{{\cal F}(\lmt_c)}}^*$ is false at 
$e_c$.  But, according to clause 2 of Lemma \ref{apr15c},   the fact that ${\elz{{\cal F}(\lmt_c)}}^*$ is false at 
$e_c$ implies that ${\cal N}_c$ loses $F^*$ against $\cal C$ on valuation $e_c$. 

Thus, no ${\cal N}_c$ computes $F^*$, meaning that  $F^*$ is not computable, because every HPM is ${\cal N}_c$ for some $c$.  Note also that, as promised in the lemma, the predicate $p^*$ (for any atom $p$) is finitary as only the value assigned to $x$ matters. 

To officially complete the present proof, it remains to show that 
\begin{equation}\label{delta}
\mbox{\em the complexity of $p^*$ is $\Delta_3$ (for any atom $p$).}
\end{equation}

Remember that an arithmetical predicate $A(c)$ (with $c$ here seen as a variable) is said to have complexity $\Sigma_3$ iff it can be written as $\exists x\forall y \exists z B(c,x,y,z)$ for some decidable predicate $B(c,x,y,z)$; and $A(c)$ is of complexity $\Delta_3$ iff both $A(c)$ and $\gneg A(c)$ are of complexity $\Sigma_3$. 
 
We defined $p^*$ as a disjunction of some $K_i$, that we now think of as unary arithmetical predicates and write as $K_i(c)$. Disjunction is known to preserve  $\Delta_3$ --- as well as $\Sigma_3$ ---
complexity, so, in order to verify (\ref{delta}), it is sufficient to show that each $K_i(c)$ ($1\mleq  i\mleq  k$) 
is of complexity $\Delta_3$. Further, we know that the limit node should be one of $d_j$ with $1\mleq  j\mleq  k$.  Hence, $\gneg K_i(c)$ is equivalent to $\mld\{K_j(c)\ |\ 1\mleq  j\mleq  k,\  j\not=i\}$. 
Consequently, in order to show that each $K_i(c)$ is of complexity $\Delta_3$, it would suffice to show that 
each $K_i(c)$ is of complexity $\Sigma_3$. 

Let, for any node $b$ of $T$, $J_b(c,t)$ mean ``in the play of  ${\cal N}_c$ against  $\cal C$ on valuation $e_c$, at time $t$, $\cal C$ jumps to node $b$''. $J_b(c,t)$ is a decidable predicate.  A decision procedure for it first 
constructs the machine ${\cal N}_c$ from number $c$. Then it lets this machine play against ${\cal C}$ on valuation $e_c$ 
as described in the proof idea for Lemma \ref{lem}. In particular, it traces, in parallel, how the configurations of the two machines evolve up to the $t$th computation step of ${\cal C}$, i.e. its $t$th configuration. Then the procedure looks at whether a jump to the node $b$ occured in that configuration, and correspondingly says ``Yes'' or ``No''.

 Looking at the meaning of $K_i(c)$, this predicate asserts nothing but that $d_i$ is the limit node of the play of ${\cal N}_c$  against $\cal C$ on valuation $e_c$. In other words, $K_i(c)$ says that, in the play of ${\cal N}_c$  against $\cal C$ on valuation $e_c$, at some time $x$ $(\exists x$), $\cal C$ jumps to the node $d_i$ and, for any $y$ ($\forall y$) greater than $x$,  if $\cal C$  jumps to a node $b$ at time $y$, then $b$ is a (proper) descendant of $d_i$, and there is a $z$ ($\exists z$) greater than  $y$ such that, at time $z$, $\cal C$ jumps to some node which is not a descendant of $b$. Looking at this characterization and taking into account the decidability of the predicate  
$J_b(c,t)$ used in it (in the form ``... jumps ...''), with  some minimal experience in dealing with arithmetical complexity, one can see that $K_i(c)$ is indeed of complexity $\Sigma_3$, as desired. \end{proof}

\section{The completeness of $\clt$}\label{s9}
This section is devoted to proof of the completeness part of Theorem \ref{thcl2}, according to which, whenever $F$ is a $\clt$-unprovable formula,  $F$ is not valid and, in particular, 
 $F^*$ is not computable  for some  interpretation 
$^*$ that interprets all elementary atoms of $F$ as finitary predicates of arithmetical complexity $\Delta_3$, and interprets all general atoms of $F$ as problems of the form
\((A^{1}_{1}\add\ldots\add A_{m}^{1})\adc\ldots\adc (A_{1}^{m}\add\ldots\add A_{m}^{m}),\)
where each $A_{i}^{j}$ is a finitary predicate of arithmetical complexity $\Delta_3$.

Outlining our proof idea, we are going to show that, if $\clt\not\vdash F$, then there is a $\clte$-formula $\lceil F\rceil$ of the same form as $F$ that is not provable in $\clte$. Precisely, ``the same form as $F$'' here means that $\lceil F\rceil$ is the result of rewriting/expanding in $F$ every general atom $P$ as a certain elementary-base formula $\check{P}_{\add}^{\adc}$. This, in view of the already known completeness of $\clte$, immediately yields non-validity for $F$. As it turns out, the above formulas  $\check{P}_{\add}^{\adc}$, that we call {\em molecules}, can be chosen to be 
as simple as sufficiently long $\adc$-conjunctions of sufficiently long $\add$-disjunctions of arbitrary ``neutral'' (not occurring in $F$ and pairwise distinct) elementary atoms, with the ``sufficient length'' 
of those conjunctions/disjunctions being bounded by the number of occurrences of general atoms in $F$. 

Intuitively, the reason why $\clte\not\vdash\lceil F\rceil$, i.e. why $\pp$ cannot win (the game represented by) $\lceil F\rceil$, is that a smart environment may start choosing different conjuncts/disjuncts in different occurrences of $\check{P}_{\add}^{\adc}$. The best that $\pp$ can do in such a play is to match any given positive or negative occurrence of $\check{P}_{\add}^{\adc}$ with one (but not more!) 
negative or positive occurrence of the same subgame --- match in the sense that $\pp$ mimics the environment's moves in order to keep the subgames/subformulas at the two occurrences identical. Yet, this 
is insufficient for $\pp$ to achieve a guaranteed success. This is so because $\pp$'s matching decisions for $\lceil F\rceil$ could be modeled by appropriate applications of the (M) rule in an attempted $\clt$-proof for $F$, and so can be --- through the remaining rules  ---
either player's decisions associated with choice, sequential and toggling connectives in the non-molecule parts of $\lceil F\rceil$. A winning strategy ($\clte$-proof) for 
$\lceil F\rceil$ would then translate into a $\clt$-proof for $F$, which, however, does not exist. What follows is an implementation of the idea we have just outlined.  

Fix a $\clt$-formula $F$. Let $\cal P$ be the set of all general atoms occurring in $F$. 
Let us fix $m$ as the total number of occurrences of such atoms in $F$;\footnote{In fact, a much smaller $m$ would be sufficient for our purposes. E.g., $m$ can be chosen to be 
such that no given general atom has more than $m$ occurrences in $F$. But why try to economize?} 
if there are fewer than two such occurrences, then we take $m\equals 2$. 

For the rest of this section, let us agree that 
\[\mbox{\em $a,b$ always range over $\{1,\ldots,m\}$.}\]

For each $P\in{\cal P}$ and each $a,b$, let us fix an  elementary atom
\begin{itemize} 
\item $\check{P}_{b}^{a}$\label{z53} 
\end {itemize}
not occurring in $F$. We assume that $\check{P}^{a}_{b}\not=\check{Q}^{c}_{d}$ as 
long as either $P\not= Q$ or $a\not=c$ or $b\not=d$. Note that the $\check{P}_{b}^{a}$ are elementary atoms despite our
``tradition" according to which the capital letters $P,Q,\ldots$ stand for general atoms.    

Next, for each $P\in{\cal P}$ and each $a$, we define

\begin{itemize}
\item $\check{P}^{a}_{\add}\ =\ \check{P}^{a}_{1}\add\ldots\add \check{P}^{a}_{m}$.
\end{itemize}

Finally, for each $P\in{\cal P}$, we define 
\begin{itemize}
\item $\check{P}^{\adc}_{\add}\ = \check{P}^{1}_{\add}\adc\ldots\adc\check{P}^{m}_{\add}$, \ i.e. \ 
\(\check{P}^{\adc}_{\add}\ =\ (\check{P}^{1}_{1}\add\ldots\add \check{P}^{1}_{m})\adc\ldots\adc (\check{P}^{m}_{1}\add\ldots\add \check{P}^{m}_{m}).\)
\end{itemize}

We refer to the above formulas $\check{P}^{a}_{b}$, $\check{P}^{a}_{\add}$ and $\check{P}^{\adc}_{\add}$ as 
{\bf molecules},
in particular, {\bf $P$-based molecules}. 
To differentiate between the three sorts of molecules, we call the molecules of the type $\check{P}^{a}_{b}$ 
{\bf small}, call the molecules of the type $\check{P}^{a}_{\add}$ 
{\bf medium}, and call the molecules of the type $\check{P}^{\adc}_{\add}$ 
{\bf large}. Thus, where $k$ is the cardinality of $\cal P$, altogether there are $k$ large molecules, $k\times m$ medium molecules and $k\times m\times m$ small molecules.

For simplicity, for the rest of this section we assume/pretend that the languages of  
$\clt$ and $\clte$ have no nonlogical atoms other than those occurring in $F$ plus the atoms $\check{P}_{a}^{b}$ 
($P\in {\cal P}$, \ $a,b\in\{1,\ldots,m\}$). This way the scope of the term ``formula'' is correspondingly redefined. 
 
An occurrence of a molecule $M$ in a formula can be {\bf positive} or {\bf negative}. While a positive occurrence literally means $M$, a negative occurrence looks 
like $\gneg M$, which --- unless $M$ is a small molecule --- should be considered a standard abbreviation. For example, a negative occurrence of  the medium 
molecule $\check{P}^{a}_{1}\add\ldots\add \check{P}^{a}_{m}$ is nothing but an (``ordinary'', positive) occurrence of $\gneg\check{P}^{a}_{1}\adc\ldots\adc\gneg \check{P}^{a}_{m}$. One should be especially careful when applying the terms ``positive occurrence'' and ``negative occurrence'' to small molecules, as here the meaning of our terminology somewhat diverges from its earlier-used meaning for atoms. Specifically, a positive occurrence of a small molecule $\check{P}^{a}_{b}$ means --- as expected --- an occurrence that comes without $\gneg$ in the formula. As for a negative occurrence of the {\em molecule} $\check{P}^{a}_{b}$ (as opposed to the {\em atom} $\check{P}^{a}_{b}$), it means an occurrence of the subformula 
$\gneg \check{P}^{a}_{b}$ rather than just the $\check{P}^{a}_{b}$ part of it under $\gneg$. So, for example, the result of replacing the negative occurrence of the molecule  
$\check{P}^{a}_{b}$ by $Q$ in the formula $E\mld \gneg \check{P}^{a}_{b}$ is the formula $E\mld Q$ rather than $E\mld\gneg Q$, as $\gneg$ was a part of what we call 
a ``negative occurrence of $\check{P}^{a}_{b}$''.   

Let us say that a (positive or negative) occurrence of a molecule in a given $\clte$-formula is {\bf independent} iff it is not a part of another (``larger") molecule. For example, the negative occurrence of the medium molecule $\check{P}_{1}^{1}\add\ldots\add\check{P}_{m}^{1}$     in the following formula is independent while its positive occurrence is not:
\[(\gneg\check{P}_{1}^{1}\adc\ldots\adc\gneg\check{P}_{m}^{1}) \mld \bigl((\check{P}_{1}^{1}\add\ldots\add\check{P}_{m}^{1})\adc\ldots\adc(\check{P}_{1}^{m}\add\ldots\add\check{P}_{m}^{m})\bigr).\] 
Of course, semisurface occurrences of molecules are always independent, and so 
are any --- semisurface or non-semisurface --- occurrences of large molecules.

Let $E$ be a $\clte$-formula. By an {\bf isolated} small molecule of $E$ (or {\bf $E$-isolated} small molecule, or a small molecule {\bf isolated in $E$}) 
we will mean a small molecule that has exactly one independent occurrence in $E$. We will say that such a molecule is {\bf positive} or {\bf negative} depending on whether its independent occurrence in $E$ is positive or negative. 

Next, the {\bf floorification} of $E$, denoted 
  \[\lfloor E\rfloor,\]
is the result of replacing in $E$ every positive (resp. negative) independent occurrence of every $P$-based (each $P\in{\cal P}$) large, medium and $E$-isolated small molecule\footnote{Remember what was said earlier about the meaning of ``negative occurrence'' for small molecules.} by the general literal $P$ (resp. $\gneg P$).

\begin{lemma}\label{april12u}
Assume $E$ is a quasielementary formula provable in $\clte$, and $E'$ is the result of replacing in $E$ some (positive or negative) isolated small molecules by $\tlg$. Then $\clte\vdash E'$.
\end{lemma}

\begin{proof} Induction on the lengths of proofs. Assume the conditions of the lemma. The rules  $(\hspace{-2pt}\sqc\hspace{-2pt}\adc)$, $(\add)$ and $(\hspace{-2pt}\sqd\hspace{-2pt})$ can only derive non-quasielementary formulas.  So, the last rule used in the proof of $E$ should be either $(\hspace{-2pt}\tgd\hspace{-2pt})$ or $(\hspace{-2pt}\tgc\hspace{-2pt})$. We correspondingly consider the following two cases.

{\em CASE 1:} {\em $E$ is derived by $(\hspace{-2pt}\tgd\hspace{-2pt})$} from a premise $G$. Obviously then $E'$ follows by the same rule $(\hspace{-2pt}\tgd\hspace{-2pt})$ from a quasielementary formula $G'$ which is the result of replacing in $G$ some  isolated small molecules by $\tlg$. And,  
by the induction hypothesis,  $\clte\vdash G'$. Hence, $\clte\vdash E'$. 

{\em CASE 2:} {\em $E$ is derived by $(\hspace{-2pt}\tgc\hspace{-2pt})$} from premises $G_1,\ldots,G_n$ (possibly $n\equals 0$). 
It is obvious that, as long as $E'$ is stable, it follows by the same rule $(\hspace{-2pt}\tgc\hspace{-2pt})$ from formulas $G'_1,\ldots,G'_n$, where each  
$G'_i$ is the result of replacing in $G_i$ some isolated small molecules by $\tlg$. And, 
by the induction hypothesis, each $G'_i$ is provable.  So, it remains to show that $ E'$ is stable, i.e., that $\elz{ E'}$ is a tautology. Since $E$ is derived by $(\hspace{-2pt}\tgc\hspace{-2pt})$, $E$ is stable, i.e., $\elz{E}$ is a tautology. But notice that the only difference between 
$\elz{E}$ and $\elz{E'}$ is that, wherever the former has (occurrences of) isolated small --- positive or negative --- molecules, the latter has $\tlg$ instead. It is known from classical logic that replacing isolated literals (literals that contain an atom that has a single occurrence in the formula) by whatever formulas does not destroy tautologicity. So, $\elz{ E'}$ is indeed a tautology.
\end{proof}

\begin{lemma}\label{april12}
 For any  quasielementary formula $E$, if $\clte\vdash E$, then $\clte $ and hence $\clt$ proves $\qelz{\lfloor E\rfloor}$.
\end{lemma}

\begin{proof} Assume $E$ is a quasielementary formula. Observe that then $\qelz{\lfloor E\rfloor}$ is nothing but the result of replacing in $E$ all isolated small molecules by $\tlg$. Therefore, by Lemma \ref{april12u}, if $\clte\vdash E$, we also have $\clte\vdash \qelz{\lfloor E\rfloor}$.
\end{proof}

We say that a $\clte$-formula $E$ is {\bf good} iff the following conditions are satisfied:
\begin{description}
\item[Condition (i):] $E$ contains at most $m$ independent occurrences of molecules.
\item[Condition (ii):] Only large molecules (may) have independent non-semisurface occurrences in $E$.  
\item[Condition (iii):] Each small molecule has at most one positive and at most one negative 
independent occurrence in $E$. 
\item[Condition (iv):] For each medium molecule $\check{P}^{a}_{\add}$, $E$ has at most one positive independent occurrence of 
$\check{P}^{a}_{\add}$, and when $E$ has such an occurrence, then for no $b$ does $E$ have a positive independent occurrence of the small molecule $\check{P}^{a}_{b}$.\vspace{2pt}
\end{description}

\begin{lemma}\label{april11}
 For any good  $\clte$-formula $E$, if $\clte\vdash E$, then $\clt\vdash\lfloor E\rfloor$.
\end{lemma}

\begin{proof}
Assume $E$ is a good  $\clte$-formula, and $\clte\vdash E$. By induction on the length of the 
$\clte$-proof of $E$, we want to show that $\clt\vdash\lfloor E\rfloor$. We need to consider the following five cases, depending on which of the rules of $\clte$  was used (last) to derive $E$.\vspace{8pt}

{\em CASES 1-2:} $E$ is derived by either $(\hspace{-2pt}\tgc\hspace{-2pt})$ or $(\hspace{-2pt}\tgd\hspace{-2pt})$. Then $E$ is quasielementary and, by Lemma \ref{april12}, $\clt\vdash\qelz{\lfloor E\rfloor}$. But since $E$ is quasielementary,  ($E$ and hence) $\lfloor E\rfloor$ does not contain any sequential or choice operators. This means that $\lfloor E\rfloor$ follows from $\qelz{\lfloor E\rfloor}$ (no junior premises) by $(\hspace{-2pt}\sqc\hspace{-2pt}\adc)$. So, $\clt\vdash \lfloor E\rfloor$.

{\em CASE 3:} $E$ is derived by  $(\hspace{-2pt}\sqc\hspace{-2pt}\adc)$. Let us fix the set $\vec{H}$ of junior premises of $E$. Each formula $H\in\vec{H}$ is provable in $\clte$. Hence, by the induction hypothesis, we have:

\begin{equation}\label{may25}
\mbox{\em For any $H\in\vec{H}$, if $H$ is good, then $\clt\vdash\lfloor H\rfloor$.}
\end{equation}

We consider the following three subcases. The first two subcases are not mutually exclusive, and either one can be chosen when both of them apply. Specifically, Subcase 3.1 (resp. 3.2) is about when $E$ has a positive (resp. negative) semisurface occurrence of a large (resp. medium) molecule. Then, as we are going to see, replacing that molecule by a ``safe'' $\adc$-conjunct of it, corresponding to a smart environment's possible move, yields a good formula $H$ from $\vec{H}$ such that $\lfloor E\rfloor=\lfloor H\rfloor$. This,  by (\ref{may25}), automatically means the $\clt$-provability of $\lfloor E\rfloor$. The remaining Subcase 3.3 is about when all semisurface occurrences of large (resp. medium) molecules in $E$ are negative (resp. positive). This will be shown to imply that $\lfloor E\rfloor$ follows from its quasielementarization and the floorifications of some elements of $\vec{H}$ by $(\hspace{-2pt}\sqc\hspace{-2pt}\adc)$ for ``the same reasons as'' $E$ follows from its quasielementarization and $\vec{H}$.\vspace{5pt}  

{\em Subcase 3.1:} $E$ has a positive semisurface occurrence of a large molecule $\check{P}_{\add}^{\adc}$, i.e., an occurrence of 
\[\check{P}_{\add}^{1}\adc\ldots\adc \check{P}_{\add}^{m}.\]
Pick any $a\in\{1,\ldots,m\}$ such that neither the medium molecule 
$\check{P}_{\add}^{a}$ nor any small molecule $\check{P}_{b}^{a}$ (whatever $b$) have independent occurrences in $E$. Such an $a$ exists, for otherwise we would have at least $m\plus 1$ independent occurrences of molecules  in $E$ (including the occurrence of $\check{P}_{\add}^{\adc}$), which violates Condition (i) of  the definition of ``goodness''. Let $H$ be the result of replacing in $E$
the above occurrence of $\check{P}_{\add}^{\adc}$ by $\check{P}_{\add}^{a}$. Clearly $H\in\vec{H}$. Observe that when transferring from $E$ to $H$, we just ``downsize" $\check{P}_{\add}^{\adc}$ and otherwise do not create any additional independent occurrences of molecules, so Condition (i) continues to be satisfied for $H$. Neither do we introduce any new non-semisurface occurrences of molecules or any new independent occurrences of small molecules, so Conditions (ii) 
 and (iii) also continue to hold for $H$. And 
our choice of $a$ obviously guarantees that so does Condition (iv). To summarize, $H$ is good. Therefore, by (\ref{may25}),  $\clt\vdash \lfloor H\rfloor$. 
Finally, note that, when floorifying a given formula, both $\check{P}_{\add}^{\adc}$ and $\check{P}_{\add}^{a}$ get replaced by the same atom $P$; and, as the only difference between $E$ and $H$ is that $H$ has $\check{P}_{\add}^{a}$ where $E$ has $\check{P}_{\add}^{\adc}$, 
obviously $\lfloor H \rfloor=\lfloor E\rfloor$. Thus, $\clt\vdash \lfloor E\rfloor$.\vspace{5pt} 

{\em Subcase 3.2:} $E$ has a negative semisurface occurrence of a medium molecule $\check{P}_{\add}^{a}$ --- that is, an occurrence of 
\[\gneg\check{P}_{a}^{1}\adc\ldots\adc \gneg\check{P}_{a}^{m}.\] Pick any $b$ such 
that $E$ does not have an independent occurrence of $\check{P}_{b}^{a}$. Again, in view of Condition (i), such a $b$ exists. Let $H$ be the result of replacing in $E$
the above occurrence of $\gneg\check{P}_{a}^{1}\adc\ldots\adc \gneg\check{P}_{a}^{m}$ by $\gneg\check{P}_{b}^{a}$. Certainly $H\in\vec{H}$. Conditions 
(i) and (ii) continue to hold for $H$ for the same reasons as in Subcase 3.1. In view of our choice of $b$, Condition (iii) is also inherited by $H$ from $E$. And so is Condition (iv), because $H$ has the same positive occurrences of (the same) molecules as $E$ does.
Thus, $H$ is good. Therefore, by (\ref{may25}),  $\clt\vdash \lfloor H\rfloor$. 
It remains to show that $\lfloor H\rfloor=\lfloor E\rfloor$.  Note that when floorifying $E$, $\check{P}_{\add}^{a}$ gets replaced by $P$. But so does $\check{P}_{b}^{a}$ when floorifying $H$ because, by our choice of $b$, $\check{P}_{b}^{a}$ is an isolated small molecule of $H$.
Since the only difference between $H$ and $E$ is that $H$ has $\check{P}_{b}^{a}$ where $E$ has $\check{P}_{\add}^{a}$, it is then obvious that indeed $\lfloor H\rfloor=\lfloor E\rfloor$.\vspace{5pt}

{\em Subcase 3.3:} Neither of the above two conditions is satisfied. This means that in $E$ all semisurface occurrences of large molecules are negative, and all semisurface occurrences of medium molecules  are positive. Every such occurrence is an occurrence of a $\add$-formula which, as we remember, gets replaced by $\tlg$ when transferring from $E$ to $\qelz{E}$; but the same happens to the corresponding occurrences of $\gneg P$ or $P$ in $\lfloor E\rfloor$ when transferring from $\lfloor E\rfloor$ to $\qelz{\lfloor E\rfloor}$. Based on this observation, with a little thought we can see that 
$\qelz{\lfloor E\rfloor}$ is ``almost the same'' as $\qelz{E}$; specifically, the only difference between these two formulas 
is that $\qelz{\lfloor E\rfloor}$ has $\tlg$ where $\qelz{E}$ has isolated small molecules (positive or negative). And this, in turn, obviously means that 
$\qelz{\lfloor E\rfloor}=\qelz{\lfloor\qelz{E}\rfloor}$. 
As $E$ is derived by $(\hspace{-2pt}\sqc\hspace{-2pt}\adc)$, we have $\clte\vdash \qelz{E}$. Hence, by Lemma \ref{april12}, $\clt\vdash \qelz{\lfloor\qelz{E}\rfloor}$ and, since $\qelz{\lfloor\qelz{E}\rfloor}=\qelz{\lfloor E\rfloor}$, we have

\begin{equation}\label{mar2}
\clt\vdash \qelz{\lfloor E\rfloor}.
\end{equation} 

Now consider an arbitrary formula $H'$ that is the result of replacing in $\lfloor E\rfloor$ a semisurface occurrence of a subformula 
$G'_1\adc\ldots\adc G'_n$ by $G'_i$ for some $i\in \{1,\ldots,n\}$. Our goal is to show that 
\begin{equation}\label{oct28a}
\mbox{\em $\clt\vdash H'$ (arbitrary $H'$ satisfying the above condition)}.
\end{equation}
The logical structure of $E$ is the same as that of $\lfloor  E\rfloor$, with the only difference that, wherever $\lfloor  E\rfloor$ has general literals, $E$ has  molecules. Hence $E$ has an  
occurrence of a subformula $G_1\adc\ldots\adc G_n$ where  $\lfloor E\rfloor$ has the above occurrence of $G'_1\adc\ldots\adc G'_n$. Let then 
$H$ be the result of replacing $G_1\adc\ldots\adc G_n$ by $G_i$ in $E$. Of course $H\in\vec{H}$. 
 So, in view of (\ref{may25}), it would suffice to show (in order to verify (\ref{oct28a}))  that $H$ is good and 
$H'=\lfloor H\rfloor$. Let us first see that $H$ is good. When transferring from $E$ to $H$, Condition (i) is inherited by $H$ for the same or a similar reasons as in all of the previous cases. So is Condition (ii) because we are not creating any new non-semisurface occurrences. Furthermore, notice that $G_1\adc\ldots\adc G_n$ 
is not a molecule, for otherwise in $\lfloor E\rfloor$ we would have a general literal rather than $G'_1\adc\ldots\adc G'_n$.
Hence, in view of Condition (ii), $G_i$ is not a small or medium molecule. This means that, when transferring from $E$ to $H$, we are not creating new 
independent/semisurface occurrences of any small or medium molecules, so that 
Conditions (iii) and (iv) are also inherited by $H$ from $E$. To summarize, $H$ is indeed good. 
Finally, it is also rather obvious that $H'=\lfloor H\rfloor$. The only case when we might have $H'\not=\lfloor H\rfloor$
would be if there was a small molecule $\check{P}_{b}^{a}$ isolated in $E$ but not in $H$, or vice versa (so that the independent occurrence of that molecule in $E$ 
would become $P$ in  $\lfloor E\rfloor$ and hence in $H'$ but stay $\check{P}_{b}^{a}$ in $\lfloor H\rfloor$, or vice versa). But, as we   observed just a little  while ago, $E$ and $H$ do not differ in their independent/semisurface occurrences of small molecules.

Next, consider an arbitrary formula $H''$ that is the result of replacing in $\lfloor E\rfloor$ a semisurface occurrence of a subformula 
$G''_1\sqc G''_2\sqc\ldots\sqc G''_n$ by $G''_2\sqc\ldots\sqc G''_n$. Our goal is to show that 
\begin{equation}\label{oct28b}
\mbox{\em $\clt\vdash H''$ (arbitrary $H''$ satisfying the above condition)}.
\end{equation}
This case is very similar to the case handled in the previous paragraph. The logical structure of  $E$ the same as that of $\lfloor  E\rfloor$, so
$E$ has an  
occurrence of a subformula $G_1\sqc G_2\sqc\ldots\sqc G_n$ where  $\lfloor E\rfloor$ has the above occurrence of $G''_1\sqc G''_2\sqc \ldots\sqc G''_n$. Let then 
$H$ be the result of replacing $G_1\sqc G_2\sqc\ldots\sqc G_n$ by $G_2\sqc\ldots\sqc G_n$ in $E$. Of course $H\in\vec{H}$. Continuing arguing as in the previous paragraph, we find that $H$ is good and that  $H''=\lfloor H\rfloor$, which, by (\ref{may25}), implies the desired (\ref{oct28b}).

Based on (\ref{mar2}), (\ref{oct28a}) and (\ref{oct28b}), we find that $\lfloor E\rfloor$ is derivable in $\clt$ by $(\hspace{-2pt}\sqc\hspace{-2pt}\adc)$.\vspace{8pt}

The remaining two CASES 4 and 5 are about when $E$ is derived by $(\add)$ or $(\hspace{-2pt}\sqd\hspace{-2pt})$ from a premise $H$. Such an $H$ turns out to be good and hence (by the induction hypothesis) its floorification $\clt$-provable. And, ``almost always'', either $\lfloor E\rfloor=\lfloor H\rfloor$, or  $\lfloor E\rfloor$ follows from $\lfloor H\rfloor$ by $(\add)$ or $(\hspace{-2pt}\sqd\hspace{-2pt})$ for the same reasons as $E$ follows from $H$. An exception is the special case of $(\add)$ when $H$ is the result of replacing in $E$ a positive occurrence of a medium molecule 
$\check{P}_{\add}^{a}$ by one of its disjuncts $\check{P}_{b}^{a}$ such that $E$ has a negative independent occurrence of $\check{P}_{b}^{a}$. Using our earlier terms, this is a step signifying $\pp$'s (final) decision to ``match'' the two $P$-based molecules. In this case, while $\lfloor E\rfloor$ is neither the same as $\lfloor H\rfloor$ nor does it  follow from $\lfloor H\rfloor$ by $(\add)$, it {\em does} follow from $\lfloor H\rfloor$ by (M). The secret is that the two $P$-based molecules are non-isolated small molecules in $H$ and hence remain elementary literals in $\lfloor H\rfloor$, while they turn into general literals in $\lfloor E\rfloor$.\vspace{8pt} 
 
{\em CASE 4:} $E$ is derived by $(\add)$. That is, we have $\clte\vdash H$, where $H$ is the result of replacing in
$E$ a  semisurface occurrence of a subformula $G =G_1\add\ldots\add G_n$ by $G_i$ for some $i\in\{1,\ldots,n\}$. Fix these formulas and this number $i$. 
Just as in CASE 3 (statement (\ref{may25})), based on the induction hypothesis, we have:
\begin{equation}\label{may26}
\mbox{\em If $H$ is good, then $\clt\vdash\lfloor H\rfloor$.}
\end{equation}

We need to consider the following three subcases that cover all possibilities:\vspace{5pt}

{\em Subcase 4.1:} $G$ is not a molecule. Reasoning (almost) exactly as we did when justifying (\ref{oct28a}), 
we find that $H$ is good. Therefore, by (\ref{may26}), $\clt\vdash\lfloor H\rfloor$. Now, a little thought can 
convince us that $\lfloor E\rfloor$ follows from $\lfloor H\rfloor$ by $(\add)$, so that $\clt\vdash\lfloor E\rfloor$.\vspace{3pt}

{\em Subcase 4.2:} $G$ is a negative large molecule $\gneg\check{P}_{\add}^{1}\add\ldots\add \gneg\check{P}_{\add}^{m}$. So, 
$G_i=\gneg\check{P}_{\add}^{i}$. A (now already routine for us) examination of Conditions (i)-(iv) reveals that 
each of these four conditions is inherited by $H$ from $E$, so that $H$ is good. Therefore, by (\ref{may26}), 
$\clt\vdash\lfloor H\rfloor$. Now, $\lfloor H\rfloor$ can be easily seen to be the same as $\lfloor E\rfloor$,
and thus $\clt\vdash\lfloor E\rfloor$.\vspace{3pt}

{\em Subcase 4.3:} $G$ is a positive medium molecule $\check{P}_{1}^{a}\add\ldots\add \check{P}_{m}^{a}$. So, $G_i=\check{P}_{i}^{a}$. There are two subsubcases to consider:

{\em Subsubcase 4.3.1:} $E$ contains no independent occurrence of $\check{P}_{i}^{a}$. One can easily verify that $H$ is good and that
$\lfloor H\rfloor =\lfloor E\rfloor$. By (\ref{may26}), we then get the desired $\clt\vdash\lfloor E\rfloor$.

{\em Subsubcase 4.3.2:} $E$ has an independent occurrence of $\check{P}_{i}^{a}$.  Since $E$ also has a positive independent occurrence of $\check{P}_{\add}^{a}$, 
Condition (iv) implies that the above occurrence of $\check{P}_{i}^{a}$ in $E$ is negative. This, in conjunction with Condition (iii), means that $E$ does not have any other independent occurrences of $\check{P}_{i}^{a}$, and thus $H$ has exactly two --- one negative and one positive --- independent occurrences of $\check{P}_{i}^{a}$. 
This guarantees that Condition (iii) is satisfied for $H$, because $H$ and $E$ only differ in that $H$ has $\check{P}_{i}^{a}$ where $E$ has $\check{P}_{\add}^{a}$. Conditions (i) and (ii) are straightforwardly inherited by $H$ from $E$. Finally, Condition (iv) also transfers from $E$ to $H$ because, even though $H$ --- unlike $E$ --- has a positive independent occurrence of $\check{P}_{i}^{a}$, it no longer has a positive independent occurrence of $\check{P}_{\add}^{a}$ (which, by the same Condition (iv) for $E$, was unique in $E$). Thus, $H$ is good and, by (\ref{may26}), $\clt\vdash\lfloor H\rfloor$. Note that since $H$ is good, by 
Condition (ii), both of the independent occurrences of $\check{P}_{i}^{a}$ in it 
are semisurface occurrences. The same, of course, is true for the corresponding occurrences of $\check{P}_{i}^{a}$ and 
$\check{P}_{\add}^{a}$ in $E$. Let us now compare $\lfloor E\rfloor$ with $\lfloor H\rfloor$. According to our earlier observation, $\check{P}_{i}^{a}$ only has one independent occurrence in $E$, i.e. $\check{P}_{i}^{a}$ is $E$-isolated. 
Hence the independent occurrence of $\check{P}_{i}^{a}$, just like that of $\check{P}_{\add}^{a}$, gets replaced by $P$ when floorifying $E$. On the other hand, $\check{P}_{i}^{a}$ is no longer isolated in $H$, so the two 
independent occurrences of it stay as they are when floorifying $H$. Based on this observation, we can easily see that the only difference between $\lfloor E\rfloor$ and $\lfloor H\rfloor$ is that 
$\lfloor E\rfloor$ has the general atom $P$ where $\lfloor H\rfloor$ has the (two occurrences of) elementary atom $\check{P}_{i}^{a}$. Since 
 $\lfloor E\rfloor$ does not contain $\check{P}_{i}^{a}$ (because the only independent occurrence of it in $E$, as well as all  large and medium $P$-based molecules, got replaced by $P$ when floorifying $E$), and since we are talking about two --- one positive and one negative --- semisurface occurrences of $P$ in $\lfloor E\rfloor$, we find that $\lfloor E\rfloor$ follows
from $\lfloor H\rfloor$ by (M). We already know that $\clt\vdash\lfloor H\rfloor$. Hence 
 $\clt\vdash\lfloor E\rfloor$.\vspace{5pt}

{\em CASE 5:} $E$ is derived by $(\hspace{-2pt}\sqd\hspace{-2pt})$. That is, we have $\clte\vdash H$, where $H$ is the result of replacing in
$E$ a  semisurface occurrence of a subformula \(G_1\sqd G_2\sqd  \ldots\sqd  G_k\) by 
\(G_2\sqd \ldots \sqd G_k.\)
Just as in CASES 3 and 4, based on the induction hypothesis, we have:
\begin{equation}\label{may264}
\mbox{\em If $H$ is good, then $\clt\vdash\lfloor H\rfloor$.}
\end{equation}
Reasoning as in the previous cases, we further find that  
 $H$ is good, and thus,  by (\ref{may264}), $\clt\vdash\lfloor H\rfloor$. Now, a moment's thought convinces us that $\lfloor E\rfloor$ follows from $\lfloor H\rfloor$ by $(\hspace{-2pt}\sqd\hspace{-2pt})$, so that $\clt\vdash\lfloor E\rfloor$.
\end{proof}

Now we are very close to finishing our completeness proof for $\clt$. Assume $\clt\not\vdash F$. 
Let $\lceil F\rceil$ be the result of replacing in $F$ all occurrences of each general atom $P\in{\cal P}$ by $\check{P}_{\add}^{\adc}$. Obviously $\lceil F\rceil$ is good. Clearly we also have $\lfloor\lceil 
F\rceil\rfloor=F$, so that $\clt\not\vdash \lfloor\lceil 
F\rceil\rfloor$. Therefore, by Lemma \ref{april11}, $\clte\not \vdash\lceil F\rceil$. Hence, by 
Lemma \ref{oct5}, there is an interpretation $^\dagger$ that interprets every elementary atom as a finitary predicate 
of arithmetical complexity $\Delta_3$, such that 
\begin{equation}\label{K}
\mbox{\em $\lceil F\rceil^\dagger$ is not computable.}
\end{equation} 

Let $^*$ be an interpretation such that:
\begin{itemize}
\item $^*$ agrees with $^\dagger$ on all elementary atoms;
\item $^*$ interprets each general atom $P\in{\cal P}$ as $(\check{P}_{\add}^{\adc})^{\dagger}$.\vspace{2pt}
\end{itemize}
 
Clearly $^*$ interprets atoms as promised in Theorem \ref{thcl2}. It is also obvious that 
$F^* = \lceil F\rceil^\dagger$. Therefore, by 
(\ref{K}), \ $F^*$ is not computable. The completeness part of Theorem \ref{thcl2} is now proven.

\appendix

\section{Proof of Theorem \ref{static}}

\begin{lemma}\label{l14} \ 

1. Assume $A_1,\ldots,A_n$ ($n\geq 2$) are constant static games, $\Omega$ is a $\xx$-delay of $\Gamma$, and $\Omega$ is a $\xx$-illegal run of $A_1\tgc\ldots\tgc A_n$. Then $\Gamma$ is also a $\xx$-illegal run of $A_1\tgc\ldots\tgc A_n$.

2. Similarly for $A_1\sqc \ldots \sqc A_n$.

3. Similarly for $A_1\tgc A_2\tgc A_3\tgc\ldots$ and $A_1\sqc A_2\sqc A_3\sqc\ldots$.
\end{lemma}
\begin{proof} We will prove this lemma by induction on the length of the shortest illegal initial segment 
of $\Omega$. 

{\em CLAUSE 1}. Assume the conditions of clause 1 of the lemma. We want to show that $\Gamma$ is a $\xx$-illegal run of $A_1\tgc\ldots\tgc A_n$. Let $\seq{\Psi,\xx\alpha}$ be the shortest ($\xx$-) illegal initial segment of $\Omega$. Let $\seq{\Phi,\xx\alpha}$ be the shortest initial segment of $\Gamma$ containing all  $\xx$-labeled moves\footnote{In this context, different occurrences of the same labmove count as different labmoves. So, a more accurate phrasing would be ``as many $\xx$-labeled moves as...'' instead ``all the $\xx$-labeled moves of ...''.} of $\seq{\Psi,\xx\alpha}$. If $\Phi$ is a $\xx$-illegal position of $A_1\tgc\ldots\tgc A_n$, then so is $\Gamma$ and we are done. Therefore, for the rest of the proof, we assume that 
\begin{equation}\label{654}
\mbox{\em $\Phi$ is not a $\xx$-illegal position of $A_1\tgc\ldots\tgc A_n$.}
\end{equation}

Let $\Theta$ be the sequence of those $\pneg\xx$-labeled moves of $\Psi$ that are not in $\Phi$. Obviously
\begin{equation}\label{e141}
\mbox{\em  $\seq{\Psi,\xx\alpha}$ is a $\xx$-delay of $\seq{\Phi,\xx\alpha,\Theta}$.}
\end{equation}
We also claim that
\begin{equation}\label{e142}
\mbox{\em $\Phi$ is a legal position of $A_1\tgc\ldots\tgc A_n$.}
\end{equation}
Indeed, suppose this was not the case. Then, by (\ref{654}), $\Phi$ should be $\pneg\xx$-illegal. This would make $\Gamma$  a $\pneg\xx$-illegal run of $A_1\tgc\ldots\tgc A_n$ with $\Phi$ as an illegal initial segment which is shorter than $\seq{\Psi,\xx\alpha}$. Then, by the induction hypothesis, any run for which $\Gamma$ is a $\pneg\xx$-delay, would be $\pneg\xx$-illegal. But, as observed in Lemma 4.6 of \cite{Jap03}, the fact that $\Omega$ is a $\xx$-delay of $\Gamma$ implies that $\Gamma$ is a $\pneg\xx$-delay of $\Omega$. So, $\Omega$ would be $\pneg\xx$-illegal, which is a contradiction because, according to our assumption, $\Omega$ is $\xx$-illegal.

We are continuing our proof. There are three possible reasons to why $\seq{\Psi,\xx\alpha}$ is an illegal (while 
$\Psi$ being legal)  position of $A_1\tgc\ldots\tgc A_n$:

{\em Reason 1}: $\alpha$ does not have the form $i$ or $i.\beta$ for some $i\in\{1,\ldots,n\}$. Then, in view of (\ref{e142}), 
$\seq{\Phi,\xx\alpha}$ is a $\xx$-illegal position of $A_1\tgc \ldots\tgc A_n$. As $\seq{\Phi,\xx\alpha}$ happens to be an initial segment of $\Gamma$, the latter then is a $\xx$-illegal run of $A_1\tgc\ldots\tgc A_n$.

{\em Reason 2}: $\alpha=i$ for some $i\in\{1,\ldots,n\}$, but  $\xx=\pp$. With (\ref{e142}) in mind, $\seq{\Phi,\xx\alpha}$ can be seen to be a $\xx$-illegal position of $A_1\tgc \ldots\tgc A_n$.  Hence,    as $\seq{\Phi,\xx\alpha}$ is an initial segment of $\Gamma$, the latter is a $\xx$-illegal run of $A_1\tgc\ldots\tgc A_n$. 

{\em Reason 3}: $\alpha=i.\beta$ for some $i\in\{1,\ldots,n\}$, but $\seq{\Psi^{i.},\xx\beta}$ is not a legal run of $A_i$. That is,
$\seq{\Psi,\xx\alpha}^{i.}$ is a $\xx$-illegal position of $A_i$. (\ref{e141}) obviously implies that 
 $\seq{\Psi,\xx\alpha}^{i.}$ is a $\xx$-delay of $\seq{\Phi,\xx\alpha,\Theta}^{i.}$. Therefore, since 
$A_i$ is static, clause 1 of Lemma \ref{may19} yields that $\seq{\Phi,\xx\alpha,\Theta}^{i.}$ is a $\xx$-illegal position of $A_i$. Notice that $\seq{\Phi,\xx\alpha,\Theta}^{i.}=\seq{\Phi^{i.},\xx\beta,\Theta^{i.}}$.
A $\xx$-illegal position will remain $\xx$-illegal after removing a block of $\pneg\xx$-labeled moves (in particular, $\Theta^{i.}$) at the end of it. Hence,  $\seq{\Phi^{i.},\xx\beta}$ is a $\xx$-illegal position of $A_i$. In view of (\ref{e142}), this implies that  $\seq{\Phi, \xx \alpha=\xx .\beta}$ is not a legal position of $A_1\tgc \ldots\tgc A_n$, so that  
$\seq{\Phi,\xx\alpha}$ is a $\xx$-illegal position of $A_1\tgc\ldots\tgc A_n$, and then so is $\Gamma$ because $\seq{\Phi,\xx\alpha}$  is an initial segment of it.

{\em CLAUSE 2}. The reasoning here is the same as in the proof of clause 1, with the only difference that one more possible reason should be considered 
to why $\seq{\Psi,\xx\alpha}$ is an illegal (while 
$\Psi$ being legal)  position of $A_1\sqc\ldots\sqc A_n$:

{\em Reason 4}: $\alpha=i$ for some $i\in\{1,\ldots,n\}$ and $\xx=\oo$, but either (1) $\Psi$ does not contain switch moves and $i\not= 2$, or (2) the last switch move of $\Phi$ is not $i-1$. In either case, with (\ref{e142}) in mind, $\seq{\Phi,\xx\alpha}$ can be seen to be a $\xx$-illegal position of $A_1\tgc \ldots\tgc A_n$.  Hence,    as $\seq{\Phi,\xx\alpha}$ is an initial segment of $\Gamma$, the latter is a $\xx$-illegal run of $A_1\tgc\ldots\tgc A_n$. 

{\em CLAUSE 3}. The reasoning here is virtually the same as in (the corresponding) clause 1 or clause 2, with only ``$i\in\{1,2,\ldots\}$'' instead of $i\in\{1,\ldots,n\}$''.
\end{proof}

Since $\tgd$ and $\sqd$ are expressible in terms of $\tgc$\hspace{-2pt}, $\sqc$ and $\gneg$ (with $\gneg$ already known to preserve the static property), in order to prove Theorem \ref{static}, considering only $\tgc$ and $\sqc$ is sufficient. For simplicity, here we restrict ourselves to the $n$-ary case of $\tgc$\hspace{-2pt}. Adapting our argument to the $n$-ary case of $\sqc$\hspace{-2pt}, or the infinite cases of $\tgc$ and $\sqc$\hspace{-2pt}, does not present any problem. 

Assume  $A_1,\ldots,A_n$ are static constant games, $\Gamma$ is a $\xx$-won run of $A_1\tgc\ldots\tgc A_n$, and $\Omega$ is a $\xx$-delay of $\Gamma$. Our goal is to show that $\Omega$ is also a $\xx$-won run of $A_1\tgc\ldots\tgc A_n$. 

If $\Omega$ is a $\pneg\xx$-illegal run of $A_1\tgc\ldots\tgc A_n$, then it is won by $\xx$ and we are done. So, assume that 
$\Omega$ is not $\pneg\xx$-illegal. According to the earlier mentioned Lemma 4.6 of \cite{Jap03}, if $\Omega$ is a $\xx$-delay of $\Gamma$, then $\Gamma$ is a $\gneg\xx$-delay of $\Omega$. So, by Lemma \ref{l14}, our $\Gamma$ cannot be $\pneg\xx$-illegal, for otherwise so would be $\Omega$. 
$\Gamma$ also cannot be $\xx$-illegal, because otherwise it would not be won by $\xx$. Consequently, $\Omega$ cannot be $\xx$-illegal either, for otherwise, by Lemma \ref{l14}, $\Gamma$ would be $\xx$-illegal. Thus, we have narrowed down our considerations to the case when both $\Gamma$ and $\Omega$ are legal runs of $A_1\tgc\ldots\tgc A_n$.

The fact that $\Gamma$ is a legal, $\xx$-won run of $A_1\tgc\ldots\tgc A_n$ implies that, where $A_i$ is the active (last-chosen) component of $A_1\tgc\ldots\tgc A_n$ in $\Gamma$,   
$\Gamma^{i.}$ is a $\xx$-won run of $A_i$. Taking into account that $\Omega^{i.}$ is obviously a $\xx$-delay of $\Gamma^{i.}$ and that $A_i$ is 
 static, the above, in turn, implies that $\Omega^{i.}$ is a $\xx$-won run of $A_i$, which, taking into account that $\Omega$ is a legal run of  $A_1\tgc\ldots\tgc A_n$ and that the active component of the latter in $\Omega$ is obviously the same $A_i$,  means nothing but that $\Omega$ is a $\xx$-won run of $A_1\tgc\ldots\tgc A_n$.

\end{document}